\documentclass[fleqn,usenatbib]{mnras}

\usepackage{newtxtext,newtxmath}

\usepackage[T1]{fontenc}
\usepackage{xspace}

\DeclareRobustCommand{\VAN}[3]{#2}
\let\VANthebibliography\thebibliography
\def\thebibliography{\DeclareRobustCommand{\VAN}[3]{##3}\VANthebibliography}

\newcommand{\COBEF}{{\it COBE/FIRAS}\xspace}
\newcommand{\PIXIE}{{\it PIXIE}\xspace}

\usepackage{graphicx}	% Including figure files
\usepackage{amsmath}	% Advanced maths commands

\usepackage[english]{babel}
\usepackage{graphicx}
\usepackage{latexsym}
\usepackage{floatflt}
\usepackage{float}
\usepackage{units}
\usepackage{hyperref}
\usepackage{amsmath}
\usepackage{caption}
\usepackage{multicol}
\usepackage{gensymb}
\usepackage{mathdots}
\usepackage{color}
\usepackage{xcolor}
\usepackage{mathtools}
\usepackage{lipsum}

\newcommand{\id}{{\rm d}}

\title[Spectral constraints on cosmic strings]{Constraints on the spectral signatures of superconducting cosmic strings}

\author[B. Cyr et al.]{
Bryce Cyr$^{1}$\thanks{E-mail: bryce.cyr@manchester.ac.uk},
Jens Chluba$^{1}$ %\thanks{E-mail: jens.chluba@manchester.ac.uk}
and
Sandeep Kumar Acharya$^{1}$
\\
$^{1}$Jodrell Bank Centre for Astrophysics, School of Physics and Astronomy, The University of Manchester, Manchester M13 9PL, U.K.
}

\date{Accepted XXX. Received YYY; in original form ZZZ}

\pubyear{2023}

\begin{document}
\label{firstpage}
\pagerange{\pageref{firstpage}--\pageref{lastpage}}
\maketitle

\begin{abstract}
If they exist, networks of superconducting cosmic strings are capable of injecting copious amounts of electromagnetic energy into the background over a broad range of frequencies. We study this injection both analytically, as well as numerically using the thermalization code \texttt{CosmoTherm}. With our refined analytic formalism, we update constraints from CMB spectral distortions by following the injection of entropy, as well as energy, on the amplitude of the $\mu$-distortion, leading to a significant improvement in those limits. Furthermore, we utilize the full shape of the distorted spectrum from \texttt{CosmoTherm} to include constraints from non-$\mu$, non-$y$ type distortions. Additionally, we use the outputs for the ionization history and global 21cm signal to derive and update constraints on string model parameters using measurements from other datasets. Analysis of CMB anisotropies provides the most stringent constraints, though with a slightly modified shape and strength when compared to previous results. Modifications of the reionization history provide new bounds in the high current domain, and we also find that the observations of the low-frequency radio background probe a small region of parameter space not explored by other datasets. We also analyze global $21$-cm constraints, and find that the inclusion of soft photon heating plays a crucial role, essentially removing any constraints in the considered parameter domain. Spectral distortion measurements from \COBEF are covered by other constraints, but our conservative forecast shows that a \PIXIE-type satellite would probe important unexplored regions of parameter space.
\end{abstract}


%%%%%%%%%%%%%%%%%%%%%%%%%%%%%%%%%%%%%%%%%%%%%%%%%%

%%%%%%%%%%%%%%%%% BODY OF PAPER %%%%%%%%%%%%%%%%%%

%--------------------------------------------
\section{Introduction}
%--------------------------------------------
We are well into an age of precision cosmology, with detailed observations being taken over much of the electromagnetic spectrum. At radio frequencies, an anomalous background is being uncovered \citep{ARCADE2011, DT2018} with no known astrophysical source. Meanwhile, data from the epoch of reionization is slowly building up, with a first claimed detection of the differential brightness temperature at cosmic dawn coming from the EDGES experiment \citep{Edges2018}.  We also have exquisite measurements of the cosmic microwave background (CMB) anisotropies from the Planck satellite \citep{Planck2018params}, as well as the frequency spectrum from \COBEF \citep{Fixsen}, which teaches us still further about our thermal history. One of the aims of these cosmological observations is to help us understand our origins. Here, we will consider how the spectral signatures of a network of superconducting cosmic strings can be probed using these observations.

Cosmic strings are a class of topological defect that may form at the interface of cosmological phase transitions if the true vacuum manifold is degenerate, and not simply connected \citep[see][for comprehensive reviews]{CSreviews3, CSreviews2, CSreviews1}. If they form, cosmic strings (also known as line defects) are nearly one dimensional objects with a small, but finite width. The interior of a string resembles the state of the universe as it was in the false vacuum, consisting of a condensate of scalar and gauge particles from this previously unbroken phase. The detection (or non-detection) of cosmic strings in the various observations give us pieces of information about our thermal history that would otherwise be very challenging to infer.

String models exhibit a property known as \textit{scaling}, where the macroscopic properties of the string network can be described by one parameter. This parameter is known as the string tension ($G\mu$), and is related to the energy scale of the phase transition ($\eta$) through $G\mu \simeq G \eta^2$, where $G$ is Newton's gravitational constant. It has also been shown that some symmetry breaking patterns can imbue the strings with superconductive properties, leading to the generation of significant currents, $\mathcal{I}$, as the string traverses through the plasma \citep[see][for seminal examples]{Witten, OstWitten}. Superconducting cosmic string models are typically described by these two parameters, $G\mu$ and $\mathcal{I}$.

As we will discuss below, a cosmic string network generally consists of a small number of long strings which run through our Hubble patch at any given time, as well as a distribution of smaller loops with curvature radii on all scales up to some $\mathcal{O}(0.1)$ fraction of the Hubble scale. Although these string loops act as seeds for density perturbations at early times, detailed observations of the microwave background have relegated them to only being a highly subdominant component of structure formation \citep{cmbBound13, cmbBound11,cmbBound12,Albrecht1997}, requiring $G\mu \lesssim 10^{-7}$. Even so, there are hints that the enhanced gravitational effects of string loops could play a role in the formation of supermassive black holes \citep{jerome, Cyr2022}.

Superconducting loops are capable of producing copious amounts of gravitational \citep{Vachaspati1984} and electromagnetic radiation \citep{Vilenkin1986, Garfinkle1987, Vachaspati2008} during every epoch after their formation. This has led to constraints on the $G\mu$-$\mathcal{I}$ parameter space from a number of observations, including primordial CMB spectral distortions \citep{Ostriker1987, Tashiro}, anisotropies \citep{TashiroIonization}, radio transient events \citep{Cai, Miyamoto}, gamma ray bursts \citep{Babul1987}, changes in the differential brightness temperature at cosmic dawn \citep{Brandenberger2018, Brandenberger2019}, and more. 

In this work, we revisit some of the constraints considered by these authors using improved analytic estimates and the thermalization code {\tt CosmoTherm} \citep{Chluba2011therm}. First, we recompute the spectral distortion signature obtained by following an approximate Green's function approach, as described in \citet{Chluba2016}. Our analytic approach improves upon the previous treatments in at least two ways. First, we compute the negative $\mu$-distortion generated by direct photon (entropy) injection into the pre-recombination plasma \citep{Chluba2015}, an effect that has been overlooked thus far but significantly alters the overall constraints. Second, we include a more precise treatment for the determination of the instantaneous spectrum of emitted photons. 

Following this, we implement the photon injection numerically into the thermalization code \texttt{CosmoTherm}. This allows us to further strengthen our constraints by analyzing the full shape of the string induced spectral distortions, accounting for non-$\mu$, non-$y$ type deviations from the \COBEF measurement \citep{Fixsen}. This implementation also allows us to efficiently compare against other independent datasets. Using the outputs from \texttt{CosmoTherm}, we generate constraints from the CMB anisotropies \citep{Komatsu2010, Planck2018params}, the radio synchrotron background (RSB) \citep{ARCADE2011, DT2018}, the EDGES experiment \citep{Edges2018}, and the optical depth to reionization as measured by the \citet{Planck2018params}. We also generate a conservative forecast to a \PIXIE-type experiment \citep{Kogut2011PIXIE, Kogut2016SPIE, Chluba2021Voyage}, assuming a fiducial sensitivity to energy release of $\Delta \rho/\rho=10^{-8}$ after foreground marginalization. 

While the constraints we find from \COBEF are less stringent than other datasets, it is important to stress that in principle, we have had access to this spectral distortion data since the late 90s. Had a more sophisticated analysis of the full shape of the \COBEF data been possible at that time, the constraints that were derived would have been dominant for many years. This highlights the legacy value that \COBEF still has today, and showcases the constraining power that a next generation space-based spectrometer such as \PIXIE could obtain for exotic energy injection scenarios.

The rest of the paper is organized as follows: in Section \ref{sec:loopNumber}, we describe the loop distribution model that we implement, and discuss the microphysics of gravitational and electromagnetic wave production. Section \ref{sec:analyticEnergy} discusses energy release rates for the string network, and reviews the simple estimates for the $\mu$ and $y$ parameters. Afterwards, in Section \ref{sec:analyticEntropy}, we refine this estimate by including entropy release contributions, which changes the analytic constraint curve significantly. Section \ref{sec:numericalImplementation} describes how to implement this source term into \texttt{CosmoTherm}, and derives a useful expression for the instantaneous injection spectrum. In Sections \ref{sec:numericalResults} and \ref{sec:constraints}, we discuss the output from \texttt{CosmoTherm} and illustrate the constraints we derive from multiple different datasets. We conclude in Section \ref{sec:conclusions}.

Throughout the analytic derivations, we use natural units where $\hbar = c = k_{\rm b} = 1$. When presenting the resultant spectra in Sections \ref{sec:numericalResults} and onwards, we use more astrophysicist-friendly units. To minimize confusion, we include a particle physics to astrophysics conversion dictionary in Appendix~\ref{sec:dictionary}.

%--------------------------------------------
\section{Number density of cosmic string loops} \label{sec:loopNumber}
%--------------------------------------------
If the universe undergoes a phase transition in which the true vacuum manifold permits cosmic string solutions, the Kibble mechanism states that a network of these defects will form \citep{Kibble1, Kibble2}. Importantly, Kibble's mechanism only guarantees the existence of long cosmic strings, with curvature radius larger than the horizon scale. A second population of smaller, sub-horizon loops is populated by the intersections and self-intersections of the long strings.

Numerical simulations in which the widths of cosmic strings are neglected (known as Nambu-Goto simulations) show that the distribution of sufficiently large loops follow a scaling solution \citep{NGsim2, NGsim3,NGsim1, NGsim4, NGsim5}. We should note that another class of simulations (the so-called Abelian-Higgs type) use field-theoretic input to resolve the cores of these strings, and observe no significant loop production \citep{AH1,AH2,AH3,AH4}. This has sparked numerous debates about the true nature of the string network on small scales, with no consensus being reached as of yet.

One should not understate the numerical challenges that arise when attempting to perform cosmological simulations over such a wide range of scales, from the string core to the horizon at a given time. For this work, we assume that a scaling loop distribution is formed as indicated by the Nambu-Goto simulations.

%--------------------------------------------
\subsection{Formation in the Radiation Era}
%--------------------------------------------
We begin by examining the distribution of loops formed before matter-radiation equality. In most regions of the $G\mu$-$\mathcal{I}$ parameter space, they are the dominant source of primordial spectral distortions, which makes them a useful case to study. Focusing now on the results from Nambu-Goto simulations, one finds that the differential number density of loops (in physical coordinates) with initial length $L_{\textrm{i}}$ is given by
%%%%%%%%%%%%%%%%%%%%%%%%%%%%%%%%%%%%%%%%%%%%%%%%%%
\begin{align}
\label{eq:dNdLi}
\frac{\id N}{\id L_{\textrm{i}}} = \begin{dcases} \frac{\alpha}{t^{4-p} L_{\textrm{i}}^{p}} \hspace{22mm} (t\leq t_{\textrm{eq}})\\
\frac{\alpha}{t_{\textrm{eq}}^{4-p} L_{\textrm{i}}^{p}} \left(\frac{t_{\textrm{eq}}}{t}\right)^2 \hspace{12.5mm} (t > t_{\textrm{eq}}).\end{dcases}
\end{align} 
%%%%%%%%%%%%%%%%%%%%%%%%%%%%%%%%%%%%%%%%%%%%%%%%%%
Simulations performed in \citet{NGsim5} yield $\alpha = 0.18$ and $p = 5/2$. For $t > t_{\textrm{eq}}$ we simply redshift the loops which formed during the radiation era, and discuss the formation and evolution of matter-dominated loops in the next subsection.

Simulations indicate that at a given time, $t$, most string loops are formed with roughly the same initial radius, given by some fraction $\beta\approx \mathcal{O}(0.1)$ of the Hubble length, i.e., $L_{\textrm{i}}(t) \approx \beta t$. After formation, the loops oscillate with period $T \approx L$, and develop transient substructures known as cusps and kinks. Their decay (in particular, the cusps) releases energy in the form of gravitational waves, photons, and exotic particles, causing the loop to shrink in size. All loops radiate gravitational waves with an oscillation-averaged power \citep{Vachaspati1984}
%%%%%%%%%%%%%%%%%%%%%%%%%%%%%%%%%%%%%%%%%%%%%%%%%%
\begin{align}
P_{\textrm{g} } \simeq \Gamma_{\textrm{g}} G \mu^2
\simeq 
1.5 \times 10^{18}\,\left[ \frac{\Gamma_{\textrm{g}}}{100}\right]
\left[ \frac{G\mu }{10^{-11}}\right]^{2} \, \textrm{GeV}^2,
\end{align}
%%%%%%%%%%%%%%%%%%%%%%%%%%%%%%%%%%%%%%%%%%%%%%%%%%
where $\Gamma_{\textrm{g}}$ is a normalization factor $\simeq \mathcal{O}(100)$.
In some symmetry breaking schemes, cosmic strings can acquire an electromagnetic current, $\mathcal{I}$, and are said to be superconducting \citep[see][for a seminal example]{Witten}. 

A proper treatment of the current generation and dissipation on a cosmological network of loops is beyond the scope of this work, and so for simplicity we assume that all string loops carry the same time-independent current. Superconducting string loops also generate sizeable electromagnetic bursts at their cusps. This leads to an additional decay channel into photons \citep{Vilenkin1986,Cai}, with
%%%%%%%%%%%%%%%%%%%%%%%%%%%%%%%%%%%%%%%%%%%%%%%%%%
\begin{align} \label{TotPow}
P_{\gamma} &\simeq \Gamma_{\gamma} \mathcal{I} \mu^{1/2} \nonumber \\
&\simeq 
3.8 \times 10^{18}\,\left[ \frac{\Gamma_\gamma}{10}\right]
\left[ \frac{G\mu }{10^{-11}}\right]^{1/2}
\left[ \frac{\mathcal{I}}{10^4\,{\rm GeV}}\right] \, \textrm{GeV}^2,
\end{align}
%%%%%%%%%%%%%%%%%%%%%%%%%%%%%%%%%%%%%%%%%%%%%%%%%%
where $\Gamma_{\gamma} \simeq \mathcal{O}(10)$ depends on the precise geometry of a loop. Superconducting strings gradually decay through these two main channels, with a total rate that is determined by
%%%%%%%%%%%%%%%%%%%%%%%%%%%%%%%%%%%%%%%%%%%%%%%%%%
\begin{align}
\Gamma\,G \mu&\simeq(P_{\textrm{g} }+P_\gamma)/\mu \simeq  \Gamma_{\textrm{g}} G \mu + \Gamma_{\gamma} \mathcal{I} \mu^{-1/2}.
\end{align}
%%%%%%%%%%%%%%%%%%%%%%%%%%%%%%%%%%%%%%%%%%%%%%%%%%
By comparing the power emitted in both gravitational and electromagnetic waves, we can define a critical current,
%%%%%%%%%%%%%%%%%%%%%%%%%%%%%%%%%%%%%%%%%%%%%%%%%%
\begin{align} 
\mathcal{I}_* = \frac{\Gamma_{\textrm{g}}}{ \Gamma_{\gamma}} G \mu^{3/2}\simeq 
3.2 \times 10^3\,{\rm GeV}\,\left[ \frac{\Gamma_{\textrm{g}}}{100}\right]
\,\left[ \frac{\Gamma_\gamma}{10}\right]^{-1}
\left[ \frac{G\mu }{10^{-11}}\right]^{3/2}.
\end{align}
%%%%%%%%%%%%%%%%%%%%%%%%%%%%%%%%%%%%%%%%%%%%%%%%%%
For a given string tension, $G\mu$, loops decay primarily into gravitational waves if $\mathcal{I} < \mathcal{I}_*$, or into photons when $\mathcal{I} \geq \mathcal{I}_*$. 
The dimensionless decay coefficient $\Gamma$ can then be expressed as
%%%%%%%%%%%%%%%%%%%%%%%%%%%%%%%%%%%%%%%%%%%%%%%%%%
\begin{align}
\Gamma = \Gamma_{\textrm{g}} \left( 1 + \frac{\mathcal{I}}{\mathcal{I}_*}\right).
\end{align}
%%%%%%%%%%%%%%%%%%%%%%%%%%%%%%%%%%%%%%%%%%%%%%%%%%
We note that $\Gamma$ is a function of both the string tension and the current.

Given the decay coefficient $\Gamma$, the loop size then shrinks as
%%%%%%%%%%%%%%%%%%%%%%%%%%%%%%%%%%%%%%%%%%%%%%%%%%
\begin{align} \label{GWdecay}
L = L_{\textrm{i}} - 
\Gamma\,G \mu(t-t_{\textrm{i}}).
%\left( \Gamma_{\textrm{g}} G\mu + \frac{ G^{1/2}\Gamma_{\gamma} \mathcal{I}}{(G\mu)^{1/2}}\right)(t-t_{\textrm{i}})
\end{align}
%%%%%%%%%%%%%%%%%%%%%%%%%%%%%%%%%%%%%%%%%%%%%%%%%%
Noting that $t_{\textrm{i}}=L_{\textrm{i}}/\beta \leq t_{\rm eq}$, we then have $L_{\textrm{i}}=(L+\Gamma\,G \mu\,t)/(1+\lambda)$, where $\lambda=\Gamma G\mu/\beta$ is the decay-rapidity parameter, a measure of how long after formation a given loop will exist before complete evaporation. 
Low values of $\lambda$ describe long-lived loops, while for $\lambda > 1/\beta$ the loops decay within one oscillation, and the expressions for the oscillation-averaged power should be re-examined. 
A loop forming at $t_{\rm i}$ decays at cosmic time $t_{\rm decay}$ given by
%%%%%%%%%%%%%%%%%%%%%%%%%%%%%%%%%%%%%%%%%%%%%%%%%%
\begin{align}
\label{eq:tdecay}
t_{\rm decay} = \left( \frac{1+\lambda}{\lambda}\right) t_i.
\end{align}
%%%%%%%%%%%%%%%%%%%%%%%%%%%%%%%%%%%%%%%%%%%%%%%%%%
From this, it is easy to derive that the total lifetime of any given loop is $t_{\rm lifetime} = t_{\rm decay} - t_{\rm i} =  t_{\rm i}/\lambda$. At any given time, the initial loop lengths that contribute to background injections are $L_{\rm i, min} \leq L \leq \beta t_{\rm i}$, where $L_{\rm i, min}=\beta \lambda t /(1+\lambda)<\beta t$. As $\lambda$ increases we find that $L_{\rm i, min}\rightarrow \beta t$, such that the largest initial loops vanish rapidly.

%meaning that loops of the corresponding initial length $L_{\textrm{i}}$ have dissolved and hence cannot contribute to the energy release anymore. %This also means that at any given time, the minimal initial loop length that contributes is $L_{\rm i, min}=\beta \lambda t /(1+\lambda)<\beta %t$. As $\lambda$ increases we find that $L_{\rm i, min}\rightarrow \beta t$, \changeJ{such} that the largest initial loops vanish rapidly.

For typical choices of our parameters, $G\mu$ and $\mathcal{I}$, the loops are long-lived, implying $t_{\rm decay}\gg t_{\textrm{i}}$, although we will also discuss more extreme cases where fast loop decay is possible.
Using $L_{\textrm{i}}=(L+\Gamma\,G \mu\,t)/(1+\lambda)$, with Eq.~\eqref{eq:dNdLi} we can then write
%%%%%%%%%%%%%%%%%%%%%%%%%%%%%%%%%%%%%%%%%%%%%%%%%%
\begin{align}
\label{eq:dNdLradiation}
\left.\frac{\id N_{\rm{loops}}}{\id L}\right|_{\rm r} = \frac{\alpha\left( 1 + \lambda\right)^{3/2}}{t^{3/2}(L+\Gamma G\mu t)^{5/2}} \times \begin{cases} 1 &(t \leq t_{\textrm{eq}})
\\
\left(\frac{t_{\textrm{eq}}}{t}\right)^{1/2} &(t>t_{\textrm{eq}}),
\end{cases}
\end{align}
%%%%%%%%%%%%%%%%%%%%%%%%%%%%%%%%%%%%%%%%%%%%%%%%%%
where the distribution is defined for $0\leq L \leq L_{\rm max}(t)$. At $t\leq t_{\rm eq}$, one has $L_{\rm max}(t)=\beta t$. However, for $t>t_{\rm eq}$, the loops last sourced at $t=t_{\rm eq}$ have shrunken to $L_{\rm max}(t)=\beta  t_{\rm eq}[1+\lambda-\lambda t/t_{\rm eq}]<\beta t_{\rm eq}$ at time $t$. %
This implies that the last loops sourced at $t_{\rm eq}$ only exists at $t\leq t_{\rm end}\equiv t_{\rm eq}(1+\lambda)/\lambda$.
Figure~\ref{fig:EvapRegions} highlights the region of parameter space which undergoes rapid decays (i.e. $\lambda \geq 1/\beta$).
For reference we also show the line for which $t_{\rm end}\simeq t_0$ (i.e., $z_{\rm end}=0$) is equal to the age of the universe, $t_0$, or $\lambda\simeq 4\times 10^{-6}$.

%--------------------------------------------
\subsection{Formation in the Matter era}
%--------------------------------------------
For loops formed in the matter-dominated era, a very similar picture can be developed. Simulations show that these evolve according to 
%%%%%%%%%%%%%%%%%%%%%%%%%%%%%%%%%%%%%%%%%%%%%%%%%%
\begin{align}
\label{eq:dNdLiM}
\frac{\id N}{\id L_{\textrm{i}}} \simeq \frac{\alpha_{\rm m}}{t^2 L_{\textrm{i}}^{2}}.
\end{align} 
%%%%%%%%%%%%%%%%%%%%%%%%%%%%%%%%%%%%%%%%%%%%%%%%%%
We will assume that the sourcing scale in the matter-dominated era is the same as in the radiation dominated era, i.e., $\beta_{\rm m} t\simeq\beta\,t$.
Then, at $t$ the {\it smallest} loop scale is $L_{\rm min}(t)=\beta[t_{\rm eq} - \lambda(t-t_{\rm eq})]$. 
This implies two regimes: at $t_{\rm eq}\leq t\leq t_{\rm end}=t_{\rm eq}(1+\lambda)/\lambda$ the minimal loop length fulfills 
$0\leq L_{\rm min}(t)\leq \beta t_{\rm eq}$,
while at $t>t_{\rm end}$ one has 
$L_{\rm min}(t)= 0$.
The loop distribution is then defined at $L_{\rm min}(t)\leq L \leq \beta t$.  

To fix the normalization, $\alpha_{\rm m}$, we assume that at $t_{\rm eq}$ the radiation-dominated loop distribution at the sourcing scale is the same as the matter-dominated one. This gives the condition
%%%%%%%%%%%%%%%%%%%%%%%%%%%%%%%%%%%%%%%%%%%%%%%%%%
\begin{align}
\label{eq:dNdLi_condition}
\frac{\alpha_{\rm m}}{t_{\rm eq}^2 L_{\textrm{i,eq}}^{2}} = \frac{\alpha}{t_{\rm eq}^{3/2} L_{\textrm{i,eq}}^{5/2}} 
\quad \rightarrow \quad \alpha_{\rm m}=\alpha/\sqrt{\beta}\approx 0.57.
\end{align} 
%%%%%%%%%%%%%%%%%%%%%%%%%%%%%%%%%%%%%%%%%%%%%%%%%%
with $L_{\textrm{i,eq}}=\beta t_{\rm eq}$. With these expressions we then have
%%%%%%%%%%%%%%%%%%%%%%%%%%%%%%%%%%%%%%%%%%%%%%%%%%
\begin{align}
\label{eq:dNdLMatter}
\left.\frac{\id N_{\rm{loops}}}{\id L}\right|_{\rm m} = \frac{\alpha_{\rm m}\left( 1 + \lambda\right)}{t^{2}(L+\Gamma G\mu t)^{2}}
\end{align}
%%%%%%%%%%%%%%%%%%%%%%%%%%%%%%%%%%%%%%%%%%%%%%%%%%
at $L_{\rm min}(t)\leq L \leq \beta t$ with $L_{\rm min}(t)=0$ at $t>t_{\rm end}$. 

%--------------------------------------------
\section{Analytic Estimates - Energy Release}
\label{sec:analyticEnergy}
%--------------------------------------------
The injection of non-thermal photons into the microwave background before recombination can lead to distortions in the frequency spectrum as the plasma attempts to thermalize this new radiation. The time-dependence of photon injection determines the spectral shape of this distortion, with roughly three eras emerging \citep{Chluba2015}. At high redshifts ($z \gtrsim z_{\rm th}$, with $z_{\rm th} \approx 2 \times 10^6$), injections are fully thermalized through a combination of Bremsstrahlung, Compton, and double Compton scattering, producing an average (unobservable) temperature increase. At lower redshifts, number changing processes freeze out. However, repeated Compton scattering with the background electrons drives the CMB photons towards kinetic equilibrium with a spectrum described by the Bose-Einstein distribution with a small chemical potential ($\mu$) for $z\gtrsim 3\times 10^5$ \citep{Illarionov1975b, Burigana1991, Hu1993}. At redshifts lower than this, Compton scattering becomes less efficient and the distorted spectrum may be analytically described by a $y$-distortion \citep{Zeldovich1969}.

The \COBEF instrument remains state-of-the-art when it comes to upper bounds on these distortion parameters, providing $|\mu| < 9 \times 10^{-5}$ and $|y| <   1.5 \times 10^{-5}$ at $2\sigma$ \citep{Fixsen}. Some improvements have been discussed in \citet{tris2, Bianchini2022}; however, only in the future can we expect significant advances with {\it PIXIE} \citep{Kogut2011PIXIE, Kogut2016SPIE}, BISOU \citep{BISOU}, COSMO \citep{Masi2021}, TMS \citep{Jose2020TMS} or a spectrometer within the ESA Voyage 2050 space program \citep{Chluba2021Voyage}.

As a network of superconducting cosmic strings decay, they are capable of injecting both significant energy, and entropy, into the background radiation. Previously, \citet{Tashiro} computed analytic bounds on the string tension and current by considering the $\mu$ and $y$ distortions caused by a pure energy injection. As we will illustrate below, their analysis neglected a factor of $\Gamma^{-1/2}$, an omission which yielded more stringent constraints at high currents than we find here. Our $\mu$-distortion estimates agree better with those presented in \citet{Miyamoto}; however, we will show that the simple energy release arguments become inaccurate at late times and for high currents, requiring a more detailed treatment of the distortion evolution using {\tt CosmoTherm}. Additional simple analytic estimates for several astrophysical and cosmological observables can be found in \citet{Miyamoto}.

In the following sections, we will further refine these analytic distortion estimates by including the resultant entropy injection, yielding a novel shape for the constraints on $G\mu$ and $\mathcal{I}$. Later on, we validate these approximations by comparing against the full numerical solution, computed with the thermalization code  \texttt{CosmoTherm}. The numerical analysis allows us to go beyond the simple $\mu$ and $y$ estimates, by comparing the full spectral shape obtained with a cosmic string network to the residuals of the \COBEF experiment.

%------------------------------------------------------------
\begin{figure}
\centering 
\includegraphics[width=\columnwidth]{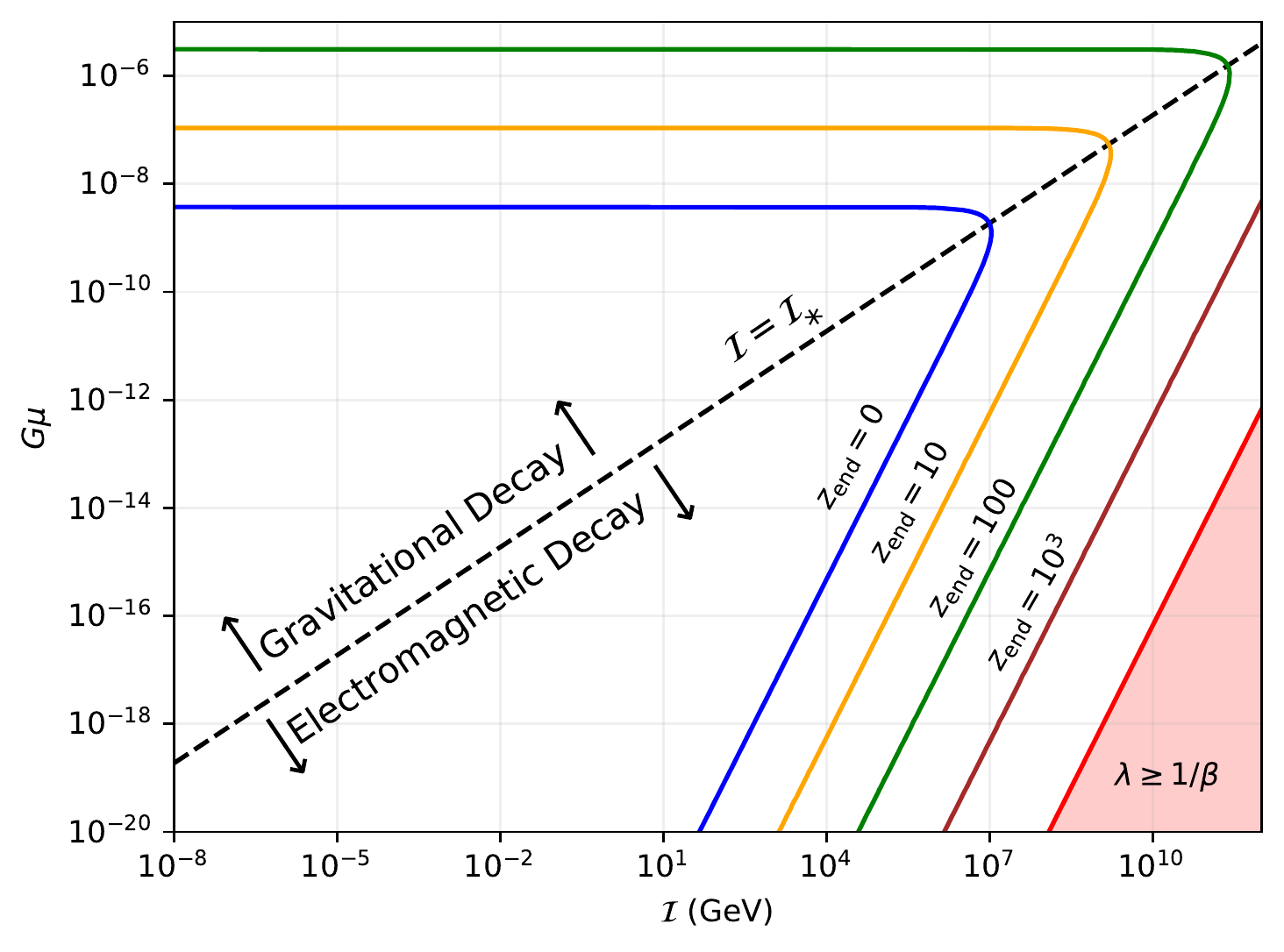}
\\
\vspace{-0mm}
\caption{Regions of parameter space affected by rapidly decaying loops. The red area corresponds to $\lambda \geq 1/\beta \approx 10$ in which loops fully evaporate within one loop oscillation time after their formation. The cusp evaporation formalism may break down in this red region, and so constraints given there should be taken with a grain of salt. The different contours correspond to the surfaces of constant $z_{\rm end}$ for a given parameter set. Regions to the right of, and above the contours represent parameter values whose radiation loops will have completely evaporated by the indicated $z_{\rm end}$.}
\label{fig:EvapRegions}
\end{figure}
%------------------------------------------------------------

%--------------------------------------------
\vspace{-2mm}
\subsection{Pure Energy Injection - Radiation Loops}
%--------------------------------------------
To obtain analytic approximations for the amplitude of these  distortion parameters, we apply a Green's function approach  \citep[method B of][]{Chluba2016} in which $\mu$ and $y$ are determined through 
%%%%%%%%%%%%%%%%%%%%%%%%%%%%%%%%%%%%%%%%%%%%%%%%%%
\begin{align} \label{DistApprox}
\mu &\approx  1.401 \int_{z_{\mu y}}^{\infty} \id z \frac{1}{\rho_{\gamma}} \frac{\id Q}{\id z} \mathcal{J}_{\rm bb}(z) \hspace{6mm} y \approx \frac{1}{4} \int^{z_{\mu y}}_{z_{\rm rec}} \id z \frac{1}{\rho_{\gamma}} \frac{\id Q}{\id z}.
\end{align}
%%%%%%%%%%%%%%%%%%%%%%%%%%%%%%%%%%%%%%%%%%%%%%%%%%
Here, we introduce the approximate distortion visibility function, $\mathcal{J}_{\rm bb}(z) \approx {\rm e}^{-(z/z_{\rm th})^{5/2}}$ with $z_{\rm th}\approx 2\times 10^6$, which models the smooth transition between the $\mu$ and thermalization eras. We approximate the transition from $\mu$ and $y$ eras by a simple step function at $z_{\mu y} = 5 \times 10^4$, the redshift below which energy redistribution becomes inefficient. 
Finally, $z_{\rm rec} \approx 10^3$ is the redshift of recombination, $\rho_{\gamma}$ is the energy density in the CMB at redshift $z$, and $\id Q/\id z$ is the rate of non-thermal energy (density) injection. 

While more accurate estimates for the transition between $\mu$ and $y$ could be obtained using various analytic methods \citep[e.g.,][]{Khatri2012mix, Chluba2013Green, Chluba2013PCA}, the precise inclusion of photon production processes necessitates full numerical treatments with the efficient scheme of {\tt CosmoTherm}.

For superconducting strings, the energy release rate is given by averaging the energy injection from one string over the entire loop distribution:
%%%%%%%%%%%%%%%%%%%%%%%%%%%%%%%%%%%%%%%%%%%%%%%%%%
\begin{subequations}
\begin{align}
\left. \frac{\id Q}{\id t} \right|_{\rm r}
&=\int_0^{L_{\rm max}} \id L \left.\frac{\id N_{\rm{loops}}}{\id L}\right|_{\rm r} \, P_{\gamma}(L)
\\[2mm]
L_{\rm max}&=
\begin{cases} \beta t &(t \leq t_{\textrm{eq}})
\\
\beta t_{\rm eq}[1+\lambda-\lambda t/t_{\rm eq}] &(t_{\rm eq} < t \leq t_{\rm end}),
\end{cases}
\end{align}
\end{subequations}
with $t_{\rm end}=t_{\rm eq}\,(1+\lambda)/\lambda$.
%%%%%%%%%%%%%%%%%%%%%%%%%%%%%%%%%%%%%%%%%%%%%%%%%%
%As we are counting the loops present at time $t$, taking $\beta t$ as the upper bound counts all loops in the distribution.  
%
%To determine $\id Q/\id t$ after matter-radiation equality, the upper bound should be replaced with $\beta t_{\textrm{eq}}$. 
%
Computation of this integral gives an energy injection rate
%%%%%%%%%%%%%%%%%%%%%%%%%%%%%%%%%%%%%%%%%%%%%%%%%%
\begin{align}
\label{eq:heatingRad}
\left. \frac{\id Q}{\id t}  \right|_{\rm r} = \frac{2}{3} \frac{\alpha \Gamma_{\gamma} } {G^{1/2}} &\frac{\mathcal{I}}{\Gamma^{3/2}G\mu} \frac{(1+\lambda)^{3/2}}{t^3}  
\\ \nonumber
&\!\!\!\!\!\!\!\!\!\!\!\!\!\!\!
\times \begin{dcases} 
%\left[ 1 - \left(\frac{\lambda}{1+\lambda}\right)^{3/2} \right]  
\left[ 1 - \left(\frac{t_{\rm eq}}{t_{\rm end}}\right)^{3/2} \right] 
& (t \leq t_{\rm eq}) 
\\ 
%\left[ 1 - \left( \frac{\lambda}{1+\lambda} \frac{t}{t_{\rm eq}}\right)^{3/2} \right] 
\left[ 1 - \left(\frac{t}{t_{\rm end}}\right)^{3/2} \right]
\left( \frac{t_{\rm eq}}{t}\right)^{1/2} & ( t_{\rm eq} < t \leq t_{\rm end}).
\end{dcases}
\end{align}
%%%%%%%%%%%%%%%%%%%%%%%%%%%%%%%%%%%%%%%%%%%%%%%%%%
for loops created in the radiation-dominated era. 
%%%%%%%%%%%%%%%%%%%%%%%%%%%%%%%%%%%%%%%%%%%%%%%%%%
\begin{figure}
\includegraphics[width=\columnwidth]{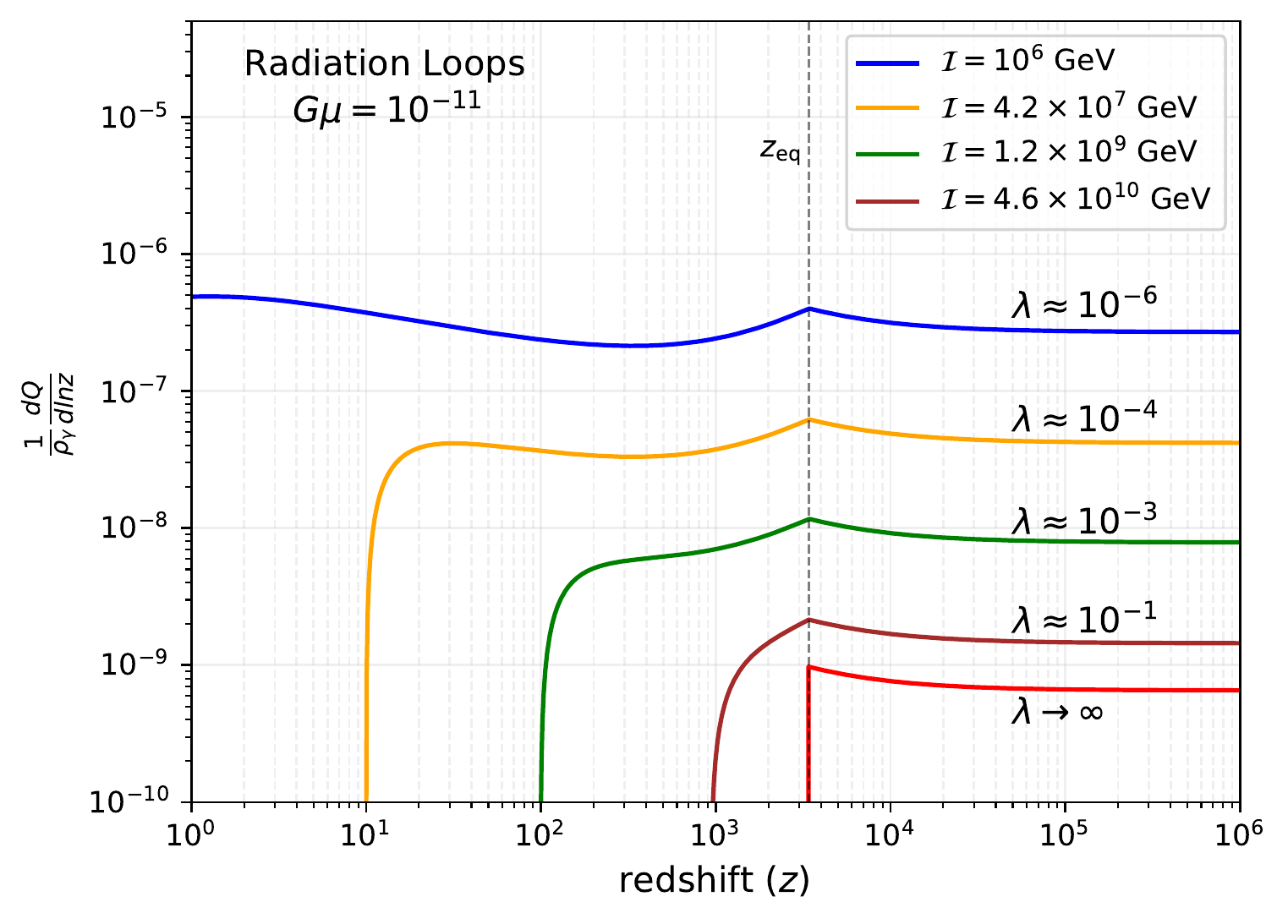}
\\
\caption{Fractional energy injection rate per logarithmic redshift bin from the decay of superconducting cosmic strings forming in the radiation era. The injection rate is roughly constant (in log-space), with an overall scaling given by $\zeta_{\rho}=\mathcal{I}/\Gamma^{3/2} G\mu$ until the loops completely evaporate. The evaporation redshift at which the last loop that formed in the radiation era will decay can be determined by solving Eq.~\eqref{eq:tdecay} with $t_i = t_{\rm eq}$ for a given $G\mu-\mathcal{I}$ pair. Deviations from an overall $\zeta_{\rho}$ scaling are expected near the evaporation redshift. The colour scheme of the contours is set to match that of Figure~\ref{fig:EvapRegions}.}
\label{fig:dQdlnz3}
\end{figure}
%%%%%%%%%%%%%%%%%%%%%%%%%%%%%%%%%%%%%%%%%%%%%%%%%%
The form of this heating rate is illustrated in Fig.~\ref{fig:dQdlnz3}. Due to the continuous sourcing of loops at $L\simeq \beta t$, the heating occurs at all redshifts until $t\simeq t_{\rm end}$, where the last loops injected at $t_{\rm eq}$ vanish, and heating ceases. In particular, in the $\mu$-distortion era, one finds $\rho_\gamma^{-1}\id Q/\id \ln z$ to be redshift-independent. It is clear from this figure that the smaller loops provide most of the heating.

As one can see from Eq.~\eqref{eq:heatingRad} and Fig.~\ref{fig:dQdlnz3}, a large value of $\lambda$ leads to a decrease in energy injection efficiency. This is because a loop population with a rapid decay rate cannot build up its number density. In other words, there are simply fewer loops around at any given time to inject energy. Naively, it looks as if the injection is amplified for large values of $\mathcal{I}$. However, we remind the reader that $\Gamma \propto \mathcal{I}$ for large currents, leading to an overall decrement in the injection. Physically, this is another consequence of a rapidly decaying population of loops.
However, in the very large current regime, one can also find regions where $\lambda\gg 1$, such that the heating rate from radiation loops becomes
%%%%%%%%%%%%%%%%%%%%%%%%%%%%%%%%%%%%%%%%%%%%%%%%%%
\begin{align}
\label{eq:heatingRadHigh}
\left. \frac{\id Q}{\id t}  \right|^{\rm high}_{\rm r} &\approx \frac{\alpha \Gamma_{\gamma} } {G^{1/2}} \frac{\mathcal{I}}{\Gamma^{3/2}G\mu} \frac{\lambda^{1/2}}{t^3}
=
\frac{\alpha \Gamma_{\gamma} } {G^{1/2}} \frac{\mathcal{I}}{\Gamma (G\mu)^{1/2}\beta^{1/2}} \frac{1}{t^3}\nonumber\\
&\approx
\frac{\alpha \Gamma_{\gamma} } {G^{1/2}} \frac{\mathcal{I}_*}{\Gamma_{\rm g} (G\mu)^{1/2}\beta^{1/2}} \frac{1}{t^3}
=
\frac{\alpha_{\rm m}}{G} \frac{G\mu}{t^3}
\end{align}
%%%%%%%%%%%%%%%%%%%%%%%%%%%%%%%%%%%%%%%%%%%%%%%%%%
independent of $\mathcal{I}$ at $t\leq t_{\rm eq}$. Here we used $\Gamma \approx \Gamma_{\rm g} \mathcal{I}/ \mathcal{I}_*$ for the decay rate and $\alpha_{\rm m}=\alpha/\sqrt{\beta}$. This expression is not valid for $t \geq t_{\rm eq}$, as for $\lambda \gg 1$ the radiation loops fully evaporate at $t_{\rm end} \approx t_{\rm eq}$.

One should note that the electromagnetic power generated by a cusp decay ($P_{\gamma}$) is an oscillation-averaged quantity, as the loop must undergo at least one oscillation for cusp formation to take place. However, an implication of $\lambda > 1/\beta$ is that the loop should fully decay before a single oscillation can even take place. We therefore urge the reader to be cautious about statements made in regions of parameter space with $\lambda > 1/\beta$, which is indicated by the red region in Fig.~\ref{fig:EvapRegions}.

Inserting the injection rate into Eq.~\eqref{DistApprox}, and noting that the background energy density (in natural units) is $\rho_{\gamma} = (\pi^2/15) T^4$, we find the following estimates for $\mu$ and $y$ from only the radiation loops
%%%%%%%%%%%%%%%%%%%%%%%%%%%%%%%%%%%%%%%%%%%%%%%%%%
\begin{subequations}
\begin{align}
\label{eq:mu}
\mu &\simeq 5.5 \times 10^{-6} \left[
\frac{\zeta_\rho}{10^{11} \, \rm GeV} \right] \hspace{20mm} (\lambda \ll 1)\\
y &\simeq 1.3 \times 10^{-6} \left[ \frac{\zeta_\rho}{10^{11} \, \rm GeV} \right] \hspace{20mm} (\lambda \ll 1),
\end{align}
\end{subequations}
%%%%%%%%%%%%%%%%%%%%%%%%%%%%%%%%%%%%%%%%%%%%%%%%%%
where we introduced the variable $\zeta_\rho=\mathcal{I}/\Gamma^{3/2} G\mu $, which describes the overall non-linear scaling of the total energy release with $\mathcal{I}$ and $G\mu$, provided the loop distribution is not at their evaporation redshift. Note that $\Gamma$ is a function of both parameters.

It can also be useful to phrase our constraints in terms of the fractional energy injection into the background, $\Delta \rho_{\gamma}/ \rho_{\gamma}$, which is given by
%%%%%%%%%%%%%%%%%%%%%%%%%%%%%%%%%%%%%%%%%%%%%%%%%%
\begin{align}
\left. \frac{\Delta \rho_{\gamma}}{\rho_{\gamma}} \right|_{\rm r} = \int_{z_{\rm rec}}^{\infty} \id z \frac{1}{\rho_{\gamma}}  \left. \frac{\id Q}{\id z}  \right|_{\rm r} \mathcal{J}_{\rm bb}(z).
\end{align}
%%%%%%%%%%%%%%%%%%%%%%%%%%%%%%%%%%%%%%%%%%%%%%%%%%
The fractional energy release from superconducting cosmic strings over the entire distortion window is therefore
%%%%%%%%%%%%%%%%%%%%%%%%%%%%%%%%%%%%%%%%%%%%%%%%%%
\begin{align} \label{eq:dRhoGamma}
\left. \frac{\Delta \rho_{\gamma}}{\rho_{\gamma}}  \right|_{\rm r} \simeq 8.9 \times 10^{-6}\left[ \frac{\zeta_\rho}{10^{11} \, \rm GeV} \right] \hspace{15mm} (\lambda \ll 1).
\end{align}
%%%%%%%%%%%%%%%%%%%%%%%%%%%%%%%%%%%%%%%%%%%%%%%%%%
This is the expression one should use in deriving constraints based purely on the total amount of energy release. However, it neglects the effects of partial Comptonization and also photon injection, which can change the character of the distortion significantly \citep{Chluba2015}.

%--------------------------------------------
\subsection{Energy Injection from Matter Loops}
%--------------------------------------------
At $t\lesssim t_{\rm eq}$, loops that are formed during the matter-dominated era can contribute.
To compute the energy release rate from these loops, we can follow the same steps as above but instead average over the loop distribution from the matter-era:
%%%%%%%%%%%%%%%%%%%%%%%%%%%%%%%%%%%%%%%%%%%%%%%%%%
\begin{subequations}
\begin{align}
\left.\frac{\id Q}{\id t} \right|_{\rm m}
&=\int^{\beta t}_{L_{\rm min}} \id L 
\left.\frac{\id N_{\rm{loops}}}{\id L}\right|_{\rm m} \, P_{\gamma}(L)
\\[2mm]
L_{\rm min}&=\left[\beta t_{\rm eq}(1+\lambda-\lambda t/t_{\rm eq}), 0\right]_>.
\end{align}
\end{subequations}
%%%%%%%%%%%%%%%%%%%%%%%%%%%%%%%%%%%%%%%%%%%%%%%%%%
Here and below, we use the shorthand notation $[a,b]_>=\max[a,b]$ and $[a,b]_< =\min[a,b]$. 
Computation of this integral gives an energy injection rate
%%%%%%%%%%%%%%%%%%%%%%%%%%%%%%%%%%%%%%%%%%%%%%%%%%
\begin{align}
\label{eq:heatingMatter}
\left.\frac{\id Q}{\id t} \right|_{\rm m}
&= \frac{\alpha_{\rm m} \Gamma_{\gamma} } {G^{1/2}} \frac{\mathcal{I}\,(G\mu)^{1/2}}{\beta} \frac{1}{t^3}  
\nonumber\\
&\qquad
\times \begin{dcases} 
\left[ \frac{t}{t_{\rm eq}}-1\right]
&\quad (t_{\rm eq}\leq t\leq t_{\rm end}) 
\\ 
\left[ \frac{t_{\rm end}}{t_{\rm eq}}-1\right] &\quad (t_{\rm end}<t),
\end{dcases}
\end{align}
%%%%%%%%%%%%%%%%%%%%%%%%%%%%%%%%%%%%%%%%%%%%%%%%%%
with $t_{\rm end}=t_{\rm eq}\,(1+\lambda)/\lambda$. Note that for $\lambda \lesssim 4 \times 10^{-6}$, all matter loops survive until today ($t_{\rm end} \geq t_0$).

%%%%%%%%%%%%%%%%%%%%%%%%%%%%%%%%%%%%%%%%%%%%%%%%%%
\begin{figure}
\includegraphics[width=\columnwidth]{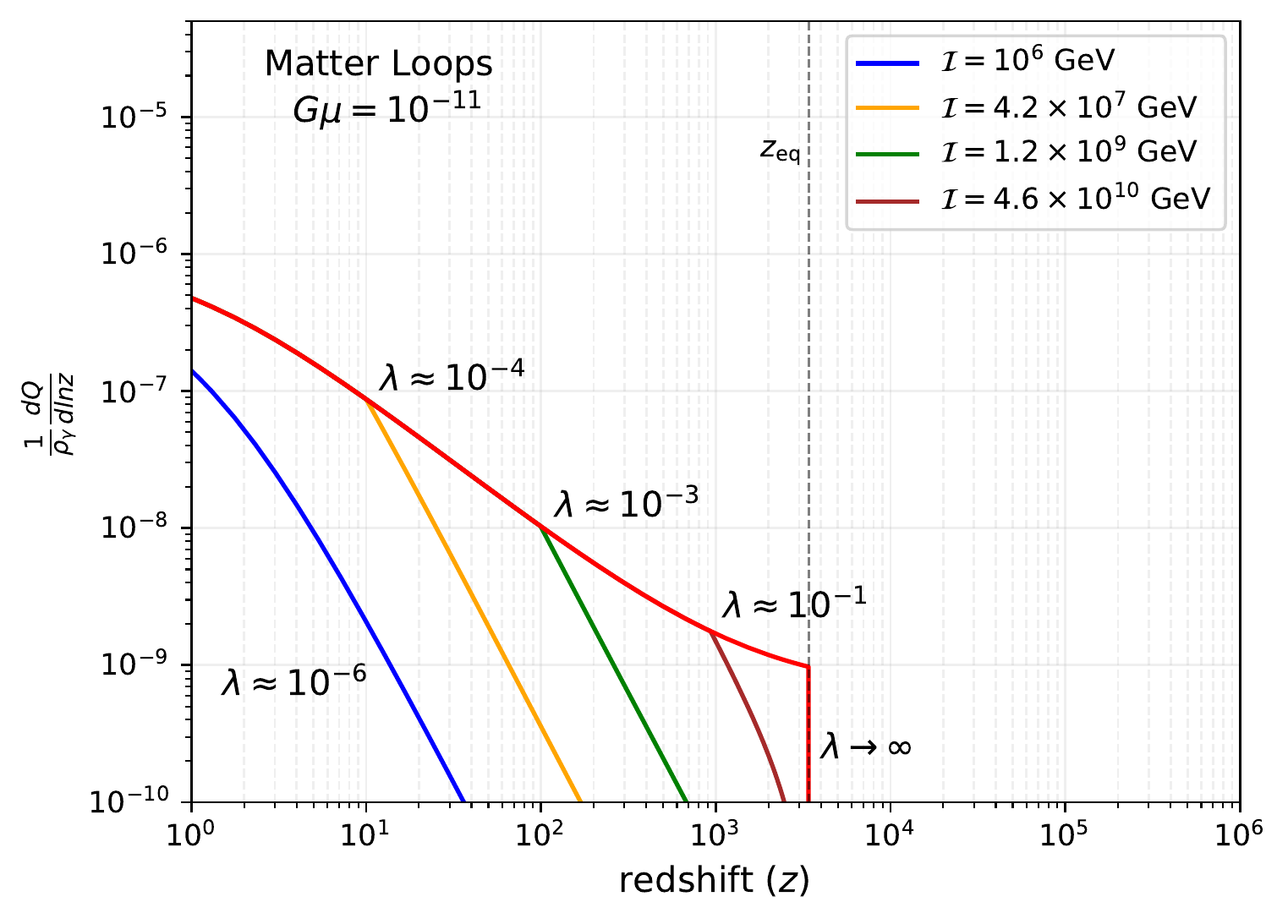}
\caption{Energy injection rate from loops which form in the matter dominated era. The amplitude of the signal is primarily governed by the number density of matter loops at any given time. A ``knee" develops when $t_{\rm end}$ is crossed for a given parameter set. The slowing of the growth of the injection rate below that redshift is because the distribution now exhibits a source-sink behaviour as the smallest loops fully evaporate, in contrast to just a source before that point.}
\label{fig:dQdlnzMatter}
\end{figure}
%%%%%%%%%%%%%%%%%%%%%%%%%%%%%%%%%%%%%%%%%%%%%%%%%%

In Fig.~\ref{fig:dQdlnzMatter} we illustrate the contribution to the heating rate from cosmic string loops created during the matter dominated era. The overall amplitude of the signal is governed primarily by two factors: the number density of the distribution, and the smallest loop size, $L_{\rm min}$, as is evident from Eq.~\eqref{eq:heatingMatter}. After matter radiation equality, $L_{\rm min}$ decreases as the loops oscillate and decay, leading to a strong increase in the injection rate as can be seen on the right hand side of the knees in Fig.~\ref{fig:dQdlnzMatter}. At $t_{\rm end}$ (the position of the knee), matter loops begin leaving the distribution as they fully evaporate, leading to a slowing of the injection rate. Since the distribution redshifts like matter, their contribution to the background energy density grows linearly in $z$ relative to the CMB.

For large currents, we can again approximate the heating rate in the $\lambda \gg 1$ regime:
%%%%%%%%%%%%%%%%%%%%%%%%%%%%%%%%%%%%%%%%%%%%%%%%%%
\begin{align}
\label{eq:heatingM}
\left. \frac{\id Q}{\id t}  \right|^{\rm high}_{\rm m} &\approx \frac{\alpha_{\rm m} \Gamma_{\gamma} } {G^{1/2}} \frac{\mathcal{I} \sqrt{G\mu }}{\beta \, t^3} \frac{1}{\lambda} 
\approx
\frac{\alpha_{\rm m}}{G} \frac{G\mu}{t^3}.
\end{align}
%%%%%%%%%%%%%%%%%%%%%%%%%%%%%%%%%%%%%%%%%%%%%%%%%%
This is identical to Eq.~\eqref{eq:heatingRadHigh}, which is expected since fundamentally nothing special happens to the loop distribution at $t = t_{\rm eq}$. For $\lambda \gg 1$, all of the energy in the newly formed loops is immediately emitted. The steady rise in fractional energy injection shown in the $\lambda \rightarrow \infty$ contour of Fig.~\ref{fig:dQdlnzMatter} is obtained because the energy density in the loops redshifts as $a^{-3}$ while the CMB photons redshift as $a^{-4}$.

\begin{figure}
\includegraphics[width=\columnwidth]{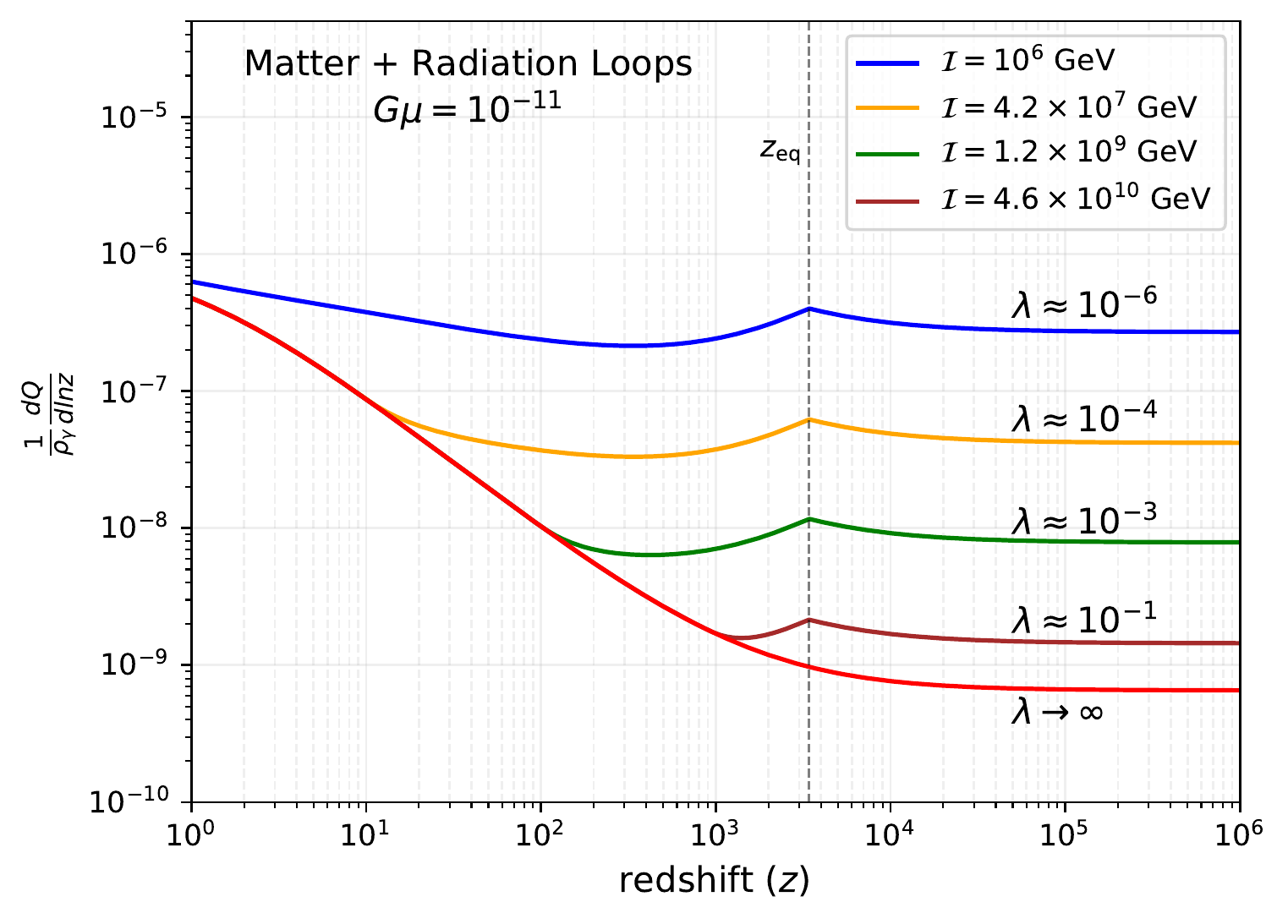}
\caption{The total energy injection rate from a string loop distribution. For a constant $G\mu$, the late time injection rate converges for $\mathcal{I} \gg \mathcal{I}_*$. In contrast, contours of current $\mathcal{I}$ will experience amplitude shifts proportional to $G\mu$ in the current-dominated regime.}
\label{fig:dQdlnzTOT}
\end{figure}

We show the total contribution of both matter and radiation loops in Fig.~\ref{fig:dQdlnzTOT}. Interestingly, the fractional injection rate converges at late times for a wide range of parameters. This behaviour persists along contours of constant $G\mu$ once $\mathcal{I} \gg \mathcal{I}_*$. In this current domination region of parameter space, adjustments to $G\mu$ will linearly shift the amplitude of the late time emission signal, while in the gravitational wave regime ($\mathcal{I} \leq \mathcal{I}_*$), the scaling of the signal is more complicated. The overall degradation of the signal with high currents is due to the fact that a smaller density of loops is available at any given time for these models. For $\mathcal{I} \gg \mathcal{I}_*$, we have we have $\zeta_{\rho} \propto \mathcal{I}^{-1/2}$, which encodes this rapid decay effect. 

%%%%%%%%%%%%%%%%%%%%%%%%%%%%%%%%%%%%%%%%%%%%%%%%%%
%%%%%%%%%%%%%%%%%%%%%%%%%%%%%%%%%%%%%%%%%%%%%%%%%%
With the injection function, we may now estimate the fractional energy release from matter loops. For small $\lambda$ (equivalently, $t_{\rm end} \gg t_{\rm eq}$), we find that from $z_{\rm rec} \leq z \leq z_{\rm eq}$, the energy release is
%%%%%%%%%%%%%%%%%%%%%%%%%%%%%%%%%%%%%%%%%%%%%%%%%%
\begin{align}
\left. \frac{\Delta \rho_{\gamma}}{\rho_{\gamma}}  \right|_{\rm m} \simeq 2.9 \times 10^{-16} \left[ \frac{\mathcal{I}}{10^4 \, \rm GeV}\right] \left[\frac{G\mu }{10^{-10}}\right]^{1/2} \hspace{5mm} (\lambda \ll 1).
\end{align}
%%%%%%%%%%%%%%%%%%%%%%%%%%%%%%%%%%%%%%%%%%%%%%%%%%
Recall that small values of $\lambda$ imply a long lifetime for a given loop. Since most of the energy release comes from the smallest loops present in a given distribution, matter loops will always source a  subdominant contribution to the primordial distortion signal relative to radiation loops for $\lambda \ll 1$. As indicated in Fig.~\ref{fig:dQdlnzMatter} and \ref{fig:dQdlnzTOT}, matter loops begin to play more of a role in pre-recombination injection for $\lambda \gtrsim 0.1$. While not a factor for primordial spectral distortions, low redshift emission from the matter loops can be constrained by the evolution of the ionization fraction. This emission may also play a role in our understanding of the EDGES and ARCADE-2 signals, as we will see below. 

In Fig.~\ref{fig:drhoConstraint} we show the constraints on energy injection in this picture using the \COBEF data ($\Delta \rho/\rho \leq 6\times 10^{-5}$), as well as a simple forecast from a {\it PIXIE}-like instrument ($\Delta \rho/\rho \leq 2\times 10^{-8}$). For \COBEF, the quoted energy release levels are consistent with the $2\sigma$ errors on $\mu$ and $y$, while for the \PIXIE-like setup we assume that a penalty from marginalization over foregrounds \citep{abitbol_pixie, Rotti2021} is already included.

The constraints are nearly symmetric about the $\mathcal{I} = \mathcal{I}_*$ contour, with departures arising in the $\lambda \simeq 1$ regime due to rapidly decaying loops.  The work of \cite{Tashiro} presented a flat constraint in the $\mathcal{I} \geq \mathcal{I}_*$ region, a symptom of them missing a factor of $\Gamma^{-1/2}$ in their equivalent expression of Eq.~\eqref{eq:dRhoGamma}. Our energy injection constraints are roughly consistent with the analysis of \cite{Miyamoto}.
In the next section, we include the effects of entropy injection which cause further significant modifications to these constraints, as can be seen in Fig.~\ref{fig:drhodNConstraint}.

%%%%%%%%%%%%%%%%%%%%%%%%%%%%%%%%%%%%%%%%%%%%%%%%%%
\begin{figure}
\includegraphics[width=\columnwidth]{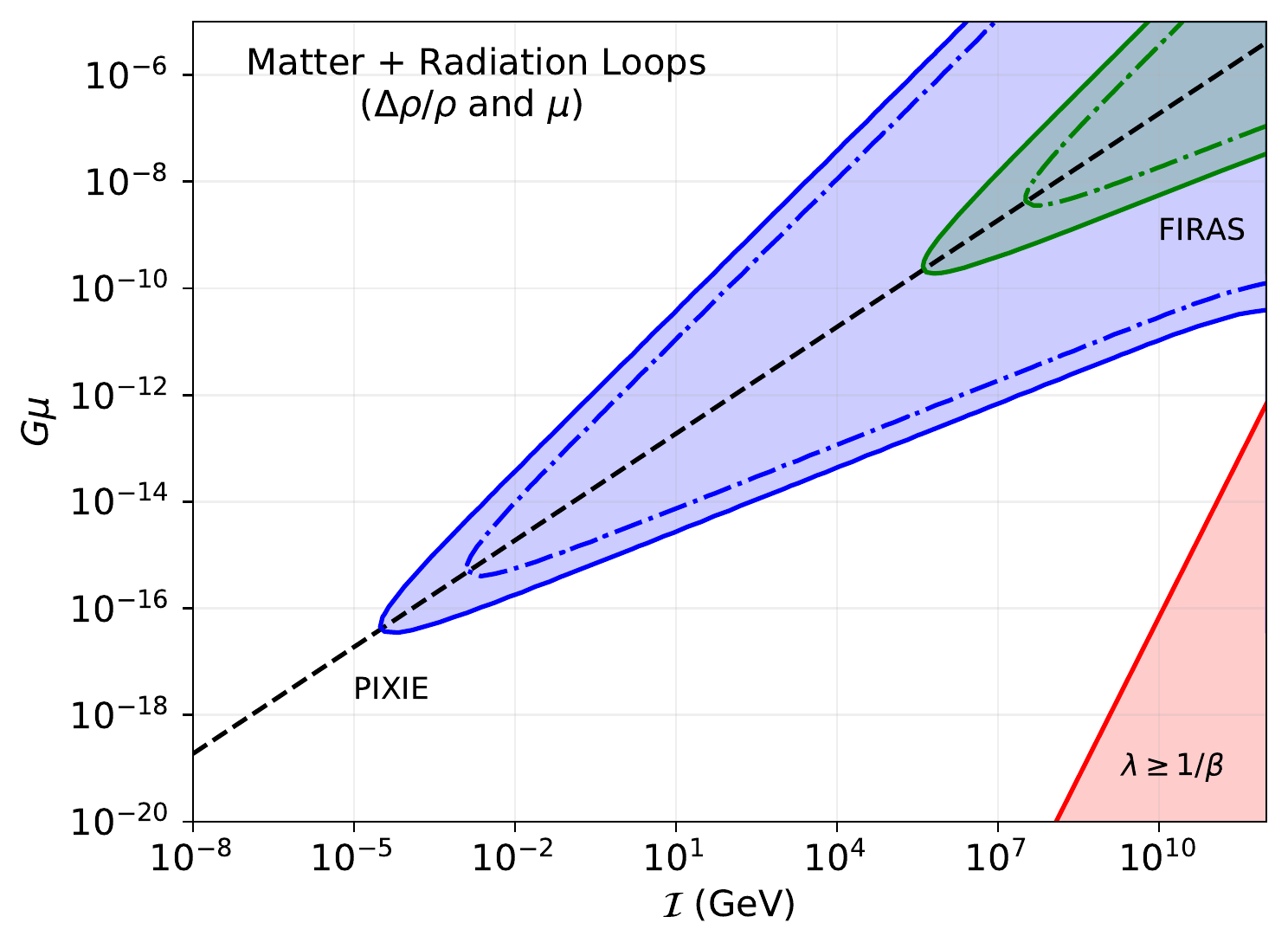}
\\
\caption{Solid lines indicate constraints on string parameters obtained by requiring $\Delta \rho/\rho \leq 6\times10^{-5}$ ($2\sigma$ for \COBEF), and $\Delta \rho/\rho \leq 2\times10^{-8}$ ($2\sigma$ forecast for a {\it PIXIE}-type instrument). Statements in the $\lambda \geq 1$ region of the plot should be viewed skeptically, as rapid decays may cause the injection framework to break down. The dash-dotted lines are constraints obtained by $\mu$ distortions over the redshift range $3\times 10^5 \leq z \leq 2 \times 10^6$, and may be directly compared to the energy+entropy constraints illustrated in Fig.~\ref{fig:drhodNConstraint}.}
\label{fig:drhoConstraint}
\vspace{-2mm}
\end{figure}
%%%%%%%%%%%%%%%%%%%%%%%%%%%%%%%%%%%%%%%%%%%%%%%%%%

%--------------------------------------------
\section{Analytic Estimate - Entropy Release}
\label{sec:analyticEntropy}
%--------------------------------------------
Energy release is not the only way one can disturb the CMB spectrum. An often overlooked fact of distortion theory is that the production of additional photons also perturbs the spectrum. In the $\mu$ era, the total distortion should be expressed as \citep{Hu1995PhD, Chluba2015}
%%%%%%%%%%%%%%%%%%%%%%%%%%%%%%%%%%%%%%%%%%%%%%%%%%
\begin{align} \label{eq:muApprox}
\mu \approx 1.401 \left[ \frac{\Delta \rho_{\gamma}}{\rho_{\gamma}} - \frac{4}{3} \frac{\Delta N_{\gamma}}{N_{\gamma}} \right].
\end{align}
%%%%%%%%%%%%%%%%%%%%%%%%%%%%%%%%%%%%%%%%%%%%%%%%%%
Therefore, models which introduce direct photon production (such as this one) must necessarily take into account both energy and entropy injection when determining constraints. In fact, this entropy injection is capable of producing a net negative distortion signature. Of course, \COBEF is only sensitive to $|\mu|$, but as we will see below, there are regions of parameter space in which the constraints of \cite{Tashiro} and \cite{Miyamoto} are significantly altered. 
There are also regions in parameter space that cannot be easily estimated analytically, since strong non-$\mu$/non-$y$ distortion signals are created \citep{Chluba2015}, requiring the numerical treatment presented below.
%%%%%%%%%%%%%%%%%%%%%%%%%%%%%%%%%%%%%%%%%%%%%%%%%%

To obtain a description for the photon source term, we adapt the work of 
\citet{Cai} who derived the power emitted in photons per unit frequency, per unit solid angle from a superconducting cosmic string cusp annihilation (averaged over an oscillation time):
%%%%%%%%%%%%%%%%%%%%%%%%%%%%%%%%%%%%%%%%%%%%%%%%%%
\begin{align} \label{diffPow}
\frac{\id^2P_{\gamma}^{\rm c}}{\id\omega \id\Omega} \approx  \left(\frac{\Gamma_{\gamma}}{3}\right)\mathcal{I}^2 L.
\end{align}
%%%%%%%%%%%%%%%%%%%%%%%%%%%%%%%%%%%%%%%%%%%%%%%%%%
The cusp is a highly relativistic object, and so the radiation is heavily beamed, with all emission in a solid angle $\Omega = (\omega L)^{-2/3}$. Here, $\omega$ is the frequency of emitted radiation. Integrating Eq.~\eqref{diffPow} over solid angle we find the instantaneous spectrum from a single loop
%%%%%%%%%%%%%%%%%%%%%%%%%%%%%%%%%%%%%%%%%%%%%%%%%%
\begin{align} \label{EnergyNum}
\frac{\id^2E_{\gamma}^{\rm c}}{\id t \id\omega} \approx \left(\frac{\Gamma_{\gamma}}{3}\right) \frac{\mathcal{I}^2 L^{1/3}}{\omega^{2/3}} , 
\hspace{15mm} \frac{\id^2N_{\gamma}^{\rm c}}{\id t \id\omega} \approx \left(\frac{\Gamma_{\gamma}}{3}\right) \frac{\mathcal{I}^2 L^{1/3}}{\omega^{5/3}}.
\end{align}
%%%%%%%%%%%%%%%%%%%%%%%%%%%%%%%%%%%%%%%%%%%%%%%%%%
Here, $E_{\gamma}^c$ and $N_{\gamma}^c$ are the total energy and number of photons produced by a given loop. We introduced the factor of $\Gamma_{\gamma}/3$ in these equations to properly match the total power produced by a loop with Eq.~(\ref{TotPow}).\footnote{This can be easily confirmed by integrating over $\omega$ between $0$ and $\omega_{\rm max}$, which yields $\id E_{\gamma}^{\rm c}/\id t\approx \Gamma_\gamma \mathcal{I}^2 L^{1/3}\,\omega_{\rm max}^{1/3}=\Gamma_\gamma \mathcal{I}\,\mu^{1/2}$, reproducing Eq.~\eqref{TotPow}.} The numerical prefactor depends on the precise shape of a loop, and so the $\Gamma_{\gamma}$ factor averages over the geometries of a network of loops. Upon integration of the differential energy spectrum, one notices that the total power is dominated by the highest frequency photons produced at a cusp. \citet{Cai} have estimated this frequency to be
%%%%%%%%%%%%%%%%%%%%%%%%%%%%%%%%%%%%%%%%%%%%%%%%%%
\begin{align} \label{eq:omegaMax}
\omega_{\rm max} \approx \frac{\mu^{3/2}}{\mathcal{I}^3 L}.
\end{align}
%%%%%%%%%%%%%%%%%%%%%%%%%%%%%%%%%%%%%%%%%%%%%%%%%%
The cusp covers a finite spatial extent on the string, and so this result is derived by requiring the total energy released in a cusp decay does not exceed the rest-mass energy of this region.

%--------------------------------------------
\subsection{Total emission from loops}
%--------------------------------------------
To obtain the number density injection from all loops, one may first determine the number of photons produced per unit time, and then average over the full loop distribution. For our purposes, it will be useful to introduce the dimensionless frequency as $x = \omega/T(z)$. The photon spectrum from cosmic strings in principle extends to arbitrarily low energies\footnote{Plasma effects (i.e., the plasma frequency or the Razin effect) become important at frequencies well below the domain of interest.}; however, low frequency photons are at risk of being absorbed by the plasma and converted into heat \citep{Chluba2015, Bolliet}. The survival probability, $\mathcal{P}_{\rm s}(x,z) \approx {\rm e}^{-x_{\rm c}/x}$ tells us how many photons survive as a true entropy injection. 

At high redshifts ($z\gtrsim 10^5$), we use the simple expression for the critical frequency \citep{Chluba2014vis}
%%%%%%%%%%%%%%%%%%%%%%%%%%%%%%%%%%%%%%%%%%%%%%%%%%
\begin{align} 
\label{eq:xcApprox}
x_{\rm c} \approx 8.6 \times 10^{-3} \sqrt{ \frac{1+z}{2 \times 10^6} } \sqrt{1 + \left[ \frac{1+z}{3.8 \times 10^5}\right]^{-2.344}}.
\end{align}
%%%%%%%%%%%%%%%%%%%%%%%%%%%%%%%%%%%%%%%%%%%%%%%%%%
This frequency is determined through the balance of Compton, double Compton (DC), and Bremsstrahlung (BR) effects. 
At later times ($z\lesssim 10^5$), the absorption probability is mainly determined by the free-free process, without much contribution from Compton scattering. In this regime, we estimate the absorption probability by finding the frequency at which the free-free optical depth is close to unity \citep{Chluba2015, Bolliet}.

\begin{figure}
\includegraphics[width=\columnwidth]{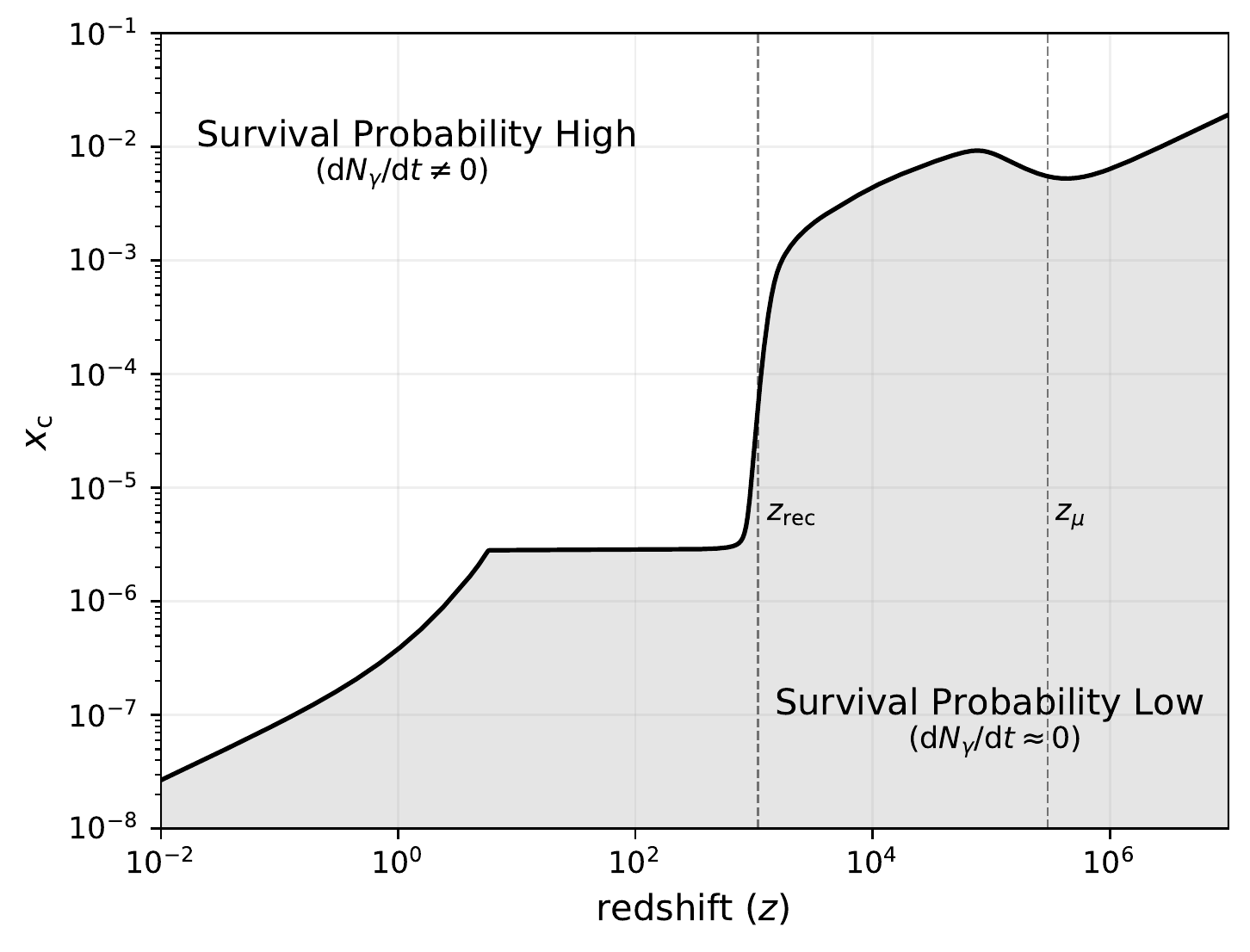}
\caption{Evolution of the dimensionless critical frequency over redshift. Photons produced with a frequency $x < x_{\rm c}$ are absorbed by the plasma quickly after their creation, and therefore do not contribute significantly to a photon number injection. The approximate expression in Eq.~\eqref{eq:xcApprox} is valid for $z \gtrsim 10^5$, while below we use the pre-tabulated free-free absorption frequency to estimate the absorption probability.}
\label{fig:critFreq}
\end{figure}

The two regimes are merged in Fig.~\ref{fig:critFreq} over the relevant redshift range.
At recombination, the fraction of free electrons and protons drops substantially, leading to a sharp decrease in the absorption probability which then remains constant until reionization begins.
%%%%%%%%%%%%%%%%%%%%%%%%%%%%%%%%%%%%%%%%%%%%%%%%%%
%%%%%%%%%%%%%%%%%%%%%%%%%%%%%%%%%%%%%%%%%%%%%%%%%%

Integrating the right hand expression of Eq.~\eqref{EnergyNum} with the factor\footnote{Although the free-free optical depth scales as $\tau_{\rm ff}\propto \ln(2.25/x)/x^2$ at low frequencies, we keep our estimates simple and just replace the critical frequency with the free-free absorption frequency. This does not significantly alter the illustrations that are presented below. For the final constraints, we explicitly compute the distortion signal without these approximations.}  $\mathcal{P}_{\rm s}(x,z) \approx {\rm e}^{-x_{\rm c}/x}$ yields the total number of photons produced per cosmic string which will contribute to an entropy injection
%%%%%%%%%%%%%%%%%%%%%%%%%%%%%%%%%%%%%%%%%%%%%%%%%%
\begin{align}
\label{eq:Ndot_c}
\frac{\id N_{\gamma}^c}{\id t} &= \frac{\Gamma_{\gamma}}{3} \frac{\mathcal{I}^2 L^{1/3}}{T^{2/3}} \int_0^{x_{\rm max}} \id x \,\frac{{\rm e}^{-x_{\rm c}/x}}{x^{5/3}} \nonumber \\
&= \frac{\Gamma_{\gamma}}{3} \frac{\mathcal{I}^2 L^{1/3}}{(x_{\rm c} T)^{2/3}} \, \Gamma\left[ \frac{2}{3}, \frac{x_{\rm c}}{x_{\rm max}} \right].
\end{align}
%%%%%%%%%%%%%%%%%%%%%%%%%%%%%%%%%%%%%%%%%%%%%%%%%%
For most reasonable choices of $G\mu$ and $\mathcal{I}$, $x_{\rm c} \ll x_{\rm max}$, where the incomplete gamma function is well approximated by $\Gamma[2/3,x_{\rm c}/x_{\rm max}] \approx \Gamma[2/3]\approx 1.354$. However, as is evident from Eq.~\eqref{eq:omegaMax}, large values of $\mathcal{I}$ and small values of $G\mu$ serve to decrease $x_{\rm max}$, implying that for those parameter choices, most of the produced photons are below $x_{\rm c}$ and hence do not contribute as entropy.

%%%%%%%%%%%%%%%%%%%%%%%%%%%%%%%%%%%%%%%%%%%%%%%%%%
\begin{figure}
\includegraphics[width=\columnwidth]{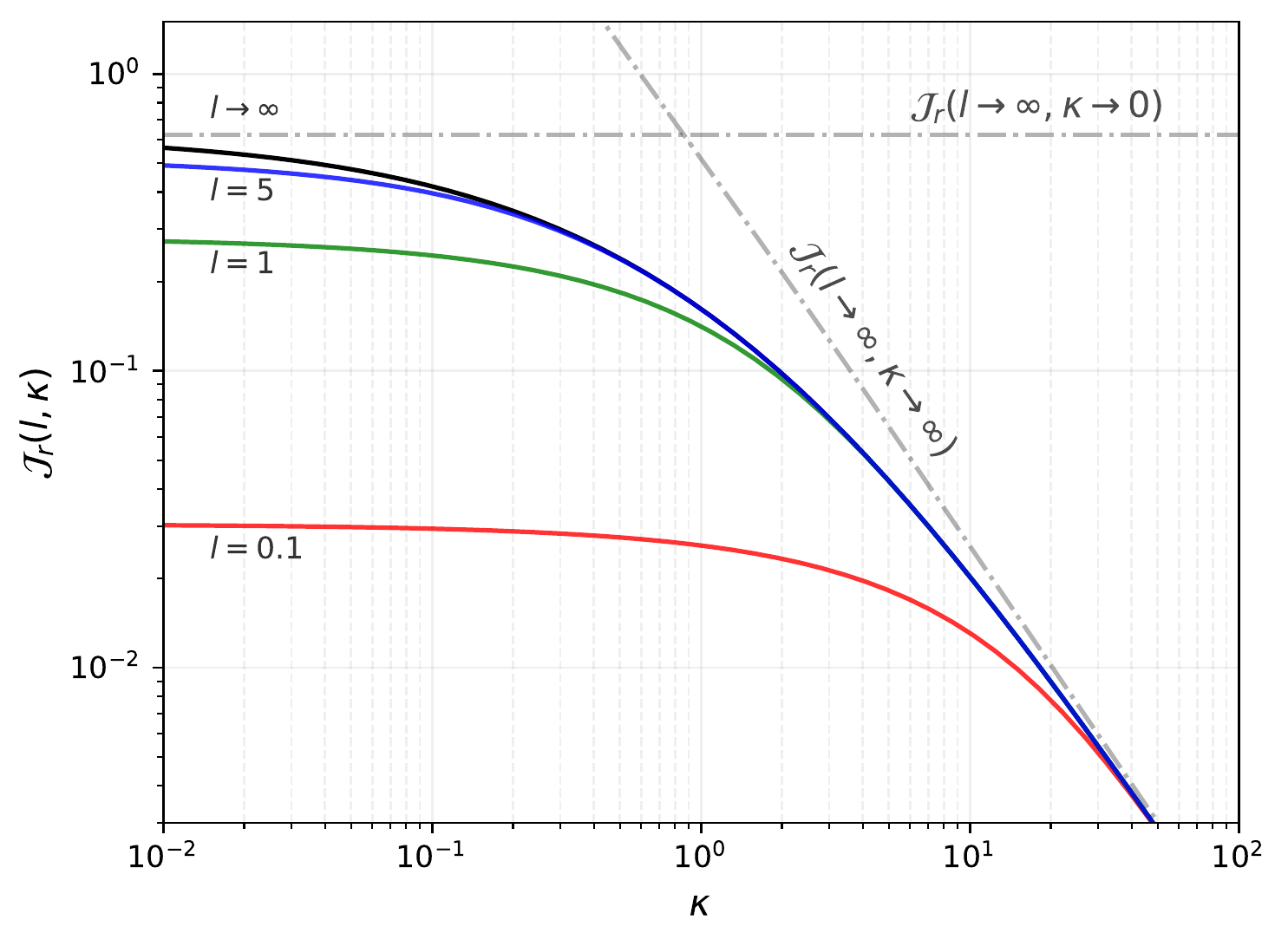}
\caption{The general functional form of $\mathcal{J}_{\rm r}(l,\kappa)$ for different $l$ values. For parameter values yielding $\lambda \ll 1$, the $l \rightarrow \infty$ contour is representative of the emission from all loops in the distribution before matter-radiation equality. Contours of smaller $l$ then show the contribution from correspondingly smaller loops, with $l$ = 1 representing emission from all loops with $L \leq \Gamma G\mu t$. The asymptotic expressions for small and large $\kappa$ can be found in Eq.~\eqref{eq:F_int} and  Eq.~\eqref{eq:Jinf}, respectively. Alternatively, for a given $\lambda$, we note that $l=1/\lambda$ yields the total emission from all loops in the distribution.}
\label{fig:JRplotCombined}
\end{figure}
%%%%%%%%%%%%%%%%%%%%%%%%%%%%%%%%%%%%%%%%%%%%%%%%%%

With Eq.~\eqref{eq:Ndot_c}, we can compute the entropy injection rate from the loop distribution in a similar way to the energy injection,
%%%%%%%%%%%%%%%%%%%%%%%%%%%%%%%%%%%%%%%%%%%%%%%%%%
\begin{align}
\frac{\id N_{\gamma}}{\id t} = \int_0^{L_{\rm up}} \id L \frac{\id N_{\rm loops}}{\id L} \frac{\id N_{\gamma}^c}{\id t}.
\end{align}
%%%%%%%%%%%%%%%%%%%%%%%%%%%%%%%%%%%%%%%%%%%%%%%%%%
%%%%%%%%%%%%%%%%%%%%%%%%%%%%%%%%%%%%%%%%%%%%%%%%%%
where $L_{\rm up}$ depends on the case of interest (i.e. matter/radiation loops at a particular redshift). For clarity, let us first focus on loops created and evolving in the radiation-dominated era (i.e., $t\leq t_{\rm eq}$). The relevant integral then becomes
%%%%%%%%%%%%%%%%%%%%%%%%%%%%%%%%%%%%%%%%%%%%%%%%%%
\begin{subequations}
\begin{align}
\frac{\id N_{\gamma}}{\id t} &= \frac{\alpha \Gamma_{\gamma}}{3} \frac{\mathcal{I}^2(1+\lambda)^{3/2}}{(\Gamma G\mu)^{7/6}\,t^{8/3}}\, \frac{\Gamma[2/3]}{(x_{\rm c} T)^{2/3}} \times \mathcal{J}_{\rm r}(l_{\rm up}, \kappa)
\\ \label{eq:Jr}
\mathcal{J}_{\rm r}(l, \kappa)&=
\int_0^{l} \id l' \, \frac{{l'}^{1/3}}{(1+l')^{5/2}} \,\frac{\Gamma\left[ \frac{2}{3}, \kappa\,l' \right]}{\Gamma[2/3]},
\end{align}
\end{subequations}
%%%%%%%%%%%%%%%%%%%%%%%%%%%%%%%%%%%%%%%%%%%%%%%%%%
where we have substituted $l = L/(\Gamma G\mu t)$ and $l_{\rm up} = L_{\rm up}/(\Gamma G\mu t)$. 
We also used $x_{\rm c}/x_{\rm max}=\kappa \, l$ with $\kappa=x_{\rm c} \,T\,\mathcal{I}^3\,\Gamma\,G\,t/\mu^{1/2}$.

\begin{figure}
\includegraphics[width=\columnwidth]{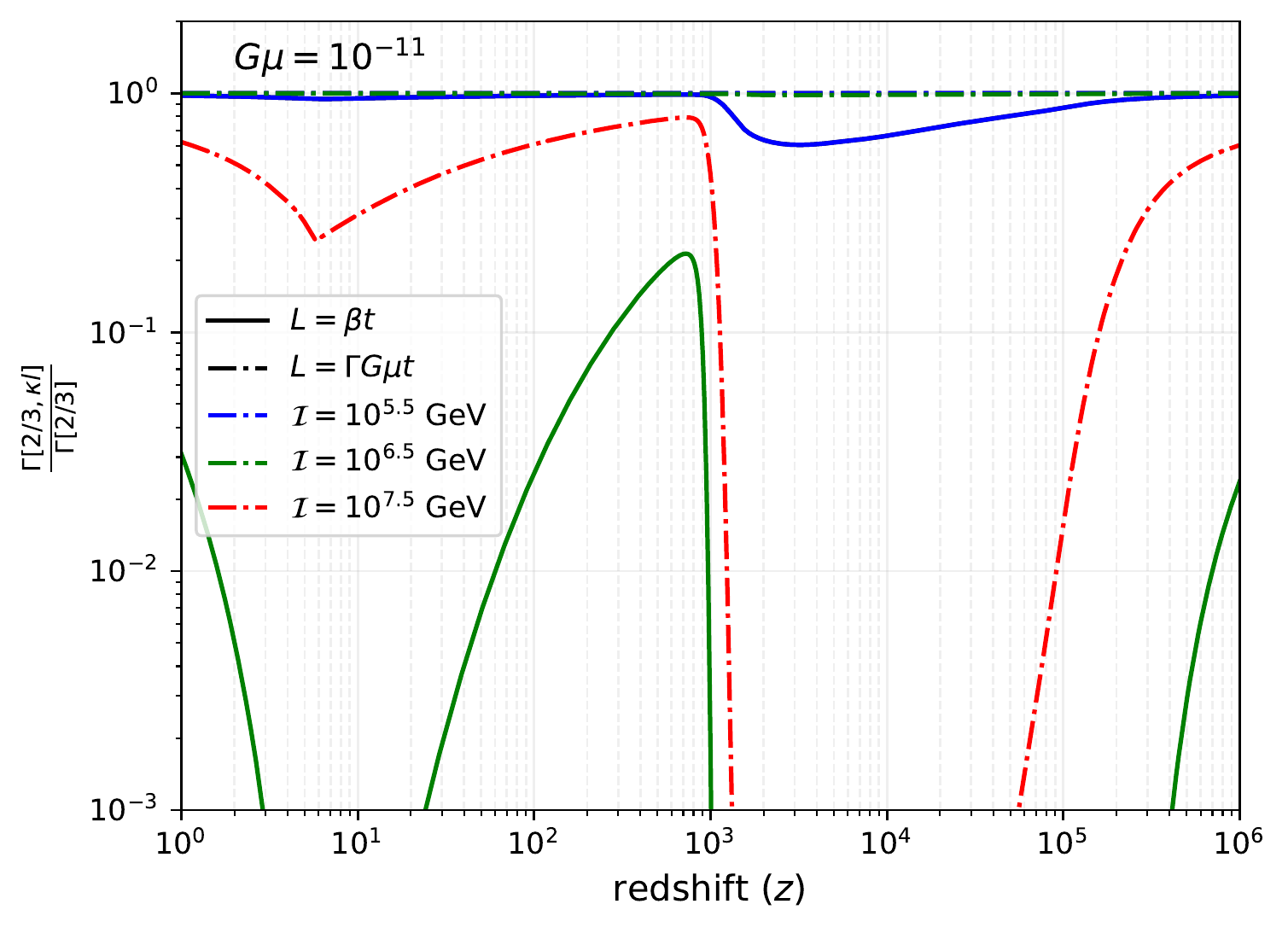}
\\
\caption{The normalized window function at different redshifts for entropy injections from cosmic strings for two different benchmark loop lengths. The solid lines characterize the largest loops in the distribution, $L = \beta t$, while the dash-dotted lines correspond to much smaller loops, with $L = \Gamma G \mu t$, who will fully evaporate within one Hubble time of any given $t$. One can clearly see that the window function for the larger loops closes off much faster than for the smaller loops once the current becomes large enough. The blue and green dash-dotted lines are overlapping in this figure, and for $\mathcal{I} = 10^{7.5}$ GeV the window function for the largest loops is completely closed. For all parameter values displayed here $\lambda \ll 1$.}
\label{fig:windowFunc}
\end{figure}

Figure~\ref{fig:JRplotCombined} shows the general form of $\mathcal{J}_r(l,\kappa)$. Heuristically, $\Gamma[2/3, \kappa l']/\Gamma[2/3]$ acts as a normalized window function which closes rapidly for $\kappa l' \gtrsim 1$. Smaller loops emit higher frequency photons, meaning that they are able to contribute to entropy injections more efficiently than the large loops. This is reflected by the fact that the argument of the window function depends on the length of any particular loop, as we illustrate in Fig.~\ref{fig:windowFunc}.
%%%%%%%%%%%%%%%%%%%%%%%%%%%%%%%%%%%%%%%%%%%%%%%%%%
%\begin{figure}
%\includegraphics[width=\columnwidth]{./eps/JRplot.pdf}
%\caption{\changeB{The general functional form of $\mathcal{J}_{\rm r}(l,\kappa)$ for $l = 1$. For $\kappa \rightarrow 0$, the integral can be directly evaluated and yields the hypergeometric function in Eq.~\eqref{eq:F_int}. Similarly for large $\kappa$, one finds the approximate form given in Eq.~\eqref{eq:Jinf}. }}
%\label{fig:JRplot}
%\end{figure}
%%%%%%%%%%%%%%%%%%%%%%%%%%%%%%%%%%%%%%%%%%%%%%%%%%
%%%%%%%%%%%%%%%%%%%%%%%%%%%%%%%%%%%%%%%%%%%%%%%%%%
%\begin{figure}
%\includegraphics[width=\columnwidth]{./eps/JRplotL.pdf}
%\caption{\changeB{$\mathcal{J}_{\rm r}(l,\kappa)$ for different values of $l$. For parameter values yielding $\lambda < 1$, the $l \rightarrow \infty$ contour is representative of the emission from all loops in the distribution before matter-radiation equality. Contours of smaller $l$ then show the contribution from correspondingly smaller loops, with $l$ = 1 representing emission from all loops with $L \leq \Gamma G\mu t$.}\todo{Do a similar plot for the matter loops?}}
%\label{fig:JRplotL}
%\end{figure}
%%%%%%%%%%%%%%%%%%%%%%%%%%%%%%%%%%%%%%%%%%%%%%%%%%
%%%%%%%%%%%%%%%%%%%%%%%%%%%%%%%%%%%%%%%%%%%%%%%%%%
%%%%%%%%%%%%%%%%%%%%%%%%%%%%%%%%%%%%%%%%%%%%%%%%%%

For $t>t_{\rm eq}$, we need to consider the evolution of loops created in the radiation dominated era as well as newly created loops. For the former, two changes occur. One is due to the change in the time-dependence of the evolution, which leads to an extra factor of $(t_{\rm eq}/t)^{1/2}$ and the other is due to the fact that the maximal loop scale now evolves from $\beta t_{\rm eq}$ at $t=t_{\rm eq}$ to $\beta  t_{\rm eq}(1+\lambda-\lambda t/t_{\rm eq})$ at $t>t_{\rm eq}$. Put together, this yields
%%%%%%%%%%%%%%%%%%%%%%%%%%%%%%%%%%%%%%%%%%%%%%%%%%
\begin{align} \label{eq:dNdtR}
\left.\frac{\id N_{\gamma}}{\id t}\right|_{\rm r} &= \frac{\alpha \Gamma_{\gamma}}{3} \frac{\mathcal{I}^2(1+\lambda)^{3/2}}{(\Gamma G\mu)^{7/6}\,t^{8/3}} \,\frac{\Gamma[2/3]}{(x_{\rm c} T)^{2/3}}
\nonumber\\
&\qquad
\times \begin{dcases} 
\mathcal{J}_{\rm r}\left(\frac{t_{\rm end}}{t_{\rm eq}}-1, \kappa\right) 
& (t \leq t_{\rm eq}) 
\\ 
 \mathcal{J}_{\rm r}\left(\frac{t_{\rm end}}{t}-1, \kappa\right) \left( \frac{t_{\rm eq}}{t}\right)^{1/2} & ( t_{\rm eq}< t \leq t_{\rm end}),
\end{dcases}
\end{align}
%%%%%%%%%%%%%%%%%%%%%%%%%%%%%%%%%%%%%%%%%%%%%%%%%%
for the loops created in the radiation-dominated era. For loops created in the matter-dominated era we similarly have
%%%%%%%%%%%%%%%%%%%%%%%%%%%%%%%%%%%%%%%%%%%%%%%%%%
\begin{subequations}
\label{eq:dNdtM}
\begin{align} 
\left.\frac{\id N_{\gamma}}{\id t}\right|_{\rm m} &= \frac{\alpha_{\rm m} \Gamma_{\gamma}}{3} \frac{\mathcal{I}^2(1+\lambda)}{(\Gamma G\mu)^{2/3}\,t^{8/3}}\,\frac{\Gamma[2/3]}{(x_{\rm c} T)^{2/3}} 
\\ \nonumber
&\qquad
\times \begin{dcases} 
\mathcal{J}_{\rm m}\left(\frac{t_{\rm end}}{t}-1, \frac{t_{\rm end}}{t_{\rm eq}}-1, \kappa\right) 
& ( t_{\rm eq}< t \leq t_{\rm end})
\\ 
 \mathcal{J}_{\rm m}\left(0, \frac{t_{\rm end}}{t_{\rm eq}}-1, \kappa\right) & 
 (t_{\rm end}<t) ,
\end{dcases}
\\ \label{eq:Jm}
\mathcal{J}_{\rm m}(l_a, l_b,\kappa)&=
\int^{l_b}_{l_a} \id l' \, \frac{{l'}^{1/3}}{(1+l')^{2}} \,\frac{\Gamma\left[ \frac{2}{3}, \kappa\,l' \right]}{\Gamma[2/3]}.
\end{align}
\end{subequations}
%%%%%%%%%%%%%%%%%%%%%%%%%%%%%%%%%%%%%%%%%%%%%%%%%%
with $1/\lambda=\frac{t_{\rm end}}{t_{\rm eq}}-1$.
Note that the integral over the loop distribution is slightly different from the one in the radiation era. Asymptotic forms for both $\mathcal{J}_{\rm r}(l,\kappa)$ and $\mathcal{J}_{\rm m}(l_{\rm a}, l_{\rm b},\kappa)$ can be found in Appendix \ref{sec:entropyInjApprox}.

\begin{figure}
\includegraphics[width=\columnwidth]{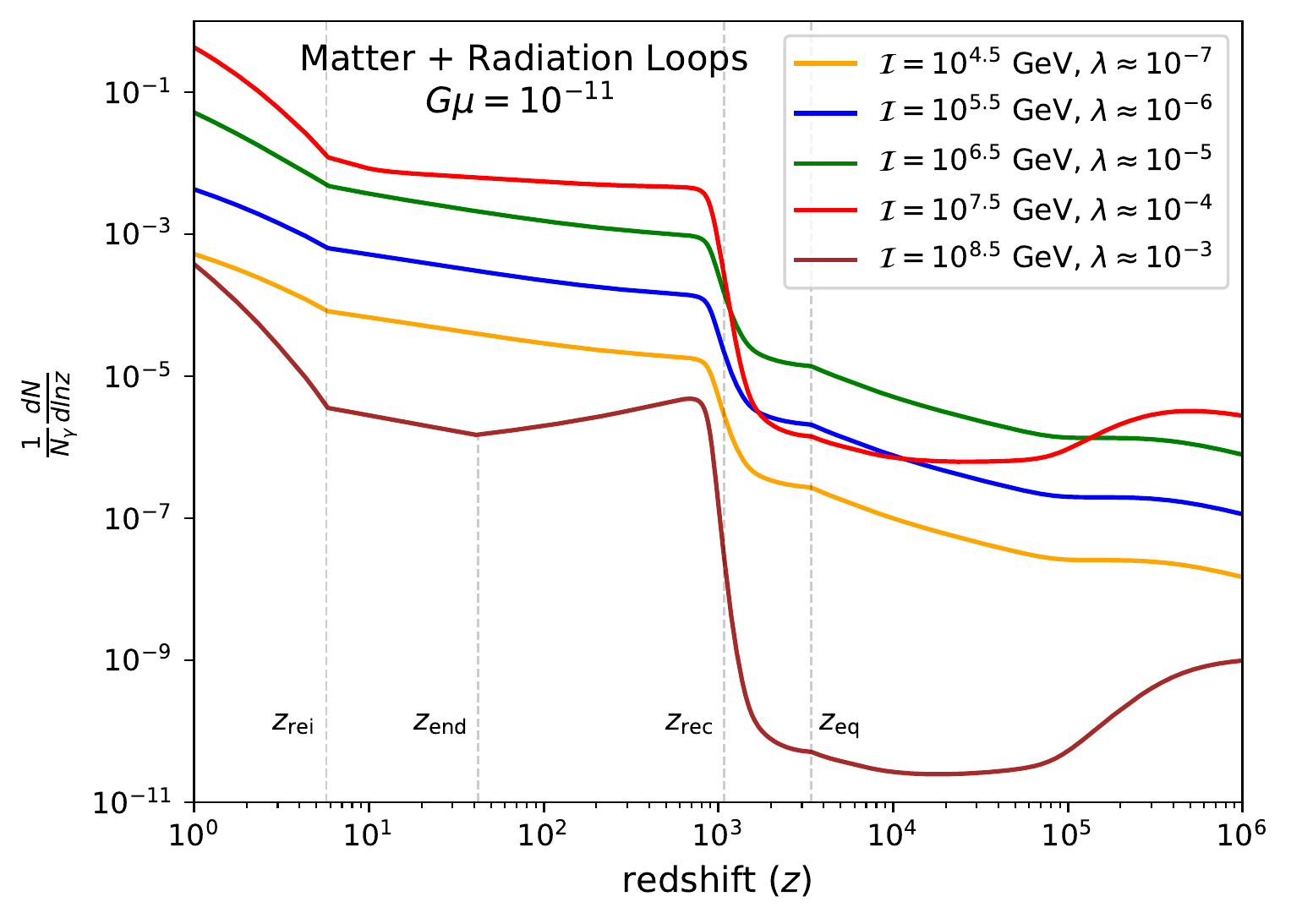}
\caption{The fractional entropy injection at different redshifts. Along contours of constant $G\mu$, increasing the current boosts the signal, until the current becomes too high and the window function (see Fig.~\ref{fig:windowFunc}) rapidly shuts off emission. For the brown contour, $z_{\rm end}$ marks the redshift at which all radiation loops fully evaporate. All other sharp features are inherited from $x_{\rm c}$.}
\label{fig:dNdlnzTOT}
\end{figure}

The total entropy injection rate is given by the sum of Eq.~\eqref{eq:dNdtR} and Eq.~\eqref{eq:dNdtM} in the relevant redshift regimes. We show the fractional injection rate in Fig.~\ref{fig:dNdlnzTOT}. One major difference between energy and entropy injection is that here we do not have a constant $\lambda \rightarrow \infty$ contour. This is because at large currents, $\kappa \sim \mathcal{I}^4$, and for large $\kappa$ the window function closes rapidly, as can be seen in Fig.~\ref{fig:windowFunc}. This shuts off all emission above $x_{\rm c}$, causing a rapid decay of the signal. For primordial spectral distortions, only the injection at $z \geq z_{\rm rec}$ is relevant. However, it is evident from the figure that large fractions of photons are injected after recombination, hinting at potential synergies with experiments detecting excess emission at low frequencies such as ARCADE-2 or 21-cm surveys.
%%%%%%%%%%%%%%%%%%%%%%%%%%%%%%%%%%%%%%%%%%%%%%%%%%
%%%%%%%%%%%%%%%%%%%%%%%%%%%%%%%%%%%%%%%%%%%%%%%%%%

\begin{figure}
\includegraphics[width=\columnwidth]{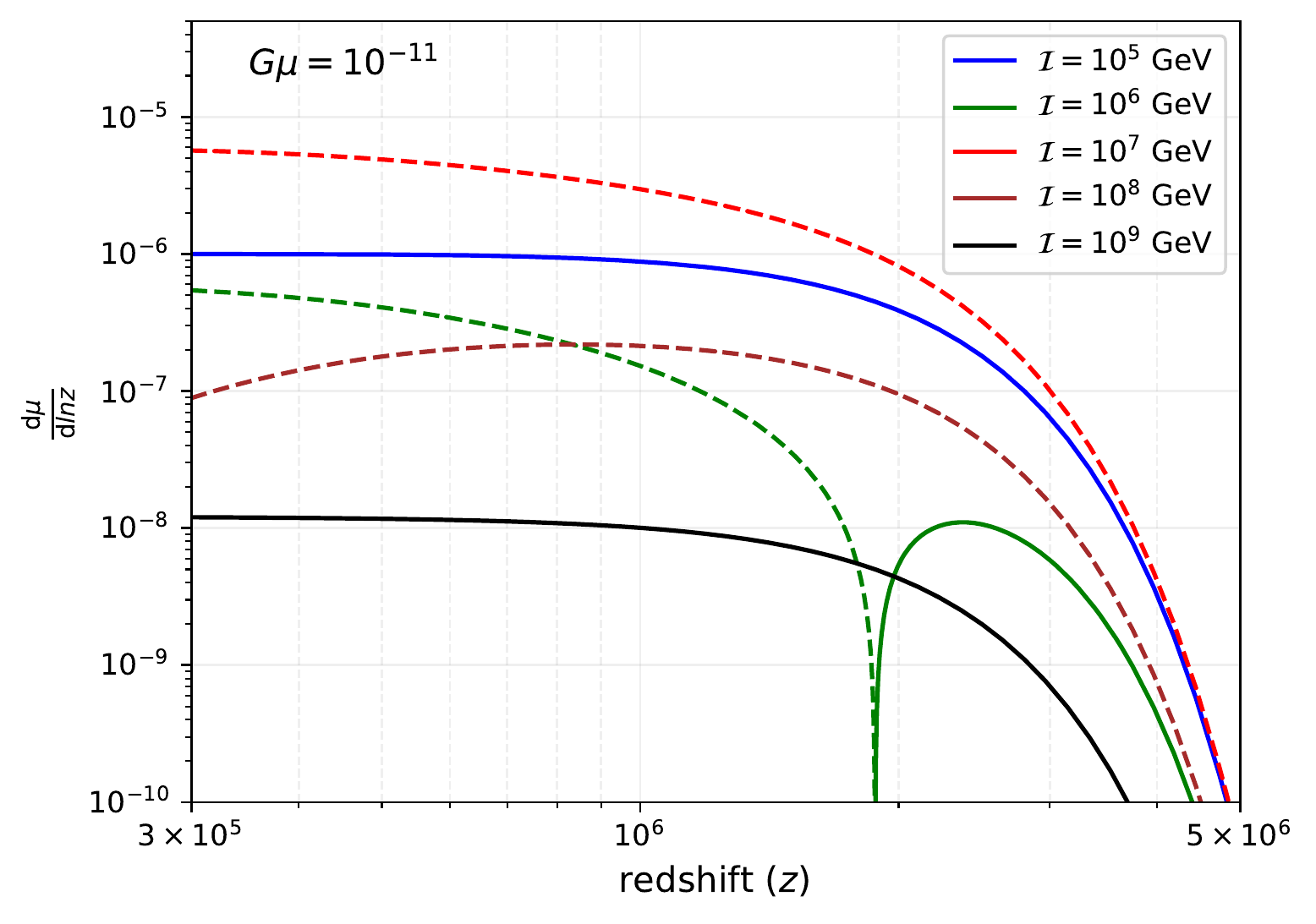}
\\
\vspace{-4mm}
\caption{The buildup of a $\mu$ distortion from energy and entropy injection for a range of cosmic string parameters. Solid lines represent positive $\mu$ (where energy injection is dominant), while dashed lines show negative $\mu$ (indicating entropy injection is the stronger process). The green contour illustrates an interesting case where a transition takes place between strong energy and strong entropy injection, which would lead to an anomalously low total $\mu$.}
\label{fig:dmu}
\end{figure}

Deep in the $\mu$ era ($3\times 10^5 \leq z \leq 2\times 10^6$) energy redistribution of the injected photons is still efficient through Compton scattering. This allows us to quantify the $\mu$ distortion by computing the relative energy and entropy injections during this era and using Eq.~\eqref{eq:muApprox}. Figure~\ref{fig:dmu} illustrates the net buildup of the $\mu$ distortion for some examples of string parameters. The figure indicates that the amplitude of a $|\mu|$ distortion can be boosted substantially compared to the case where entropy injection is neglected, leading to more stringent constraints in some regions of parameter space.
%%%%%%%%%%%%%%%%%%%%%%%%%%%%%%%%%%%%%%%%%%%%%%%%%%
%%%%%%%%%%%%%%%%%%%%%%%%%%%%%%%%%%%%%%%%%%%%%%%%%%

%%%%%%%%%%%%%%%%%%%%%%%%%%%%%%%%%%%%%%%%%%%%%%%%%%
\begin{figure}
\includegraphics[width=\columnwidth]{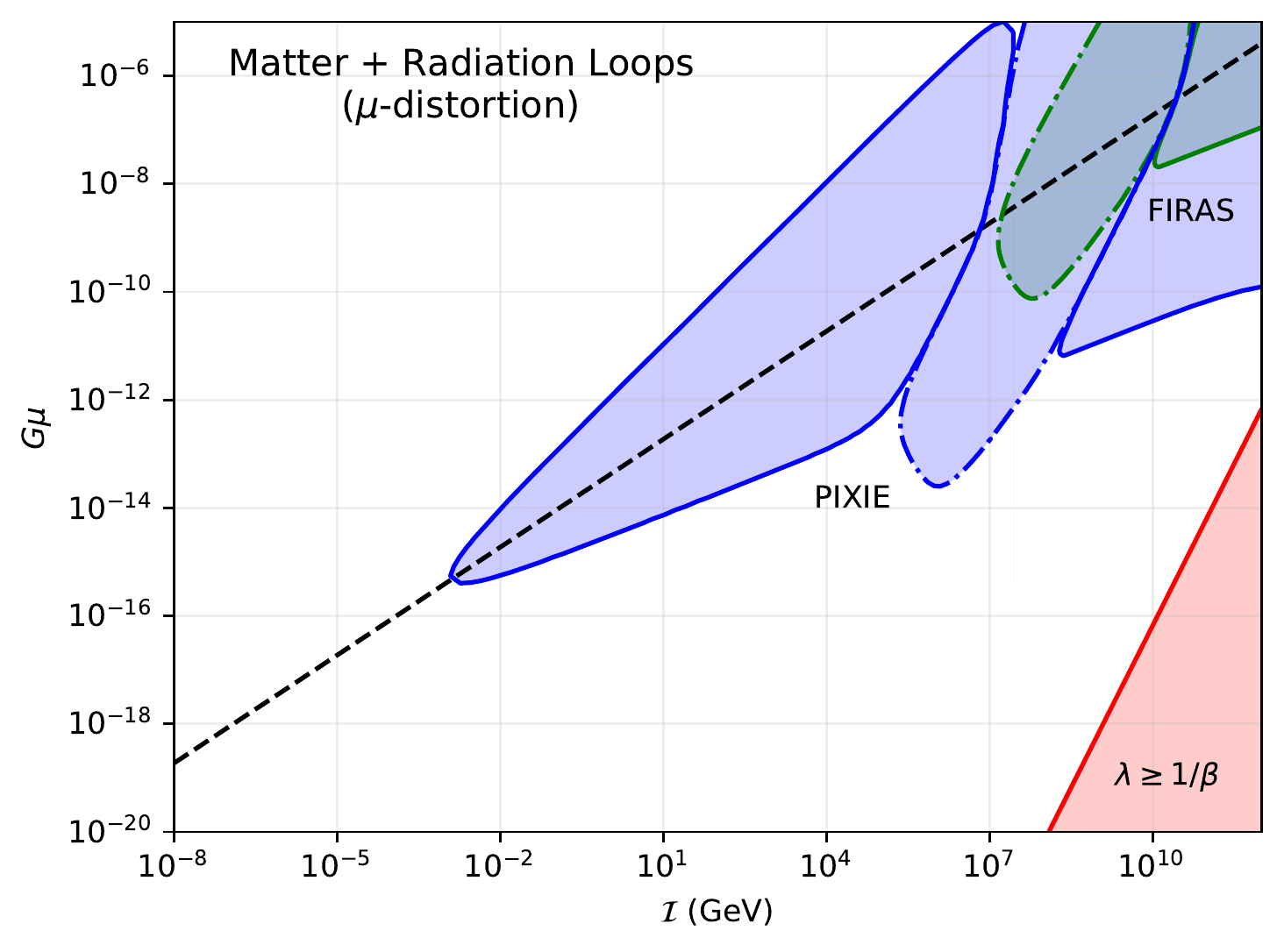}
\\
\vspace{-5mm}
\caption{Constraints ($2\sigma$) obtained requiring $|\mu| \leq 8.4\times10^{-5}$ from \COBEF, and $|\mu| \leq 2.8\times 10^{-8}$ for a {\it PIXIE}-like experiment. At high currents and string tensions, energy injection is responsible for sourcing a positive $\mu$ distortion. However, as the string parameters are lowered, we see a large strip of parameter space that is ruled out due to excessive photon production, yielding a negative $\mu$. Finally, {\it PIXIE} is capable of constraining further regions of parameter space from energy injection at still smaller currents and tensions. Dashed-dotted lines correspond to negative $\mu$ distortions.}
\label{fig:drhodNConstraint}
\vspace{-3mm}
\end{figure}
%%%%%%%%%%%%%%%%%%%%%%%%%%%%%%%%%%%%%%%%%%%%%%%%%%

In Fig.~\ref{fig:drhodNConstraint} we show the $\mu$ distortion constraints obtained by considering both entropy and energy injection from a distribution of string loops. The figure clearly shows a region of parameter space that is ruled out from entropy injection but was missed in the analysis of \cite{Tashiro} and \cite{Miyamoto}. 
For {\it PIXIE}, the left and right solid contours correspond to the usual energy injection constraints seen in Fig.~\ref{fig:drhoConstraint}, while the central contour is novel, and comes from entropy injection. For the \COBEF results, the sensitivity to $|\mu|$ is too weak to resolve the energy injection constraint at lower currents. 
The energy injection constraints appear weaker than in Fig.~\ref{fig:drhoConstraint} simply because here we consider a smaller redshift window in which the injection is active. This choice is because for $z \leq 3 \times 10^5$, injected photons are not efficiently redistributed and hence Eq.~\eqref{eq:muApprox} is not valid. To consider the spectral distortions produced below this redshift, one has to go beyond the simple analytic picture and solve the problem numerically. In the remainder of this work we compute and analyze these numerical solutions using \texttt{CosmoTherm}.

%--------------------------------------------
\section{Numerical Implementation in \texttt{CosmoTherm}}
\label{sec:numericalImplementation}
%--------------------------------------------
%%%%%%%%%%%%%%%%%%%%%%%%%%%%%%%%%%%%%%%%%%%%%%%%%%
\begin{figure*}
\centering 
\includegraphics[width=\columnwidth]{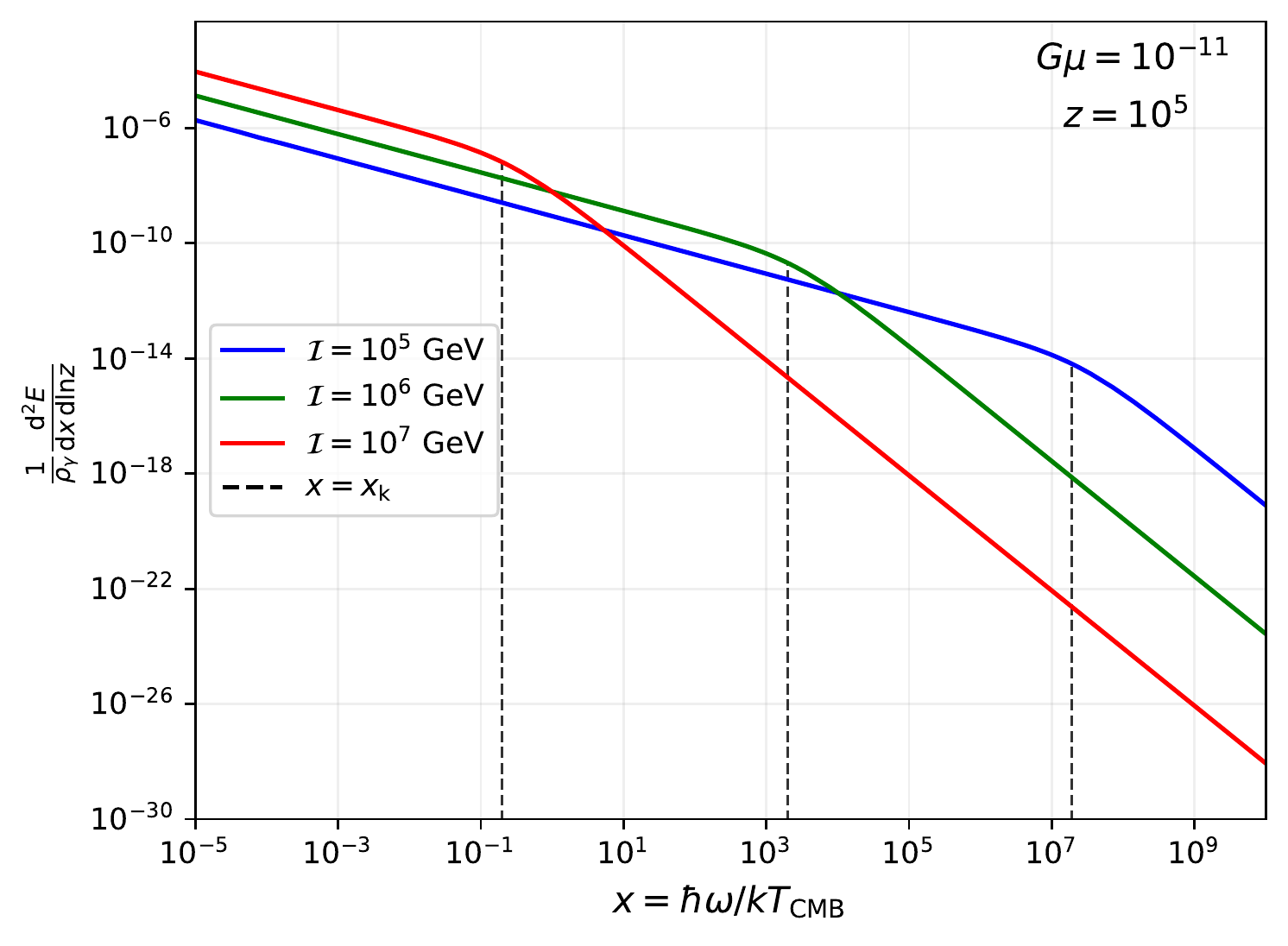}
\hspace{4mm}
\includegraphics[width=\columnwidth]{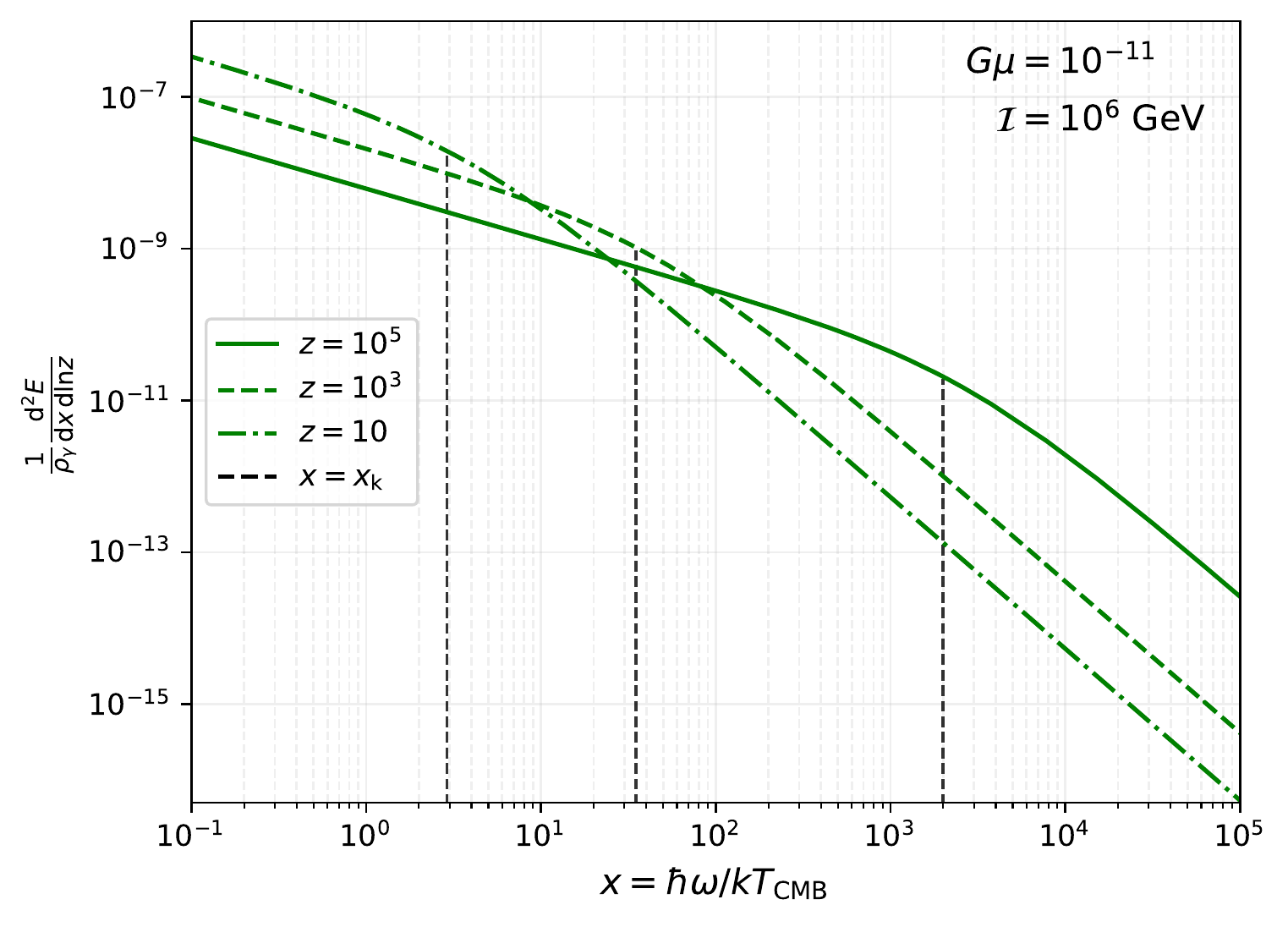}
\caption{The instantaneous spectrum of photons produced by a network of superconducting cosmic string loops as a function of the dimensionless frequency $x = \hbar \omega/k T$. The left plot shows the response of the spectrum at constant $G\mu$ and redshift, for variations of $\mathcal{I}$. On the right we fix $G\mu$ and $\mathcal{I}$ to observe how the injection spectrum varies at different redshifts. For increasing current and decreasing redshift, the position of the knee moves to lower frequencies, indicating that fewer high energy photons are produced by the network. The knee position $x_{\rm k}= \omega_{\rm H}(t)/[\lambda l_{\rm c} T]$ by indicated as dashed vertical lines.}
\label{fig:exactSpec}
\end{figure*}
%%%%%%%%%%%%%%%%%%%%%%%%%%%%%%%%%%%%%%%%%%%%%%%%%%

In order to provide robust spectral distortion constraints, as well as to validate our analytic treatment, we now aim to implement source terms from cosmic strings into
\texttt{CosmoTherm} \citep{Chluba2011therm}, which is a state of the art cosmological thermalization code that allows explicitly treating multiple energy release and photon injection scenarios.\footnote{\url{www.Chluba.de/CosmoTherm}}
Practically, \texttt{CosmoTherm} takes as input a redshift-dependent injection function describing the change in photon occupation number, and evolves it forward to determine the spectral distortion signature. To determine the total source term, we average Eq.~\eqref{EnergyNum} over the full loop distribution present at any given time. For clarity we first consider injection in the radiation era (i.e., cases with $t\leq t_{\rm eq}$), and then generalize to the matter-dominated era.
Inserting the corresponding loop distribution, we then have
%%%%%%%%%%%%%%%%%%%%%%%%%%%%%%%%%%%%%%%%%%%%%%%%%%
\begin{align}
\left.\frac{\id^2 E_{\gamma}}{\id t \id \omega}\right|_{\rm r} &= \int_0^{L_{\rm max}} \id L \, \frac{\id^2 E_{\gamma}^c}{\id t \id \omega} \, \left.\frac{\id N_{\rm loops}}{\id L}\right|_{\rm r} \nonumber\\
&=\left(\frac{\alpha \Gamma_{\gamma}}{3}\right) \frac{\mathcal{I}^2 }{t^{3/2}} \frac{\left( 1 + \lambda\right)^{3/2}}{\omega^{2/3}} \int_0^{L_{\rm up}(\omega)} \frac{\id L\, L^{1/3}}{(L+\Gamma G\mu t)^{5/2}},
\end{align}
%%%%%%%%%%%%%%%%%%%%%%%%%%%%%%%%%%%%%%%%%%%%%%%%%%
where the upper bound of the $L$ integral depends on $\omega$.
Since the maximal frequency of the emission for a loop of length $L$ is given by $\omega_{\rm max}(L)=\mu^{3/2}/[\mathcal{I}^{3} L]$ \citep{Cai}, at a given frequency $\omega$ this means that a limiting length, 
%%%%%%%%%%%%%%%%%%%%%%%%%%%%%%%%%%%%%%%%%%%%%%%%%%
\begin{align}
L_{\rm lim}(\omega)=\frac{\mu^{3/2}}{\mathcal{I}^{3} \omega},
\end{align}
%%%%%%%%%%%%%%%%%%%%%%%%%%%%%%%%%%%%%%%%%%%%%%%%%%
should not be exceeded, implying $L_{\rm up}=[\beta t, L_{\rm lim}(\omega)]_<$. With the substitution $l=L/[\Gamma G\mu t]$ we then find
%%%%%%%%%%%%%%%%%%%%%%%%%%%%%%%%%%%%%%%%%%%%%%%%%%
\begin{align} \label{eq:radSpec}
\left.\frac{\id^2 E_{\gamma}}{\id t \id \omega}\right|_{\rm r} &=\left( \frac{\alpha \Gamma_{\gamma}}{3}\right)\frac{\mathcal{I}^2\,\left( 1 + \lambda\right)^{3/2}}{(\Gamma G\mu)^{7/6}\,t^{8/3}}\times \frac{\mathcal{F}_{\rm r}(l_{\rm up})}{\omega^{2/3}},
\end{align}
%%%%%%%%%%%%%%%%%%%%%%%%%%%%%%%%%%%%%%%%%%%%%%%%%%
where $\mathcal{F}_{\rm r}(l)=\mathcal{J}_{\rm r}(l,0)$ with various approximations discussed in Appendix \ref{sec:entropyInjApprox}. Since $l_{\rm up}$ scales strongly with frequency, the emission spectrum steepens at high frequencies.

Also considering the evolution of loops from the radiation dominated era at $t_{\rm eq} \leq t \leq t_{\rm end}$ we then find
%%%%%%%%%%%%%%%%%%%%%%%%%%%%%%%%%%%%%%%%%%%%%%%%%%
\begin{align}\label{eq:radSpecMat}
\left.\frac{\id^2 E_{\gamma}}{\id t \id \omega}\right|_{\rm r} &= \left( \frac{\alpha \Gamma_{\gamma}}{3}\right)\frac{\mathcal{I}^2\,\left( 1 + \lambda\right)^{3/2}}{(\Gamma G\mu)^{7/6}\,t^{8/3}} \,\frac{1}{\omega^{2/3}}
\\ \nonumber
&\qquad
\times \begin{dcases} 
\mathcal{F}_{\rm r}\left(\left[\frac{t_{\rm end}}{t_{\rm eq}}-1, l_{\rm lim}\right]_<\right) 
& (t \leq t_{\rm eq}) 
\\ 
 \mathcal{F}_{\rm r}\left(\left[\frac{t_{\rm end}}{t}-1, l_{\rm lim}\right]_<\right)\left( \frac{t_{\rm eq}}{t}\right)^{1/2} & ( t_{\rm eq}< t \leq t_{\rm end}).
\end{dcases}
\end{align}
%%%%%%%%%%%%%%%%%%%%%%%%%%%%%%%%%%%%%%%%%%%%%%%%%%
For the loops created in the matter-dominated era, we similarly find
%%%%%%%%%%%%%%%%%%%%%%%%%%%%%%%%%%%%%%%%%%%%%%%%%%
\begin{align} \label{eq:matSpec}
\left.\frac{\id^2 E_{\gamma}}{\id t \id \omega }\right|_{\rm m} &= \frac{\alpha_{\rm m} \Gamma_{\gamma}}{3} \frac{\mathcal{I}^2(1+\lambda)}{(\Gamma G\mu)^{2/3}\,t^{8/3}} \,\frac{1}{\omega^{2/3}}
\\ \nonumber
&\quad
\times \begin{dcases} 
\mathcal{F}_{\rm m}\left(\frac{t_{\rm end}}{t}-1, \left[\frac{t_{\rm end}}{t_{\rm eq}}-1, l_{\rm lim}\right]_< \right) 
& ( t_{\rm eq}< t \leq t_{\rm end})
\\ 
 \mathcal{F}_{\rm m}\left(0, \left[\frac{t_{\rm end}}{t_{\rm eq}}-1, l_{\rm lim}\right]_< \right) & 
 (t_{\rm end}<t) ,
\end{dcases}
\end{align}
%%%%%%%%%%%%%%%%%%%%%%%%%%%%%%%%%%%%%%%%%%%%%%%%%%
with $\mathcal{F}_{\rm m}\left(l_a, l_b\right)=\mathcal{J}_{\rm m}\left(l_a, l_b, 0\right)=\mathcal{F}_{\rm m}\left(l_b\right)-\mathcal{F}_{\rm m}\left(l_b\right)$ while $l_b>l_a$ and zero otherwise. The expression for $\mathcal{F}_{\rm m}\left(l\right)$ is given in Eq.~\eqref{eq:F_int_m}. This allows us to compute all the required emission spectra. 

As a simple example, let us approximate the spectrum of photons produced before matter-radiation equality. Starting from Eq.~\eqref{eq:radSpec}, we can use the approximation for $\mathcal{F}_{\rm r}(l_{\rm up})$ found in Appendix \ref{sec:entropyInjApprox} to obtain
%%%%%%%%%%%%%%%%%%%%%%%%%%%%%%%%%%%%%%%%%%%%%%%%%%
\begin{align} 
\left.\frac{\id^2 E_{\gamma}}{\id t \id \omega}\right|_{\rm r} &\approx \left( \frac{\alpha \Gamma_{\gamma}}{3}\right) \frac{\mathcal{I}^2 (1+\lambda)^{3/2}}{\omega^{2/3} (\Gamma G\mu)^{7/6} t^{8/3}} \frac{\mathcal{F_{\infty}}}{1+(l_c/l_{\rm up})^{7/6}},
\end{align}
%%%%%%%%%%%%%%%%%%%%%%%%%%%%%%%%%%%%%%%%%%%%%%%%%%
where $\mathcal{F}_{\infty} = \Gamma\left( \frac{7}{6}\right) \Gamma \left( \frac{7}{3}\right)/\sqrt{\pi} \approx 0.6232$ and $l_{\rm c} \approx 1.314$ were determined by inspecting the asymptotic regimes of $\mathcal{F}_{\rm r}(l_{\rm up})$ The approximation is valid to within $10 \% $ for $l_{\rm up} \gtrsim 0.1$, and we discuss better approximations in the Appendix. For a given frequency and parameter set, $l_{\rm up}$ can take on different values. Specifically,
%%%%%%%%%%%%%%%%%%%%%%%%%%%%%%%%%%%%%%%%%%%%%%%%%%
\begin{align}
l_{\rm up} &= \begin{dcases} 
\frac{t_{\rm end}}{t_{\rm eq}}-1 = \frac{1}{\lambda} \hspace{10mm}
& \left(\omega \leq \omega_{\rm H} \right)
\\ 
\frac{L_{\rm lim}(\omega)}{\Gamma G\mu t}=\frac{1}{\lambda}\frac{\omega_{\rm H}}{\omega}  \hspace{10mm} & \left(\omega > \omega_{\rm H}\right),
\end{dcases}
\end{align}
%%%%%%%%%%%%%%%%%%%%%%%%%%%%%%%%%%%%%%%%%%%%%%%%%%
where, $\omega_{\rm H} \equiv \omega_{\rm max}(\beta t)= (G\mu)^{3/2}/[G^{3/2} \mathcal{I}^3 \beta t]$. For most parameter values, $\lambda \ll 1$, allowing us to simplify the spectrum as
%%%%%%%%%%%%%%%%%%%%%%%%%%%%%%%%%%%%%%%%%%%%%%%%%%
\begin{align} \label{eq:specApprox}
\left.\frac{\id^2 E_{\gamma}}{\id t \id \omega}\right|_{\rm r}  
&\approx
\left( \frac{\alpha \Gamma_{\gamma}}{3}\right) \frac{\mathcal{I}^2(1+\lambda)^{3/2}}{\omega^{2/3}\,(\Gamma G\mu)^{7/6}\,t^{8/3}}\,
\begin{cases} 
\mathcal{F}_\infty
&\; \omega \leq \omega_{\rm H}
\\[1mm]
\frac{\mathcal{F}_\infty}{1+(\omega/\omega_{\rm k})^{7/6}}
&\; \omega > \omega_{\rm H}
\end{cases}
\nonumber \\
&\approx
\left( \frac{\alpha \Gamma_{\gamma}}{3}\right)\frac{\mathcal{I}^2 (1+\lambda)^{3/2}}{\omega^{2/3}\,(\Gamma G\mu)^{7/6}\,t^{8/3}}
\frac{\mathcal{F}_\infty}{1+(\omega/\omega_{\rm k})^{7/6}}.
\end{align}
%%%%%%%%%%%%%%%%%%%%%%%%%%%%%%%%%%%%%%%%%%%%%%%%%%
Here, we defined $\omega_{\rm k} = \omega_{\rm H}/[\lambda l_{\rm c}]$. Going from the first to the second line, we note that for small $\lambda$, we have $\omega \leq \omega_{\rm H} \ll \omega_k$. At low frequencies ($\omega \ll \omega_{\rm k}$), the spectrum decays as $\omega^{-2/3}$. As we increase $\omega$, we encounter a knee in the spectrum at $\omega \approx \omega_{\rm k}$ after which the spectrum decays much more rapidly as $\omega^{-11/6}$. The position of this knee can be seen for different values of the current and redshift in Fig.~\ref{fig:exactSpec}, where we illustrate the exact spectrum which we compute numerically. 

\begin{figure}
\includegraphics[width=\columnwidth]{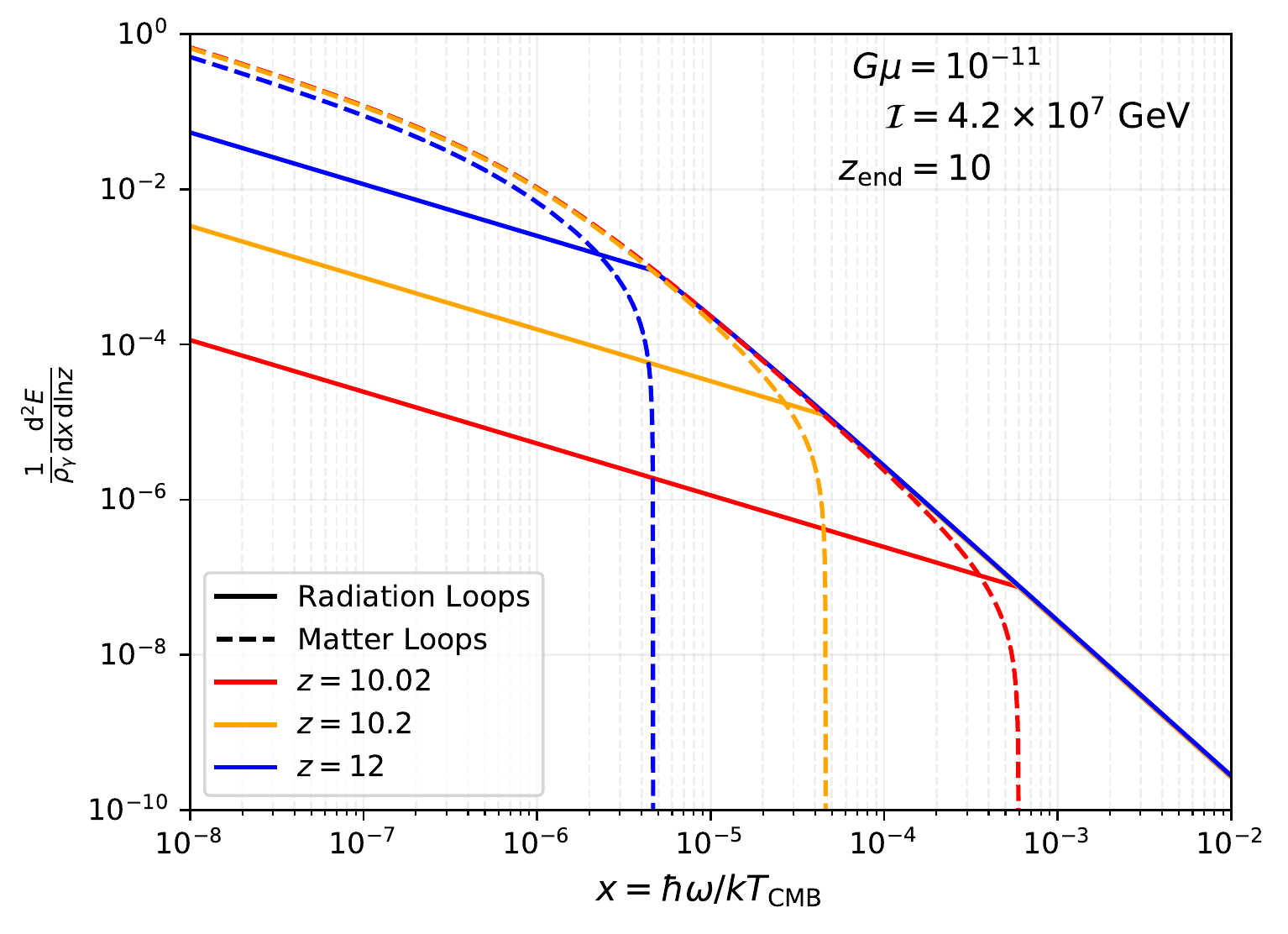}
\caption{Contributions to the emission spectrum from loops formed in the matter era (dashed) and radiation era (solid) at different redshifts near $z_{\rm end}$. At $z \simeq z_{\rm end}$, the radiation loops play a dominant role only at the highest frequencies. At low frequencies, the contribution from matter and radiation loops is roughly equal at $z \simeq 20$. Before this time, matter loops are subdominant for the entire frequency range for the chosen parameters.}
\label{fig:specComp}
\end{figure}

In Fig.~\ref{fig:specComp} we separately show the contribution to the emission spectrum from loops formed in the matter and radiation eras for $G\mu =10^{-11}$ and $\mathcal{I}=4.2\times 10^7\,{\rm GeV}$. For $z\gg z_{\rm end}\simeq 10$, the radiation loops dominate the contribution at all frequencies. As one approaches $z \simeq z_{\rm end}$, the matter loops play a more important role, except at the highest frequencies. This is because the highest frequency photons are produced by the smallest loops, which still date back to the radiation era until $z$ drops below $z_{\rm end}$.

%%%%%%%%%%%%%%%%%%%%%%%%%%%%%%%%%%%%%%%%%%%%%%%%%%
%%%%%%%%%%%%%%%%%%%%%%%%%%%%%%%%%%%%%%%%%%%%%%%%%%

To validate this approximation, we can readily integrate over frequencies to find the total energy release rate
%%%%%%%%%%%%%%%%%%%%%%%%%%%%%%%%%%%%%%%%%%%%%%%%%%
\begin{align}
\left.\frac{\id Q}{\id t }\right|_{\rm r} &= \frac{2}{3} \frac{\alpha \Gamma_{\gamma}}{G^{1/2}} \frac{\mathcal{I}}{\Gamma^{3/2} G\mu } \frac{(1+\lambda)^{3/2}}{t^3} \times \frac{3\pi \mathcal{F}_{\infty}}{7 l_{\rm c}^{1/3} \cos(3\pi/14)}.
\end{align}
%%%%%%%%%%%%%%%%%%%%%%%%%%%%%%%%%%%%%%%%%%%%%%%%%%
This matches the exact result found in Eq.~\eqref{eq:heatingRad} to better than $3 \% $. This small departure is caused by corrections to the total emission by means of our approximation to $\mathcal{F}_{\rm r}(l)$. When using a refined approximation for $\mathcal{F}_{\rm r}(l) $ in Eq.~\eqref{eq:Fr} and numerically integrating the function find agreement with the exact result at the level of $\simeq 0.06\%$. 

%%%%%%%%%%%%%%%%%%%%%%%%%%%%%%%%%%%%%%%%%%%%%%%%%%
\begin{figure*}
\centering 
\includegraphics[width=\columnwidth]{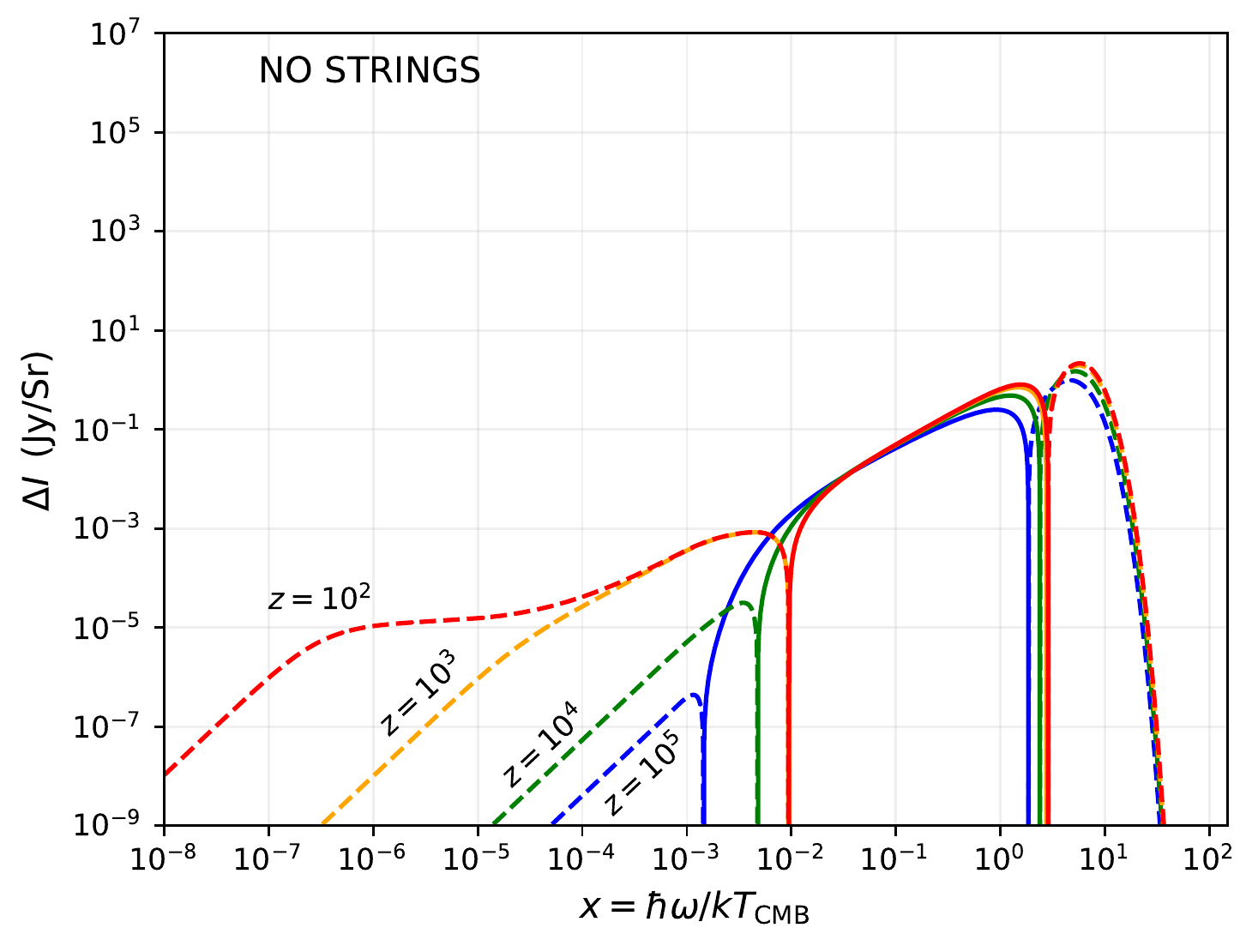}
\hspace{4mm}
\includegraphics[width=\columnwidth]{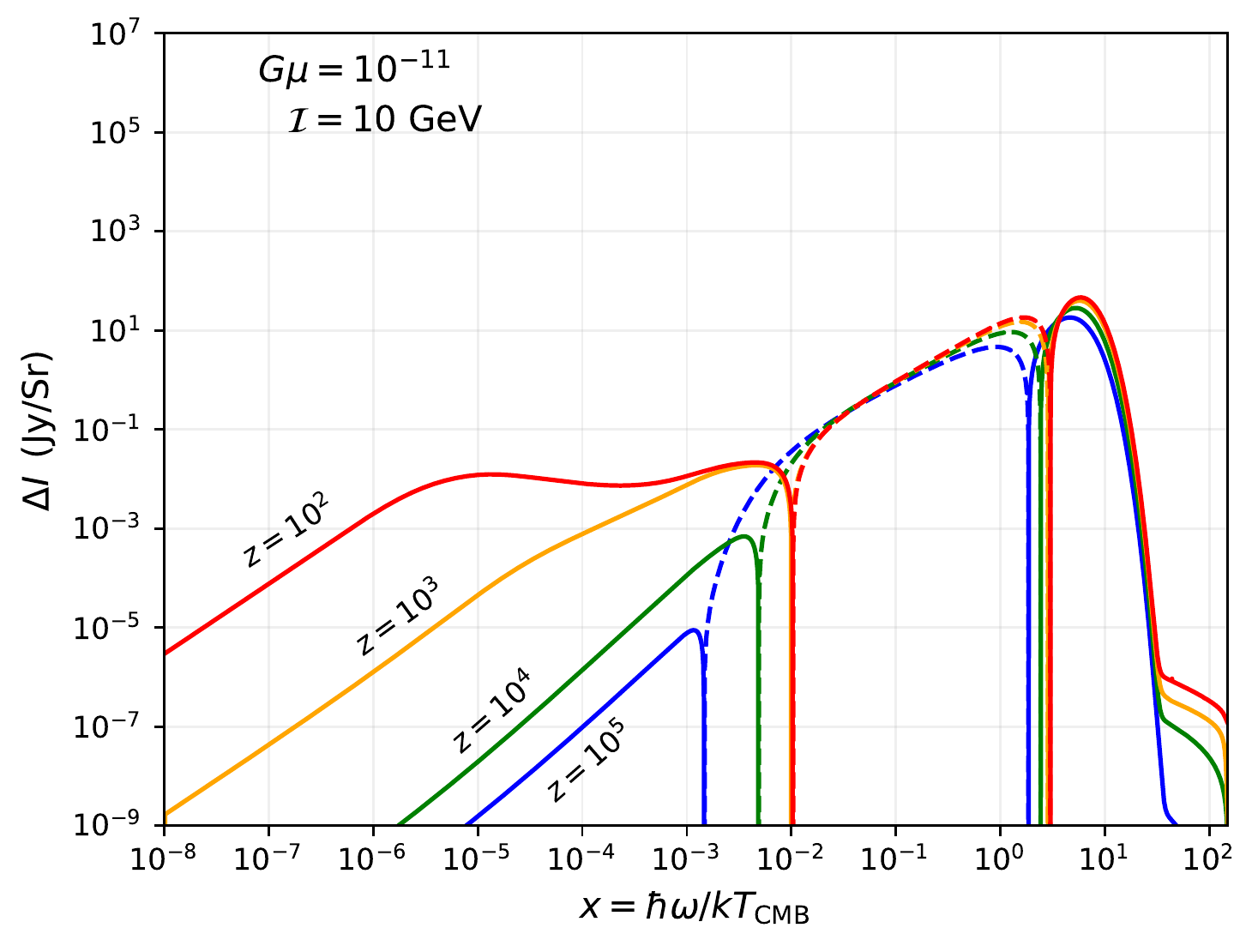}
\includegraphics[width=\columnwidth]{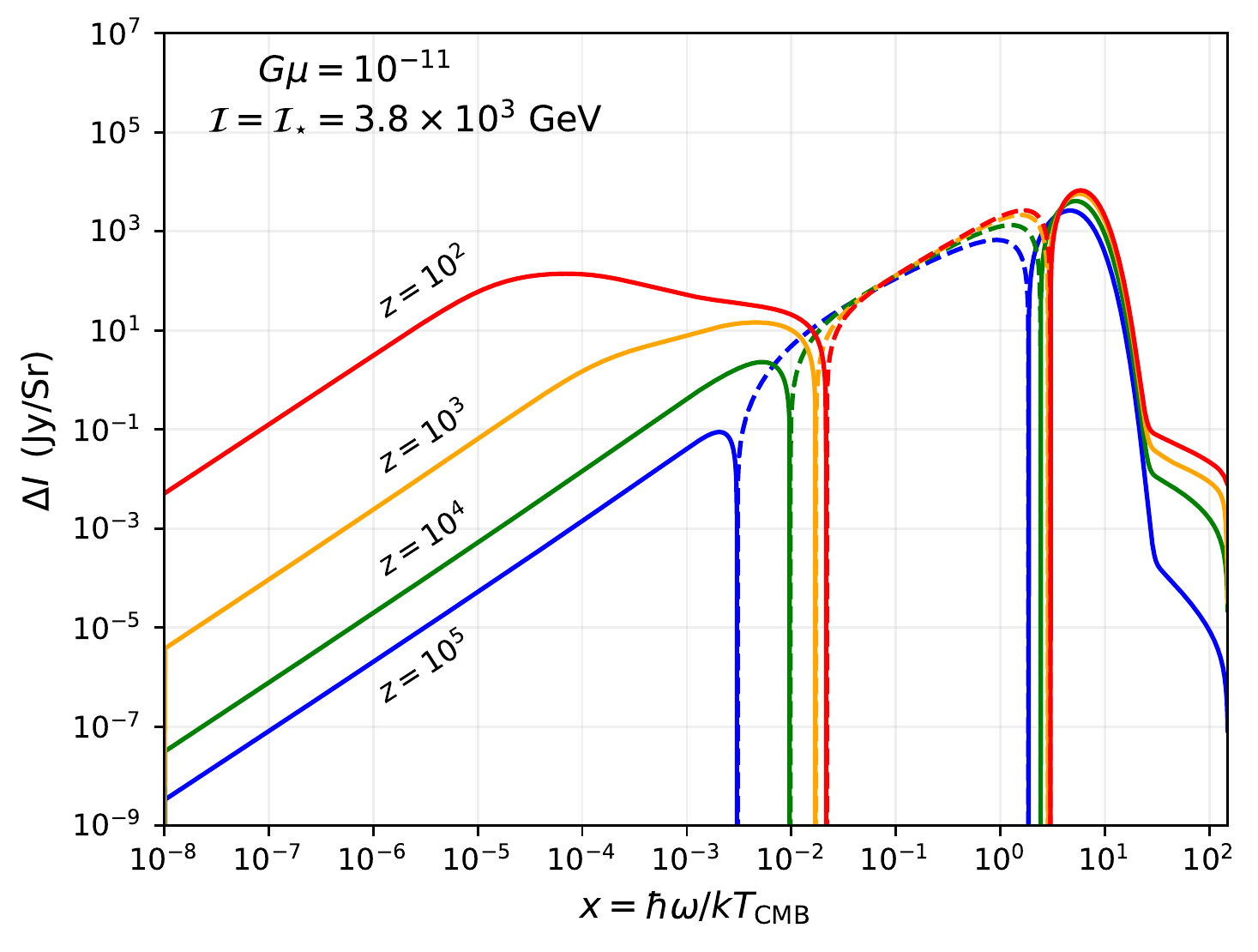}
\hspace{4mm}
\includegraphics[width=\columnwidth]{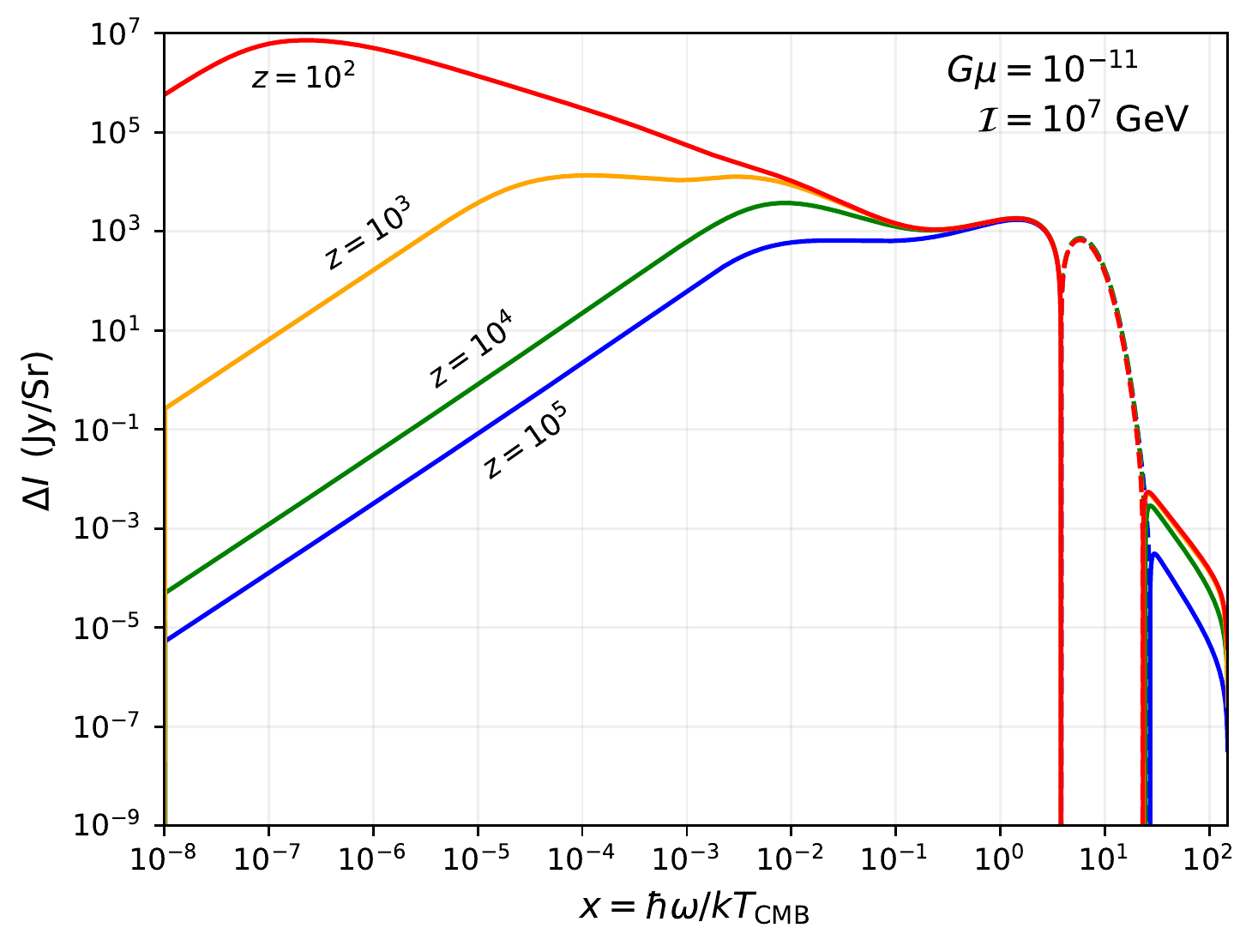}
\\
\caption{Snapshots of the distortion spectra $\Delta I = I_{\rm CT} - I_{\rm pl}$ as output from \texttt{CosmoTherm} at different redshifts. Dashed lines represent a deficit of photons (a negative distortion) when compared with a blackbody. Top left: Output with no external heating sources. A small (negative) distortion is generated through the adiabatic cooling of electrons, which continuously extracts energy from the photon background. Top right: Inclusion of a weak source term from string loops. A positive $\mu$ distortion is now generated through injection of energy at early times. At late times, a low-frequency spectrum of photons is generated. Bottom left: A string source with $\mathcal{I} = \mathcal{I}_*$, which produces a sizeable $\mu$ distortion as well as a low frequency excess. Bottom right: String source with a high current. \COBEF is capable of constraining this model based upon entropy injection.}
\label{fig:CTSpecsRedshift}
\end{figure*}
%%%%%%%%%%%%%%%%%%%%%%%%%%%%%%%%%%%%%%%%%%%%%%%%%%

For the photon number injection, the integral over frequency must be modulated by the probability factor, ${\rm e}^{-\omega_{\rm c}/\omega}$, which makes it more difficult to tract analytically. For applications we recommend simply taking the integral numerically. 
As input, \texttt{CosmoTherm} requires a time-dependent occupation number injection function. Schematically this can be done for expressions~\eqref{eq:radSpec}-\eqref{eq:matSpec} by noting
%%%%%%%%%%%%%%%%%%%%%%%%%%%%%%%%%%%%%%%%%%%%%%%%%%
\begin{align} \label{eq:sourceTerm}
\frac{\id n_{\gamma}}{\id t } = \frac{\pi^2}{\omega^3} \frac{\id^2 E_{\gamma}}{\id t \id \omega},
\end{align}
%%%%%%%%%%%%%%%%%%%%%%%%%%%%%%%%%%%%%%%%%%%%%%%%%%
where $n_{\gamma}$ is the occupation number of the injected photons. We validated our implementations for $\id^2 E_{\gamma}/\id t \id \omega$ in multiple regimes, finding excellent agreement with the exact analytic expressions.

%--------------------------------------------
\vspace{-0mm}
\section{Numerical Results}
\label{sec:numericalResults}
%--------------------------------------------
After feeding in a source term of the form Eq.~\eqref{eq:sourceTerm}, \texttt{CosmoTherm} follows the direct evolution of photons with frequency $x_{\rm min} \leq x \leq x_{\rm max}$\footnote{We usually use $x_{\rm min}=10^{-8}$ and $x_{\rm max}=150$ with 4000 grid points.}. Heuristically, this code solves the coupled evolution equations for the photons and electrons on a finely spaced grid of redshift slices, which allows us to analyze spectral data at any time. In principle, the source term can take any form, making \texttt{CosmoTherm} a powerful and flexible tool to quickly and accurately assess the validity of many beyond the standard model (BSM) scenarios, such as decaying dark matter \citep{Bolliet}.

%--------------------------------------------
\subsection{Spectral information}
%--------------------------------------------
The main data product produced by \texttt{CosmoTherm} is the differential spectra $\Delta I = I_{\rm CT} - I_{\rm pl}$, the difference between the numerically computed spectrum with source functions and a pure blackbody. Figure~\ref{fig:CTSpecsRedshift} illustrates the buildup of spectral distortions for a range of different redshifts and string parameters. In the top left panel, we note that even in the absence of an external source term, a non-zero distortion is observed. This comes from the fact that the adiabatic cooling of electrons is more rapid than the photons ($T_{\rm e} \propto a^{-2}$ vs $T_{\gamma} \propto a^{-1}$). In reality, $T_{\rm e} \simeq T_{\gamma}$ as the electrons are continuously heated by the photons through the process of Compton cooling, which is effective until $z \simeq 10^2$. This process of energy extraction from the CMB produces the observed distortion in this panel.

The remaining three panels show benchmark models with values of $\mathcal{I}$ below, equal to, and above the critical current $\mathcal{I}_{\rm c}$. In each of these three cases, we can observe the buildup of a low frequency background of photons over time. Higher currents typically lead to stronger backgrounds, though the background is only significant for $\omega \lesssim \omega_{\rm k}$ as can be seen by Fig.~\ref{fig:exactSpec}. This implies that low frequency data such as that from the ARCADE-2 \citep{ARCADE2011} experiment will not be sensitive to arbitrarily large values of $\mathcal{I}$.

\begin{figure*}
\centering 
\includegraphics[width=\columnwidth]{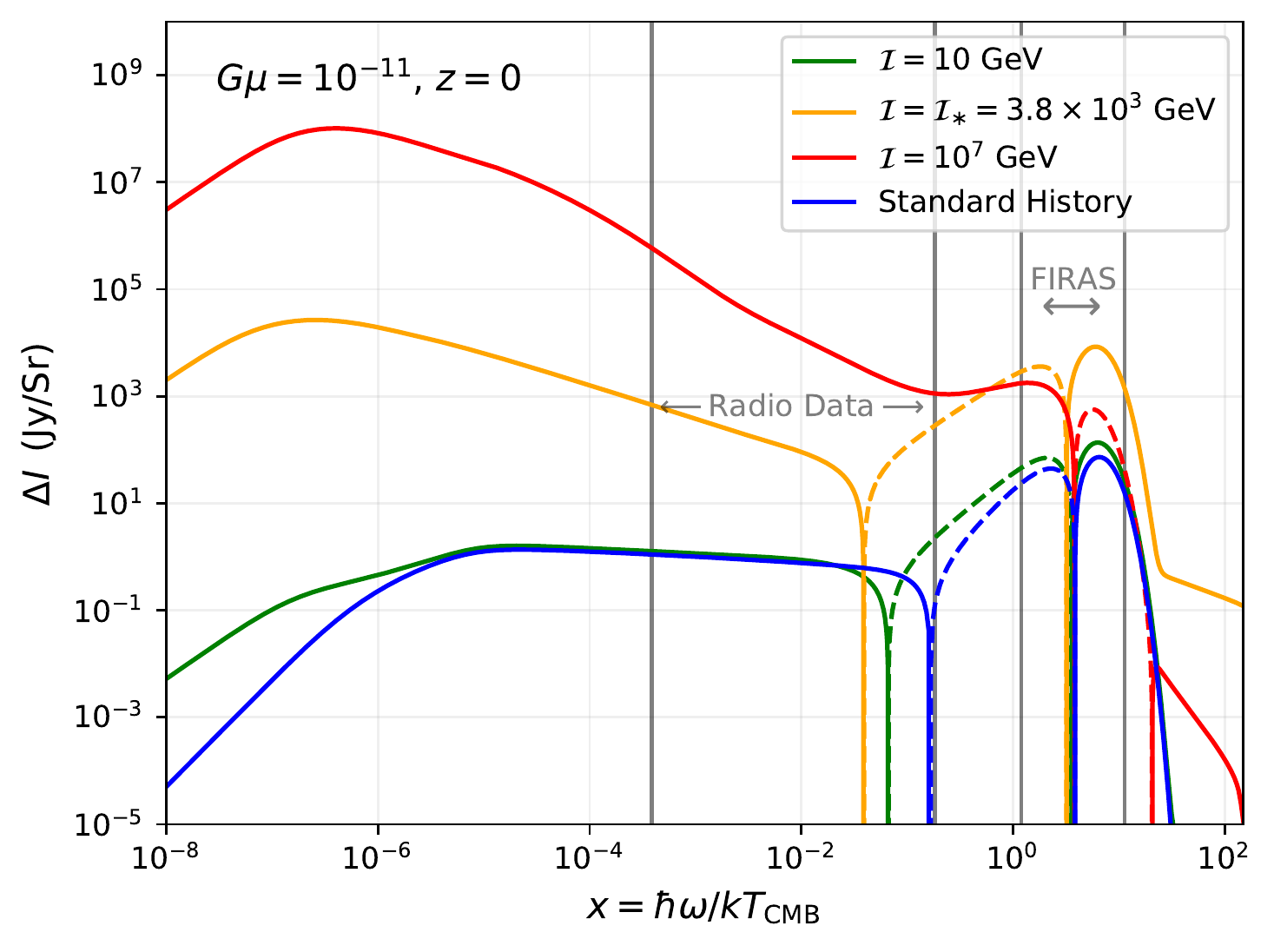}
\hspace{4mm}
\includegraphics[width=\columnwidth]{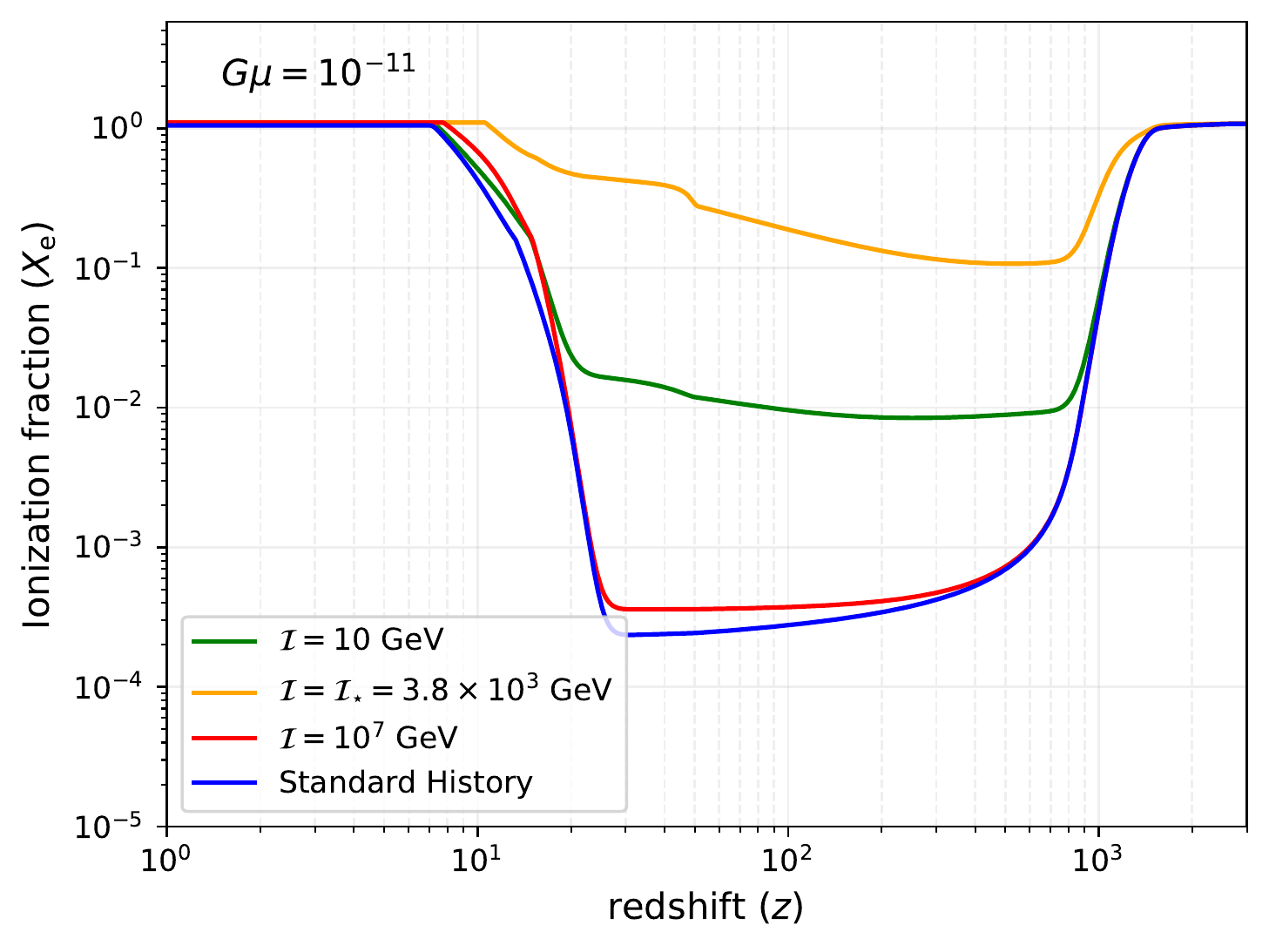}
\caption{Left: A comparison of the final spectra that we would observe today from our benchmark string parameters. Vertical lines indicate to frequencies bands probed by CMB and radio experiments. Higher currents tend to build up larger low frequency backgrounds, while parameters near the $\mathcal{I} = \mathcal{I}_*$ contour deposit more raw energy/entropy into the CMB, and are therefore more easily constrained by \COBEF. Right: Ionization histories for the parameters. Departures from the standard history are typically stronger for larger currents, until for a given set of parameters, $\omega_{\rm k} \lesssim 13.6$ eV. In that case, ionizing photons are produced with a greatly reduced efficiency, as indicated in Eq.~\eqref{eq:specApprox}. Further details on this can be found in Fig.~\ref{fig:XeConstraint}.}
\label{fig:CTSpecs}
\end{figure*}

%The left panel of Fig.~\ref{fig:CTSpecs} shows a side-by-side comparison of the distortions signatures we would observe today from these benchmark models. 
%
Positive $\mu$ and $y$ distortions are signified by an excess of photons at high frequencies, and a decrement at low frequencies when compared to a blackbody. For $\mathcal{I} = 10$ GeV and $\mathcal{I} = \mathcal{I}_{*}$, the distortion is primarily sourced by energy release, while for $\mathcal{I} = 10^7$ GeV, a strong entropy release generates a negative distortion. Radio observations and CMB experiments probe complementary regions of the induced cosmic string spectra, and with \texttt{CosmoTherm} we are able to utilize constraints from both datasets simultaneously.  See Fig.~\ref{fig:CTSpecs}, for additional illustration of the final distortion signals for these benchmark scenarios.

Throughout this work we have chosen $G\mu = 10^{-11}$ as a fiducial value for illustrations. This choice has been made because these contours highlight all of the relevant physical effects. Here, we would like to discuss the changes one observes when varying $G\mu$, and assuming that $\lambda \ll 1$. First, we note that the energy injection rate scales as $\id Q/\id t \propto \mathcal{I}/ G\mu$ in the gravitational decay regime (GDR, $\mathcal{I} \ll \mathcal{I}_{*}$), and as $\id Q/\id t \propto (G\mu)^{5/4} / \mathcal{I}^{1/2}$ in the electromagnetic decay regime (EDR, $\mathcal{I} \gg \mathcal{I}_*$). This implies that the total energy injection is maximized for $\mathcal{I} \simeq \mathcal{I}_{*}$, and that reducing $G\mu$ causes a faster decay of the signal compared to increases in $\mathcal{I}$ while in the EDR. This broadly explains the constraints presented in Fig.~\ref{fig:drhoConstraint}.

In terms of direct photon production, the scaling is slightly more complicated. In the GDR, the overall amplitude of the spectrum follows $\id^2 N_{\gamma}/\id t\, \id \omega \propto \mathcal{I}^2/(G\mu)^{7/6}$, while in the EDR we have $\id^2 N_{\gamma}/\id t\, \id \omega \propto \mathcal{I}^{5/6} (G\mu)^{7/12}$. As expected, decreases of $G\mu$ in the EDR cause a decrease in the amplitude. In contrast to the energy injection case, the total number of photons produced is not maximized on the $\mathcal{I} = \mathcal{I}_{*}$ contour, but instead increases with $\mathcal{I}^{5/6}$ in the EDR. However, the constraints on direct photon injection are sensitive to the precise frequencies of the produced photons. The position of the spectral knee determines an effective upper cutoff to the photon frequencies, and scales as $\omega_{\rm k} \propto (G\mu)^{1/2}/\mathcal{I}^3$ in the GDR, and $\omega_{\rm k} \propto (G\mu)^2/\mathcal{I}^4$ in the EDR. Thus, decreases in $G\mu$ and increases in $\mathcal{I}$ rapidly degrade this effective cutoff frequency. Once $\omega_{\rm k} \lesssim \rm{MHz}$, the vast majority of the photons are produced outside of the sensitivity range of microwave and radio observations. 

%%%%%%%%%%%%%%%%%%%%%%%%%%%%%%%%%%%%%%%%%%%%%%%%%%
\begin{figure}
\includegraphics[width=\columnwidth]{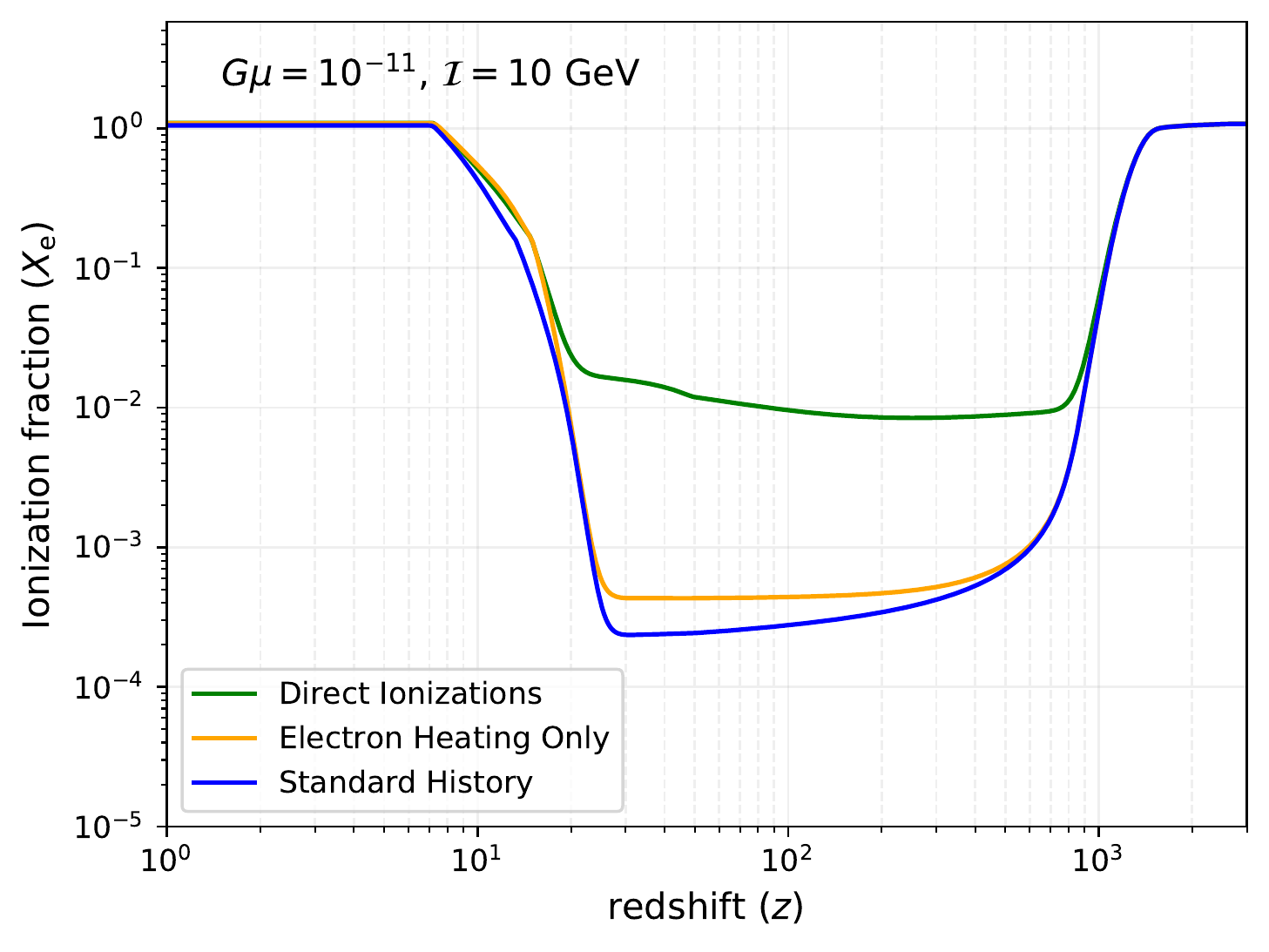}
\caption{A comparison of the free electron fraction with and without the direct production of ionizing photons from the string loops. When no ionizing photons are included, all energy is injected in the form of heat. This energy injection reduces the recombination rate and increases the collisional ionization efficiency, leading to the mild deviation from the case with no strings. When the direct production of ionizing photons is included, the $X_{\rm e}$ history is strongly modified, which leads to much stronger constraints.}
\label{fig:ionHistory}
\end{figure}
%%%%%%%%%%%%%%%%%%%%%%%%%%%%%%%%%%%%%%%%%%%%%%%%%%

%%%%%%%%%%%%%%%%%%%%%%%%%%%%%%%%%%%%%%%%%%%%%%%%%%
\begin{figure}
\includegraphics[width=\columnwidth]{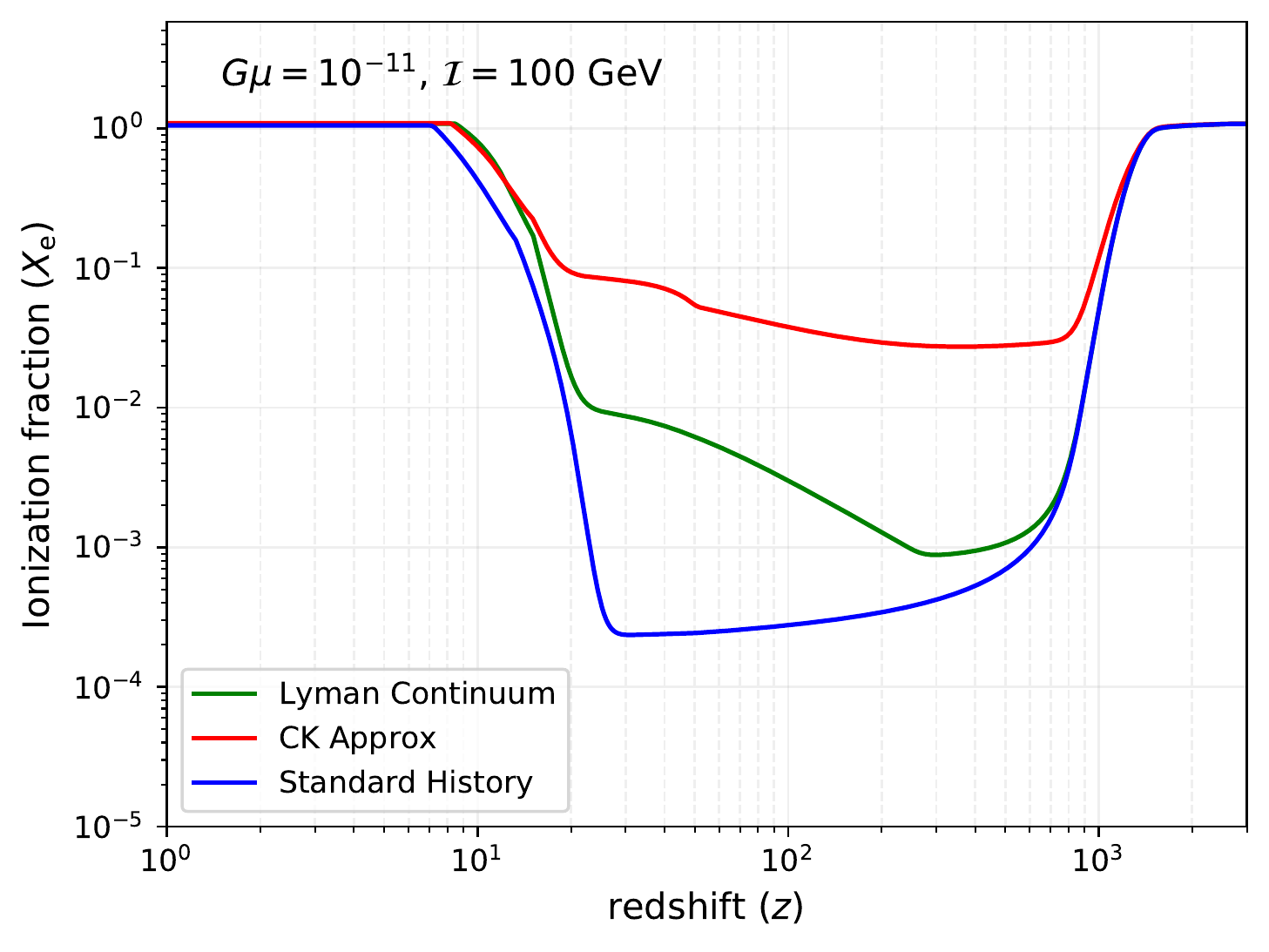}
\caption{The free electron fraction for two different treatments of ionizing photons. For the green curve, \texttt{CosmoTherm} directly computes the number of ionizing photons produced by the strings, removing them as they liberate electrons \citep[see][for details]{Bolliet}. This is computationally expensive, and can systematically underestimate the number of ionizations that take place. To produce our constraints we follow the approximate treatment described by \citet{Chen2004}.}
\label{fig:ionComp}
\end{figure}
%%%%%%%%%%%%%%%%%%%%%%%%%%%%%%%%%%%%%%%%%%%%%%%%%%

%%%%%%%%%%%%%%%%%%%%%%%%%%%%%%%%%%%%%%%%%%%%%%%%%%
%%%%%%%%%%%%%%%%%%%%%%%%%%%%%%%%%%%%%%%%%%%%%%%%%%

%--------------------------------------------
\subsection{Treating recombination history changes}
\label{sec:Xe_treat}
%--------------------------------------------
The emission produced by the cosmic string network also injects photons above the ionization thresholds of hydrogen and helium. This causes modifications to the recombination history of the Universe, which we treat approximately in this work. Specifically, photons outside the computational domain have to be added, leading to treatments discussed now.

For the first treatment, we only include the total heating from photons outside of our computational domain, accounting for photons injected at both $x\leq x_{\rm min}$ and $x\geq x_{\rm max}$, but do not consider ionizations due to photons at frequency $x$ above the atomic ionization thresholds. The energy density integrals are computed numerically at every time-step and then converted into heating of the baryonic matter. Since hotter electrons recombine less rapidly, this causes a delay of recombination.

In the second treatment, we more carefully consider the interactions of photons above the atomic energy levels. For photons injected below the Lyman-$\alpha$ line of hydrogen, we simply directly follow their evolution in the distortion domain, computing the Compton heating internally.\footnote{In our calculations we have $x_{\rm max}\simeq 150$, which means that at $z\lesssim 200$, the hydrogen Lyman-$\alpha$ line moves outside of the computational domain. We neglect the related heating of photons between $x_{\rm max}$ and $x_{\rm Ly-\alpha}$ at $z\lesssim 200$, which should have a minor effect.}
At $x\geq x_{\rm Ly-\alpha}$, we integrate the total energy density of photons and then use the method of \citet{Chen2004} with refinements according to \citet{Chluba2010a} to add heating and ionizations to the recombination problem. At $z\geq 3000$, we assume all the injected high-frequency energy is converted into heat. At $z\leq 3000$, we do not add any photons at $x\geq x_{\rm Ly-\alpha}$, assuming that these get efficiently absorbed and converted. A comparison of treatments one and two are shown in Fig.~\ref{fig:ionHistory} with the orange and green contours respectively.

In principle, we can also directly follow the evolution of photons in the Lyman continuum using \texttt{CosmoTherm}, though this also has limitations. While this process gives a more direct correspondence between the number of ionizing photons produced, and the total amount of ionizations \citep[see][for details]{Bolliet}, it misses out on an important reprocessing effect. Namely, high energy photons ($x \gg x_{\rm Ly-\alpha}$) will both ionize and heat the background. This heating of the background can introduce important secondary ionizations, particularly in the high energy regime \citep[e.g.,][]{Shull1985, Slatyer2009, Valdes2010} that can be missed otherwise. \cite{Bolliet} utilized this Lyman continuum treatment for generic decaying particle scenarios and at high energies found weaker bounds compared to different approaches employed by \citet{Capozzi2023} and \citet{Liu2020}. The approximate treatment by \citet{Chen2004} attempts to account for this by partitioning a fraction of the energy above the Lyman-$\alpha$ threshold to use for ionizations, excitation and heating. We neglect the effect of excitations, which have been found to be minor \citep{Galli2013}. A comparison of the free electron fraction for these two treatments can be found in Fig.~\ref{fig:ionComp}, where we see that the pure Lyman continuum computation generically underestimates the effect.

Prior treatments of the ionization history in the presence of strings \citep{TashiroIonization} utilized an incorrect spectral index for a large region of parameter space, which we correct here. Additionally, their analysis considered a simple photon counting procedure to derive their anisotropy constraints, which misses out on the reprocessing effect mentioned above. Ionization histories for our benchmark cases can be found in the right panel of Fig.~\ref{fig:CTSpecs}.

As mentioned above, our implementation is simplified and does not take into account complications of high-energy cascades \citep[e.g.,][]{Slatyer2009, Huetsi2009, Slatyer2015}, or most recently \citet{Liu2023I, Liu2023II}. However, it allows us to approximately follow the evolution of both the spectral distortions and the ionization fractions, which also can be used to compute the related 21-cm signals, as outlined in \citet{Acharya2022}. Our treatment could be further improved by directly treating the ionizations \citet{Bolliet} and also adding secondary energetic particles. This would also allow us to go beyond the 'on-the-spot' approximation \citep{Chen2004, Padmanabhan2005}, but we leave a more detailed exploration to future work.

%--------------------------------------------
\subsection{Evolution at late stages}
%--------------------------------------------
At $z\lesssim 500$, the effect of electron scattering on the evolution of the spectral distortions starts to become very small. We can therefore omit the broadening and shifting of distortions introduced at these late times. The photon evolution equation then simplifies, and we only have to include the $y$-distortion sources from differences in the electron and photon temperature, the emission and absorption of photons by the free-free process and the external photon source from cosmic strings. This greatly simplifies the calculation, as the evolution of the photon distribution in each frequency bin becomes independent. We confirmed that the results remain largely unaffected by this simplification.

%--------------------------------------------
\subsection{Soft photon heating and the global $21$-cm signal}
\label{subsec:21cm_section}
%--------------------------------------------
As was recently discussed in \citet{Acharya2023}, the presence of a sufficiently steep radio background produces an important backreaction effect on the $21$-cm differential brightness temperature ($\delta T_{\rm b}$) at cosmic dawn. This brightness temperature is given by
%-------------------------------
\begin{equation}
    \delta T_{\rm b} = \frac{(1-{\rm e}^{-\tau_{21}})}{1+z} (T_{\rm s} - T_{\rm R}),
\end{equation}
%-------------------------------
where $T_{\rm R}$ is the temperature of the background at $21$-cm (usually assumed to be solely the CMB), $\tau_{21}$ is the $21$-cm optical depth, and $T_{\rm s}$ is the spin temperature. The spin temperature is a measure of the ratio of hydrogen atoms in the triplet state relative to the singlet. 

Multiple prescriptions for calculations the evolution of $T_{\rm s}$ can be found in the literature \citep{F2006, Hirata2006, Venu2018}. Overall, they can be expressed as
%-------------------------------
\begin{equation}
     T_{\rm s}^{-1} = \frac{x_{\rm R} T_{\rm R}^{-1} + x_{\rm c} T_{\rm m}^{-1} + x_{\alpha} T_{\alpha}^{-1} }{x_{\rm R} + x_{\rm c} + x_{\alpha}}.
\end{equation}
%-------------------------------
Here, $T_{\rm R}$, $T_{\rm m}$, and $T_{\alpha}$ are the temperatures of the radiation, matter, and the colour temperature of the Lyman-$\alpha$ radiation. Additionally, $x_{\rm R}$, $x_{\rm c}$, and $x_{\alpha}$ are the radiative, collisional, and Wouthuysen-Field couplings respectively. 

During cosmic dawn, the dominant contribution to the spin temperature comes from the kinetic motion of the hydrogen atoms. In \citet{Acharya2023}, it was shown that if an additional radio background is present with a significantly steep spectral index, the hydrogen atoms are heated and the spin temperatures rises. As the magnitude of $\delta T_{\rm b}$ is proportional to $T_{\rm R}/T_{\rm s}$, this soft photon heating (SPH) dampens the expected signal relative to the case where SPH is neglected. 

\begin{figure}
\includegraphics[width=\columnwidth]{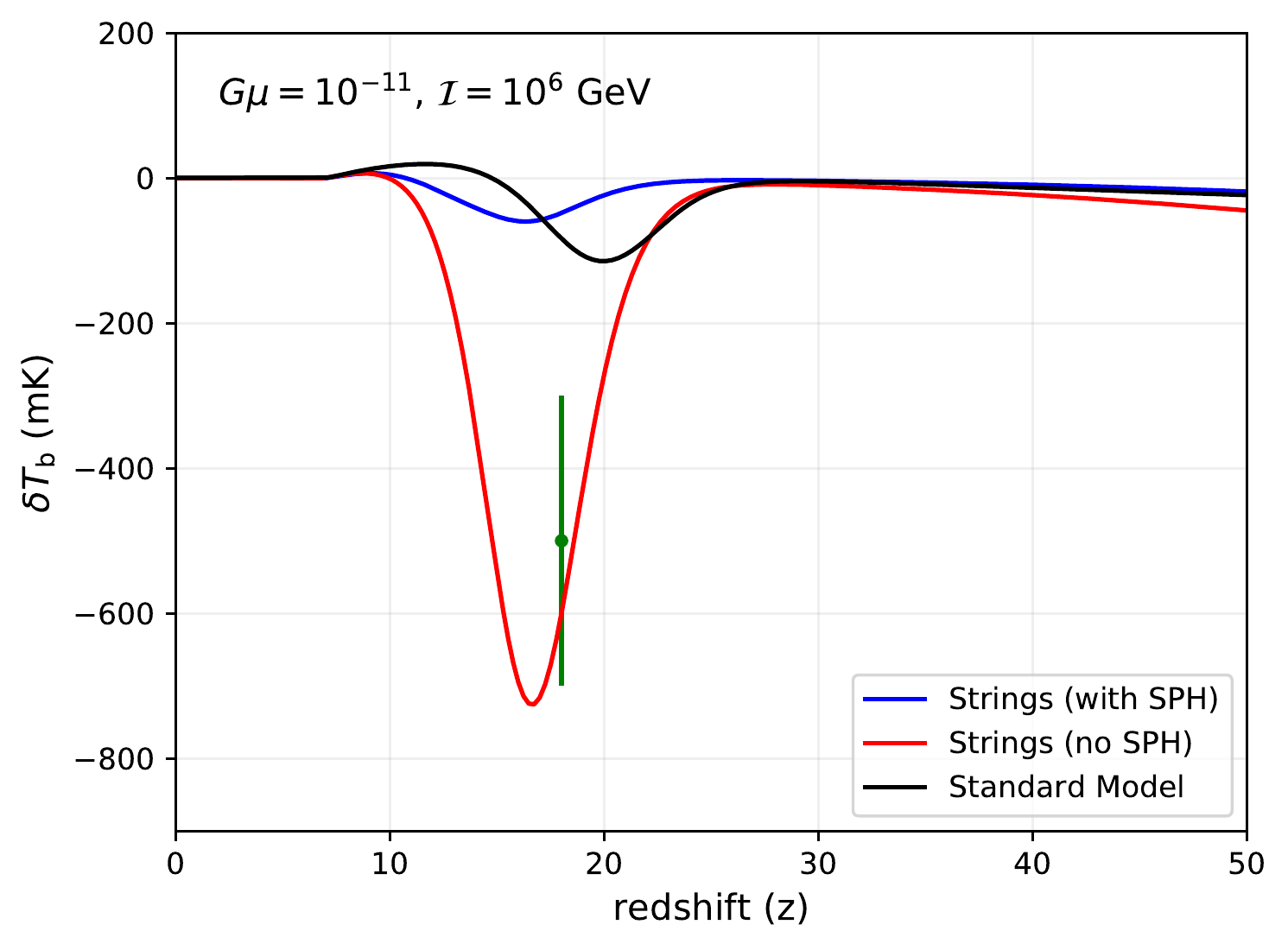}
\caption{The brightness temperature for a benchmark string model with and without the inclusion of soft photon heating (SPH). As was recently discussed in \citet{Acharya2023}, the presence of a steep radio background increases the spin temperature of the gas. This leads to a strongly subdued $\delta T_{\rm b}$ relative to the estimation one would obtain by neglecting this heating. The EDGES datapoint is plotted in green.}
\label{fig:dTb}
\end{figure}

\citet{Brandenberger2019} considered the impact of superconducting strings on $\delta T_{\rm b}$, but neglected this SPH. In addition, the estimates for the enhancement of the 21cm signal did not account for the radiative transfer effects and reduction of photons by free-free absorption. As a result, much of the string parameter space was ruled out by taking the EDGES result as a strict upper bound. Here, we find that by including soft photon heating, no string models can be ruled out by taking the EDGES upper limit alone. 
Figure~\ref{fig:dTb} shows how the inclusion of SPH prevents a particular cosmic string model from being constrained by the EDGES measurement.

We also note that \citet{Hernandez2014,Hernandez2021} has studied the effects of cosmic string wakes on the brightness temperature at $21$-cm and found that a network can amplify the signal. In the case of superconducting cosmic strings, the amplitude reduction from soft photon heating surpasses the potential amplification from the effect of the wakes, meaning that our results do not change. However, this effect is important for a network of non-superconducting strings.

%--------------------------------------------
\subsection{Reionization treatment}
%--------------------------------------------
We give a brief description of our reionization modelling in this section. We refer the readers to \cite{Acharya2022} for a more detailed discussion. 
The evolution of the ionization fraction due to hydrogen and helium is given by,
%-------------------------------------------------
%-------------------------------
\begin{subequations}
\begin{align}
    \dfrac{\id x_{\rm HII}}{\id t} &= \xi_\text{ion}(z)\dfrac{\id f_\text{coll}}{\id t} - \alpha_A C\,x_{\rm HII} \,n_{\rm e} 
    \\
    \dfrac{\id x_{\rm HeII}}{\id t} &= \xi_\text{ion}(z)
    \dfrac{\id f_\text{coll}}{\id t} - \alpha_A C\,x_{\rm HeII} \,n_{\rm e},
\end{align}
\end{subequations}
%-------------------------------
where $x_{\rm HII}=\frac{n_{\rm HII}}{n_{\rm H}}$, $x_{\rm HeII}=\frac{n_{\rm HeII}}{n_{\rm H}}$ are the hydrogen and helium ionized fractions respectively, $\xi_\text{ion}$ is the ionizing efficiency parameter, $f_\text{coll}$ is the matter collapse fraction, $\alpha_A$ is the case-A recombination coefficient, $C \equiv \langle n_{\rm e}^2\rangle/\langle n_{\rm e}\rangle^2$ is the clumping factor which is a function of gas density, and $n_{\rm e}$ is the total electron number density. We use the fitting function of \cite{SHTS2012} for the clumping factor. 
The physics of this modelling can be explained as follows. Once we have sufficiently massive dark matter halos which can form galaxies, the photons emitted by these galaxies will ionize their environment. Such massive halos are rare at higher redshifts (an effect captured by the collapse fraction), leading to a few isolated ionization bubbles. As more structure forms at lower redshifts, the number of photon sources increases and reionization proceeds rapidly.  

The ionizing efficiency parameter is given by
%-------------------------------
\begin{equation}
    \xi_\text{ion} = A_\text{He} f_\star f_\text{esc}N_\text{ion},
\end{equation}
%-------------------------------
where $A_\text{He}$ is a correction factor due to the presence of helium, $N_\text{ion}$ is the number of ionizing photons per stellar baryon, $f_\text{esc}$ is the fraction of ionizing photons escaping the host halo and $f_\star$ is the star formation efficiency. Since $\xi_\text{ion}$ is a degenerate combination of parameters, there are multiple ways to get the same reionization history using different parameter choices.  

For our fiducial model, we use the following combination of parameters ($N_\text{ion}$, $f_\star$, $f_\text{esc}$, $f_\alpha$, $f_X$) = (4000, 0.1, 0.1, 1.0, 1.0). We include the heating of electrons due to energetic X-ray photons through the expression given by \citet{F2006}, 
\begin{equation}
    \dfrac{2}{3}\dfrac{\epsilon_X}{k_\text{B}n_{\rm H} H(z)} = 10^3\text{K}~f_X \left[\dfrac{f_\star}{0.1}\right]
    \left[\dfrac{f_{X,h}}{0.2}\right]
    \left[\dfrac{\id f_\text{coll}/\id z}{0.01}\right]
    \left[\dfrac{1+z}{10}\right],
\end{equation}
where $f_{X,h} \simeq (1+2x_{\rm e})/3$ is the fraction of X-ray energy contributing to the heating \citep{Chen2004}, and $f_X$ is a scaling factor. 

%%%%%%%%%%%%%%%%%%%%%%%%%%%%%%%%%%%%%%%%%%%%%%%%%%
%--------------------------------------------
\vspace{-3mm}
\section{Constraints}
\label{sec:constraints}
%--------------------------------------------
To explore the $G\mu$-$\mathcal{I}$ parameter space, we run a finely spaced grid of models to generate our numerical data, which directly produces the likelihood values. We then obtain constraints from these \texttt{CosmoTherm} outputs using observational data from \COBEF \citep{Fixsen}, CMB anisotropies \citep{Planck2018params}, the radio synchrotron background (RSB) \citep{ARCADE2011, DT2018}, the EDGES experiment \citep{Edges2018}, and the optical depth to reionization as measured by the \citet{Planck2018params}. In addition, we also forecast constraints from $\mu$, as well as non-$\mu$, non-$y$ type distortions from a \PIXIE-type experiment \citep{Kogut2011PIXIE}. In order to make contact with these observations, we analyze the output produced by \texttt{CosmoTherm} using a rudimentary likelihood analysis. This typically involves comparing the output produced with strings, to that without, as a test of the null hypothesis. An exception to this is when comparing against measurements of the radio synchrotron background, in which we compare to a best-fit power law of the datapoints. In this section we describe in detail how we obtain the constraint curves presented in Fig.~\ref{fig:AllConstraints}

\begin{figure*}
\centering 
\includegraphics[width=\columnwidth]{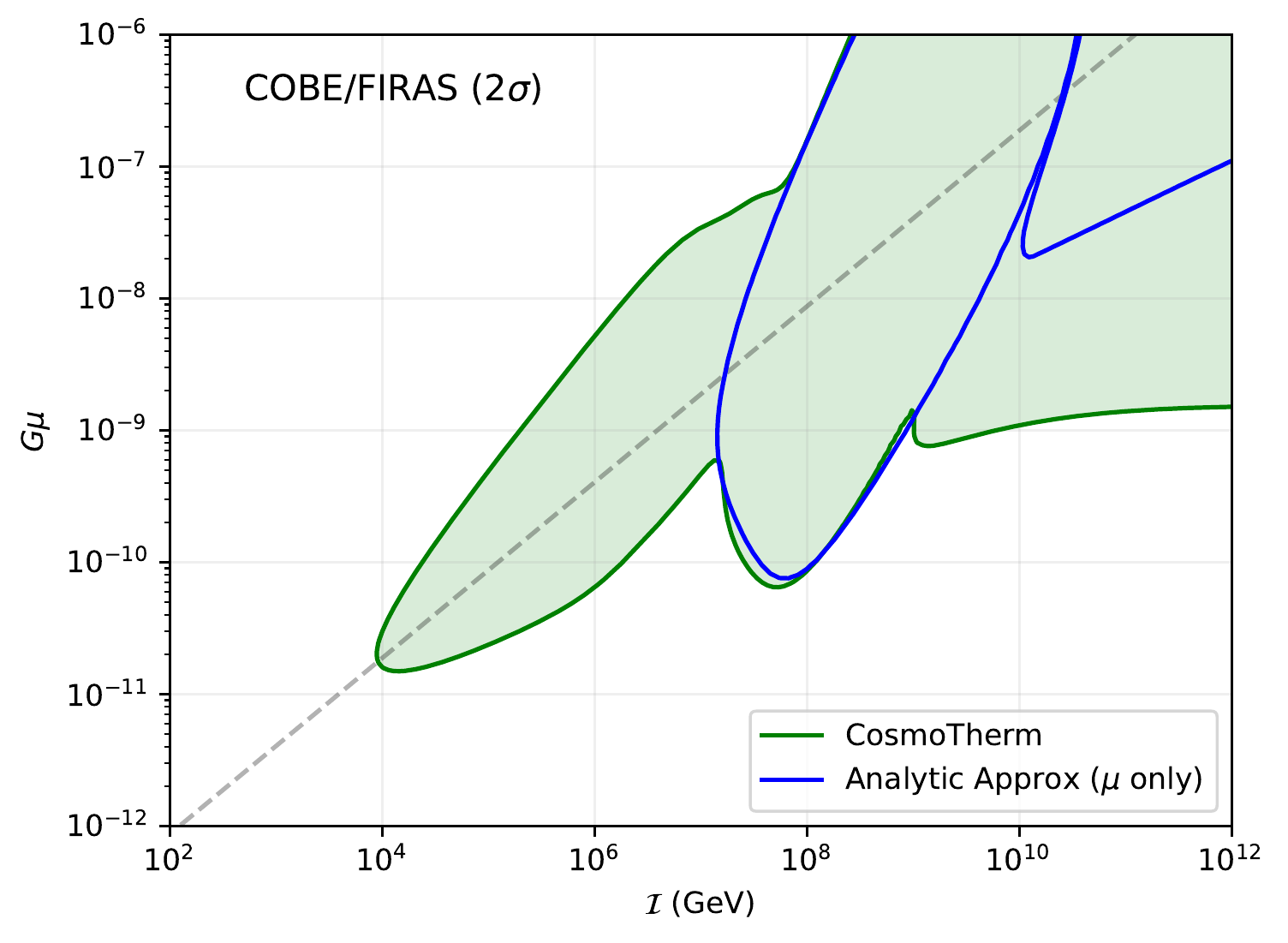}
\hspace{4mm}
\includegraphics[width=\columnwidth]{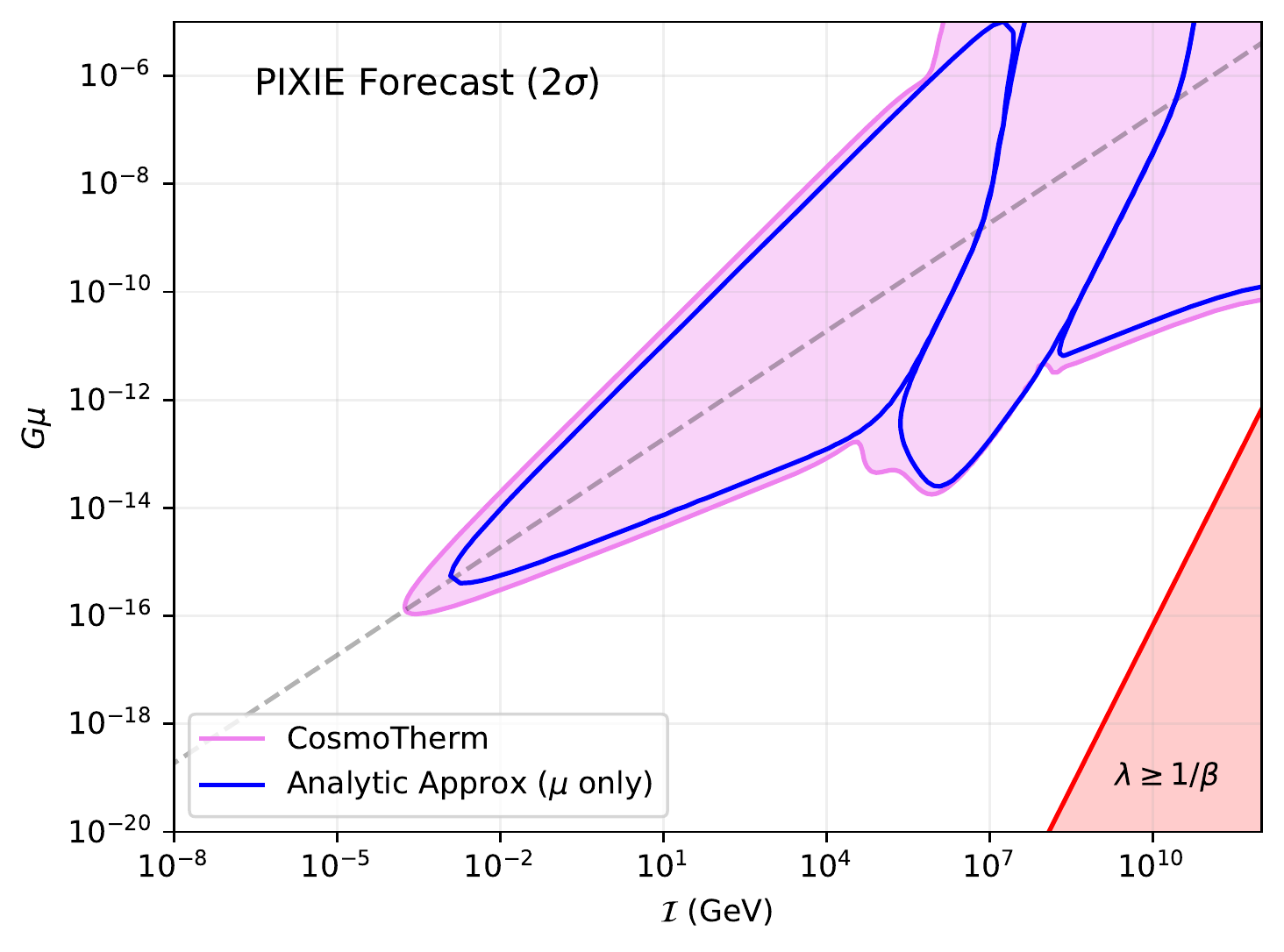}
\caption{Comparison of the $2\sigma$ constraints found from \COBEF, as well as the forecast for a \PIXIE-type instrument, to the analytic predictions seen in Fig~\ref{fig:drhodNConstraint}. A numerical treatment shows that \COBEF is capable of constraining much more than simple analytic treatments would suggest. With current specifications, \PIXIE would observe a strong $y$-type distortion from reionization. To obtain a conservative estimate, we do not consider the sensitivity of \PIXIE to a primordial $y$-distortion, and instead focus on limits from $\mu$ and non-$\mu/y$ type signatures.}
\label{fig:SDconstraints}
\end{figure*}

%--------------------------------------------
\subsection{CMB spectral distortions}
\label{sec:SD_Like}
%--------------------------------------------
The CMB spectrum can be well approximated by a Planck (blackbody) spectrum at a temperature $T_0=2.7255\,{\rm K}$, with upper limits on the distortions of the order $\Delta I/I \lesssim 10^{-5}-10^{-4}$ at frequencies $\nu \simeq 60-600\,{\rm GHz}$ from \COBEF \citep{Fixsen}. As we have seen above, electromagnetic energy injection into the baryon-photon plasma heats the electrons which in turn boosts the CMB photons creating distortions to the CMB spectrum. Additionally, direct photon injection (entropy) can create unique spectral patterns which strengthen constraints in some regions of parameter space. 

The current $2\sigma$ upper limit for the amplitude of the $\mu$ and $y$ parameters is $|\mu|\lesssim 9\times 10^{-5}$ and $|y|\lesssim 1.5\times 10^{-5}$ \citep{Fixsen}, which translates into a constraint on the energy release as $\Delta\rho_{\gamma}/\rho_{\gamma}\lesssim 6\times 10^{-5}$, where $\rho_{\gamma}$ is the CMB energy density today. Using \texttt{CosmoTherm}, we can go beyond these simple $\mu$ and $y$ parameters by comparing the full shape of the string-induced spectra to the residuals of the \COBEF measurement. This allows a more precise determination of the validity of any particular model by utilising all of the available data, therefore provides stronger and more robust constraints when compared with previous analysis such as the work by \citet{Tashiro}, or \citet{Miyamoto}. However, it assumes that the marginalization over galactic foregrounds does not further alter the constraints. In addition, we automatically deproject any contributions that are degenerate with a simple shift in the CMB temperature. This is achieved using a simple scalar product of the signal vector on the \COBEF bands weighted by the inverse covariance matrix.

The \COBEF experiment measured the monopole of the background temperature to exquisite precision. After subtracting the best fit blackboody from this measurement, one is left with a series of residuals that are mostly consistent with $0$ at the $1\sigma$ level. The lowest order monopole distortion expected in standard cosmology is a $y$-type that comes from reionization at the level of $\Delta I/I \simeq 10^{-7}-10^{-6}$ \citep{Hill2015}. This is far below the sensitivity of \COBEF, and so we can be confident that any distortion which exceeds the residuals comes from the cosmic strings and would therefore be constrained. To compute the likelihood, we assume that the \COBEF datapoints are uncorrelated, and perform two $\chi^2$ evaluations using the residuals. The first using the output spectrum from \texttt{CosmoTherm} with a cosmic string source term ($\chi^2_{\rm CS}$), and the second without the source ($\chi^2_0$). Finally, we compute $\Delta \chi^2 = \chi^2_{\rm CS} - \chi^2_{0}$ and require $\Delta \chi^2 \leq 2$ for our $2\sigma$ constraint curves. 

The sensitivity of a \PIXIE-type instrument is high enough that it would see the reionization $y$-distortion at more than $100\sigma$ \citep{abitbol_pixie}. In principle, this means that in order to claim a $y$-distortion signature of cosmic strings from \PIXIE, we would need to subtract off the reionization signal at very high precision. The uncertainties in the model of reionization implemented in \texttt{CosmoTherm} make this a tricky procedure to pull off successfully. Therefore, we also choose to deproject the $y$-distortion from the forecasted \PIXIE residuals, and search for distortions of the $\mu$ type, and the non-$\mu$, non-$y$ type that are much cleaner. We perform our likelihood analysis in the same way as for \COBEF, but with the deprojected data. For the forecasting, we assume \PIXIE has null residuals with an effective (foreground-marginalized) sensitivity of $\Delta I = 5 \, \rm{Jy/Sr}$, which roughly reproduces a $1\sigma$ sensitivity to $\mu\simeq 1.4\times 10^{-8}$.

We show the results of this analysis in Fig.~\ref{fig:SDconstraints}, and compare with the analytic approximations computed in Section \ref{sec:analyticEntropy}. It is important to note that the analytic approximations consider only an integration from $3\times 10^5 \leq z \leq z_{\rm th}$, as it is impossible to analytically follow the entropy injection constraints after this point. In contrast, the numerical result considers the full evolution in the CMB distortion band down to redshift $z = 0$. We therefore expect some natural increase in the contours between the analytic and numerical results. In general, the approximate constraints computed from the negative $\mu$-distortion due to entropy release are rather consistent with the numerical result. It is clear that the analysis of the \COBEF residuals provides us with significantly increased constraining power.

%--------------------------------------------
\vspace{-3mm}
\subsection{CMB anisotropies}
\label{sec:CMB_Xe_like}
%--------------------------------------------
The addition of energetic electromagnetic particles will ionize and heat the background electrons during and after recombination, modifying the standard recombination history. As a result, temperature anisotropies are damped while polarization anisotropies are boosted \citep{ASS1999,Chen2004, Padmanabhan2005, Galli2009}, and these effects have been measured to great precision by CMB anisotropy experiments \citep{Komatsu2010,Planck2018params}. There have been several studies in past years which have improved our knowledge of recombination physics  \citep{Zeldovich68,Peebles68,Seager2000,Sunyaev2009,chluba2010b,Yacine2010c} and one can now compute the recombination history accurately using publicly available codes such as {\tt CosmoRec}
\citep{chluba2010b}.
%%%%%%%%%%%%%%%%%%%%%%%%%%%%%%%%%%%%%%%%%%%%%%%%%%
\begin{figure}
\includegraphics[width=\columnwidth]{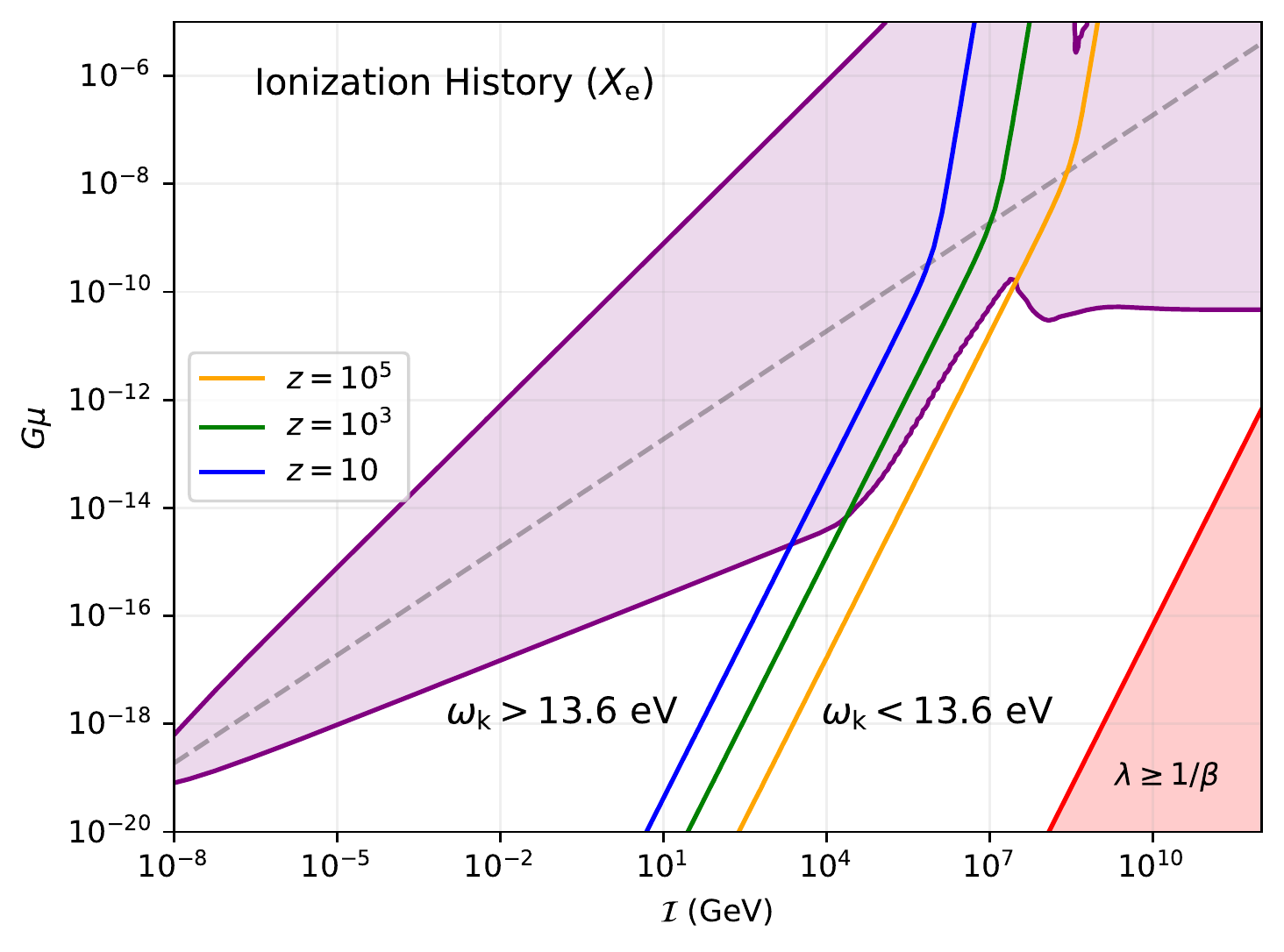}
\caption{$2\sigma$ limits on cosmic string parameters from the induced changes to the free electron fraction $x_{\rm e}$. The coloured lines indicate when $\omega_k = 13.6$~eV. To the left of the lines, the spectrum of ionizing photons produced by the loop network decays as $\omega^{-5/3}$, while to the right it falls off much faster as $\omega^{-17/6}$. Consequently, the constraints are greatly relaxed in the latter region.}
\vspace{-3mm}
\label{fig:XeConstraint}
\end{figure}
%%%%%%%%%%%%%%%%%%%%%%%%%%%%%%%%%%%%%%%%%%%%%%%%%%

To avoid the time-consuming computations that individual samples of the CMB likelihood would entail, in this work, we use a direct projection method developed in \cite{PCA2020}. This is a principal component analysis (PCA) method following the works developed in \citet{Farhang2011, Farhang2013}. For our case, with energy injections from string decay,  we compute the changes to the standard recombination history of the universe, $\xi(z)=\Delta x_{\rm e}/x_{\rm e}$, using the recombination module in {\tt CosmoTherm}. We then compute the first three principal component coefficients by projecting $\xi(z)$ onto the eigenmodes, $E_i(z)$, with the integral
%--------------------------------------------------
\begin{equation}
\mu_i=\int \xi(x)E_i(z) \,{\rm d}z.
\end{equation}
%-------------------------------------------------
We use the covariance matrix of the $\mu_i$ obtained in \cite{PCA2020} and compute the likelihood of the model assuming Gaussian statistics. We do not include modifications to reionization in this setup, therefore, the eigenmodes $E_i(z)$ are sensitive to changes in the ionization history only at high redshifts $100\lesssim z \lesssim 4\times 10^3$ \citep{PCA2020}. We treat the modification to reionization history separately which is described below.

We show the $2\sigma$ constraints from CMB anisotropies in Fig.~\ref{fig:XeConstraint}. Of the datasets we consider, variations in the electron recombination history are the most wide-sweeping and stringent. As mentioned above, the photons produced from the strings heat the electrons (increasing collisional ionization rates and quenching recombinations), as well as directly ionize hydrogen and helium. Of these two effects, the production of photons with $\omega \geq 13.6$ eV are the most potent (see Fig.~\ref{fig:ionHistory}). The spectrum of injected photons follows a broken power law given by Eq.~\eqref{eq:specApprox}, which falls off very rapidly for $\omega \gtrsim \omega_{\rm k}$. Therefore, a given parameter set with $\omega_{\rm k} \lesssim 13.6$ eV will be much less efficient at ionizing the background.
However, as discussed in Sec.~\ref{subsec:21cm_section}, soft photons can be efficiently absorbed for $x\lesssim 10^{-5}$ (Fig.~\ref{fig:critFreq}), a process that heats the electrons and in turn ionize the neutral hydrogen. Since the total energy of photons within $x\lesssim 10^{-5}$ is similar for certain choices of parameters, we see the almost horizontal feature in $x_{\rm e}$ constraint at high currents. We show contours of constant $\omega_{\rm k}$ in the figure, noting that constraints are greatly reduced for sufficiently low values of the knee frequency. 

%--------------------------------------------
\subsection{Radio synchrotron background (RSB) data}
%--------------------------------------------

We use the RSB data from ARCADE experiment \citep{ARCADE2011} and \cite{DT2018}. The ARCADE-2 experiment measured a RSB between 3-90 GHz. They reanalyzed earlier data at lower frequencies and compiled it in their Table 4. With this, they found a best fit power law with spectral index 2.6 and temperature $T\simeq 24$~K at 310 MHz. In \cite{DT2018}, the authors redid this analysis using independent data points around $\simeq 40-80$ MHz and found the best fit slope to be consistent with ARCADE but with slightly higher normalization with $\simeq$ 30K at 310 MHz. In this paper, we use the ARCADE data points within 3-90 GHz and the independent data points of \cite{DT2018} within the 40-80 MHz band to compute the likelihood.

In the analysis of these datasets, the contribution from resolved extra-galactic sources were not taken into account. To isolate the contribution to the radio background from string decay, we add an irreducible extra-galactic background to our solution obtained from {\tt CosmoTherm}. The fitting function to the minimal extra-galactic background (MEG) is given by \citep{GTZBS2008},
%---------------------------------------------------------
\begin{equation}
    T_{\rm bg}(\nu)\simeq 0.23\,{\rm K}\left(\frac{\nu}{\rm GHz}\right)^{-2.7}. 
    \label{eq:extra-galactic}
\end{equation}
%---------------------------------------------------------
In contrast to the spectral distortion constraints, we compare the induced radio background from a network of strings not to the \texttt{CosmoTherm} output with no source term, but to a best-fit power law to the RSB data. With the inclusion of the MEG, this power law is given by 
%---------------------------------------------------------
\begin{equation}
    T_{\rm RSB}(\nu)\simeq 1.230\,{\rm K}\left(\frac{\nu}{\rm GHz}\right)^{-2.555}.
    \label{eq:bestFitRSB}
\end{equation}
%---------------------------------------------------------
With this as our null hypothesis, we once again perform a $\chi^2$ analysis against the string induced RSB as described in Sect.~\ref{sec:SD_Like} to obtain our constraints. This allows us to treat the RSB as a strict upper bound on the amount of radio emission which can be produced by the string network. We note, however, that for the RSB limits presented below, we do not add a penalty if the total background is not reproduced by the sum of the MEG and our distortion outputs. The excluded region is illustrated in Fig.~\ref{fig:AllConstraints}.

%--------------------------------------------
\subsection{Global $21$-cm measurements}
%--------------------------------------------
We use the claimed detection by the EDGES collaboration \citep{Edges2018} as a figure of merit to constrain the energy injection process from cosmic strings. EDGES has claimed a detection of a 21-cm absorption feature with $\delta T_{\rm b}\simeq-500$~mK originating from $z\approx 18$ and a 1$\sigma$ error of 200 mK. Recently, SARAS, an independent experiment, could not reproduce this result \citep{Saras2022}. Therefore, our discussions on constraints from global 21-cm measurements are broadly qualitative.

To constrain our energy injection cases, we demand that $-500~{\rm mK}\lesssim \delta T_{\rm b} \lesssim 0$ at $z=18$. For $\delta T_{\rm b} \leq -500$ mK, we use a Gaussian likelihood with an error of 200 mK to quantify the tension with this data. We also penalize cases with $\delta T_{\rm b}>0$ at $z=18$ by using a gaussian likelihood with an error of 84 mK which is the value of $\delta T_{\rm b}$ at $z=18$ for our fiducial 21-cm model without any energy injection. 

We find that from the EDGES measurement alone, none of the models exhibit a $2\sigma$ tension with the data. This is why the left panel of Fig.~\ref{fig:AllConstraints} does not show a constraint curve. We comment that the soft photon heating effect described in \citet{Acharya2023} indeed eliminates any regions in tension with the EDGES measurement. Interestingly, we do find regions of parameter space where $\delta T_{\rm b} \geq 0$, implying that the signal may be in emission in the presence of a string network. This is also a direct consequence of the soft photon heating effect. While we do make the choice to penalize these models, the affected regions of parameter space have already been ruled out at more than $2\sigma$ by other datasets.

%--------------------------------------------
\subsection{Optical depth constraints}
%--------------------------------------------
We use the optical depth measurement of \cite{Planck2018params} with $\tau=0.0544\pm 0.0073$ to constrain our energy injection cases. The significantly lower value of measured $\tau$ has sparked great interest and has resulted in a shift of our understanding of the reionization epoch \citep{KKHBPCA2019}. However, these works use detailed hydrodynamical simulations and include several physical effects such as a non-homogeneous ionizing photon background which are difficult to capture in a simple analytic setup.

\begin{figure*}
\centering 
\includegraphics[width=0.99\columnwidth]{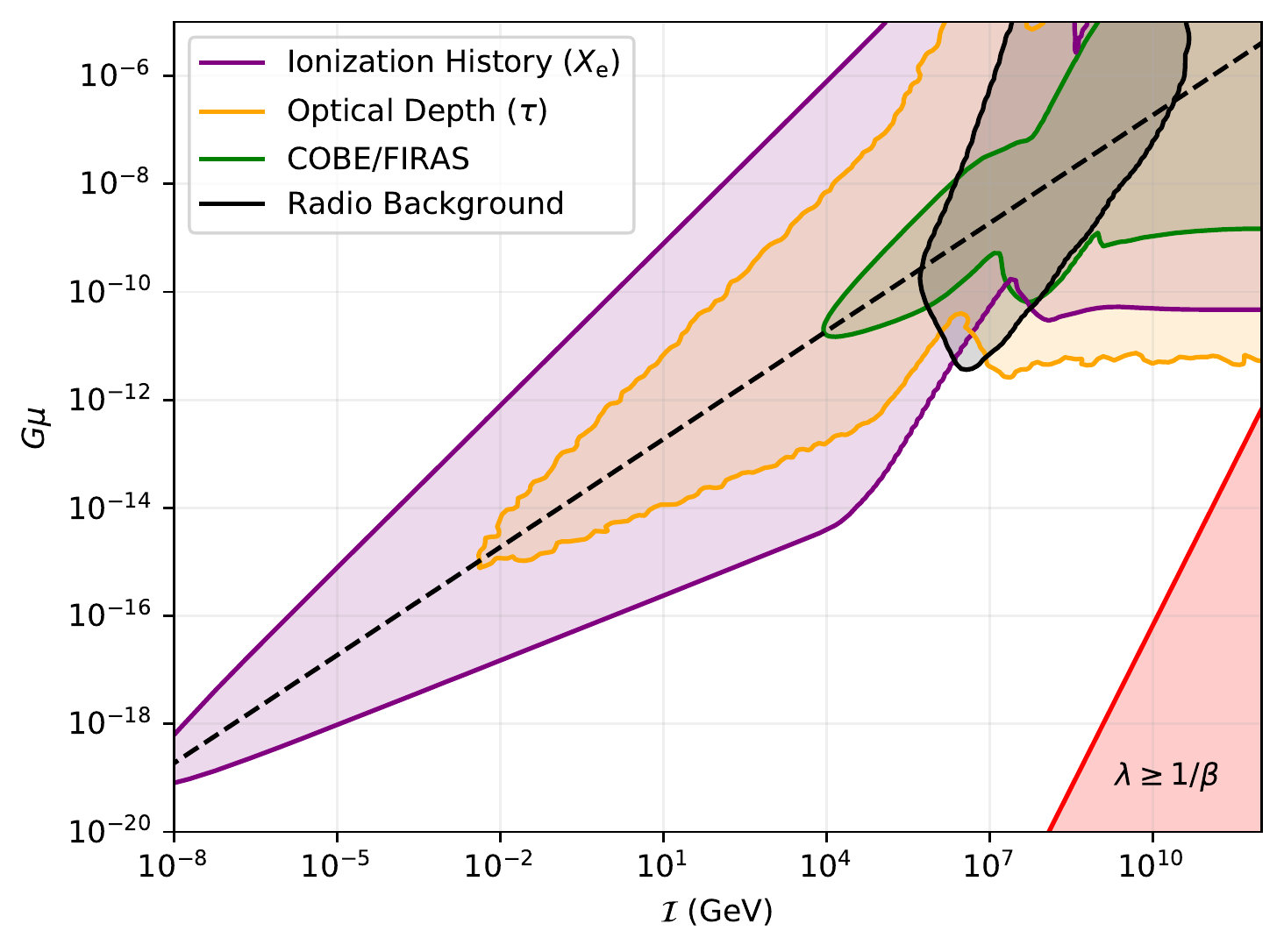}
\hspace{4mm}
\includegraphics[width=0.99\columnwidth]{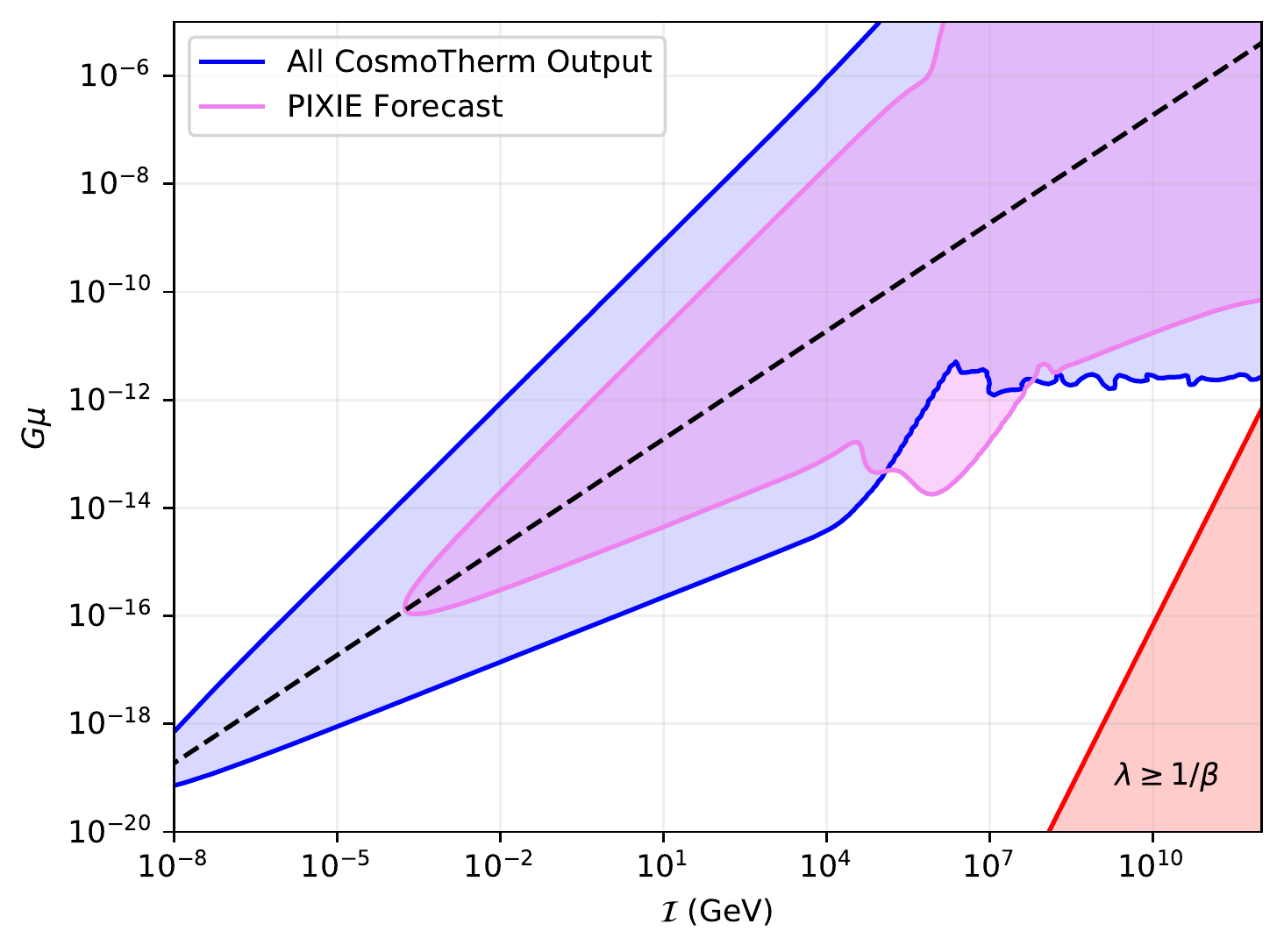}
\\
\caption{Left: A breakdown of the individual ($2\sigma$) constraints analyzed by \texttt{CosmoTherm}. While changes to the ionization history are dominant for the majority of the parameter space, late time effects such as an excess radio background and changes to the optical depth to reionization can become important for high currents. We also utilize the observation from the EDGES \citep{Edges2018} experiment as a strict upper bound on the brightness temperature at cosmic dawn, though we find no constraints on strings at the $2\sigma$ level from that. Right: A combination of the constraints, as well as a conservative forecast for a \PIXIE-like instrument. Such an instrument would probe a new wedge of parameter space by searching for a negative $\mu$-distortion sourced by a strong entropy injection from the strings at early times.}
\label{fig:AllConstraints}
\end{figure*}

In this work, we assume that energy injections modify the reionization history, or $\tau$, only perturbatively. Since our fiducial reionization model gives a $\tau=0.078$,\footnote{For this we assume the starting redshift of reionization be $z=30$, though it does not change significantly with small change to this starting redshift.} we compute the difference of the optical depth obtained with cosmic strings included, and then add a simple Gaussian penalty to constrain the model. We checked that tuning the reionization model parameters to more closely reproduce the measured $\tau$ value, does not alter the constraints much.
%
%We obtain $\Delta\tau$ with our fiducial reionization model as reference, and compute the likelihood using the error bar of the measurement reported by Planck. We have checked that small changes to our fiducial reionization history do not change the constraints significantly. 
%
In Fig. \ref{fig:AllConstraints}, we show the optical depth constraints which becomes dominant at high currents, i.e., $I\gtrsim 10^7$ GeV. As we discussed above, at these high currents, the heating due to soft photons is an important process that changes the ionization history of the universe appreciably.   

%%%%%%%%%%%%%%%%%%%%%%%%%%%%%%%%%%%%%%%%%%%%%%%%%%
%%%%%%%%%%%%%%%%%%%%%%%%%%%%%%%%%%%%%%%%%%%%%%%%%%

%--------------------------------------------
\subsection{Summary of constraints}
%--------------------------------------------
In the left panel of Fig.~\ref{fig:AllConstraints} we show the $2\sigma$ constraints obtained through our likelihood analysis of the \COBEF, CMB anisotropy, radio synchrotron background, EDGES, and optical depth measurements. It is clear that of the datasets we analyzed, limits coming from CMB anisotropies are by far the most stringent.
Our updated limits using a full spectral analysis of the \COBEF data are superseded by more recent observational results. It is important to realize, however, that the \COBEF data has been available since the mid 90s. If it had been used to its full extent at the time, these constraints would have been relevant to the parameter space of superconducting cosmic string models for many years.

Reionization and RSB constraints cover a region of high currents, which is perhaps unsurprising as strong photon emitters in the late-time universe are easier to detect. Interestingly, regions on the boundary of the RSB data may offer a viable solution to this observed radio excess, as we discuss in a forthcoming publication.

The right panel of Fig.~\ref{fig:AllConstraints} presents a joint likelihood analysis of the datasets analyzed by \texttt{CosmoTherm}, alongside our simple forecast for a \PIXIE-type instrument. As described above, we make a conservative choice when forecasting by considering the \PIXIE sensitivity only to $\mu$, and non-$\mu$, non-$y$ type distortions to avoid having to perform a careful subtraction of the $y$-distortion induced by reionization. With proper foreground subtraction and removal of this $y$-distortion, we expect a marginal improvement in the \PIXIE sensitivity to this cosmic string scenario. Importantly, \PIXIE would be capable of constraining an important region of parameter space that is currently not covered.

Finally, in Fig.~\ref{fig:FinalConstraints} we present a more complete illustration of the open regions of parameter space for superconducting cosmic string models. In addition to our work, we add constraint curves from \citet{Miyamoto} from pulsar timing array data, big-bang nucleosynthesis, and radio transients that could be observed by the Parkes array. As a word of caution we mention that they utilize a slightly different set of input parameters ($\alpha = 0.1$ and $\Gamma_{\rm g} = 50$) when computing their results, and so their curves here should be taken as a rough estimate rather than a robust boundary. They also consider a third form of energy release for the string network, through so-called plasma dissipation. The analysis of the plasma dissipation efficiency depends on many uncertain parameters, such as the velocity of any given loop, the local plasma viscosity and more. We choose not to model these effects in our work, but note that in \citet{Miyamoto}, that deviations from our results seem to appear at very low values of the string tension ($
G\mu \simeq 10^{-18}$).

%--------------------------------------------
\section{Discussion and Conclusions}
\label{sec:conclusions}
%--------------------------------------------

%%%%%%%%%%%%%%%%%%%%%%%%%%%%%%%%%%%%%%%%%%%%%%%%%%
\begin{figure*}
\centering 
\includegraphics[width=1.5\columnwidth]{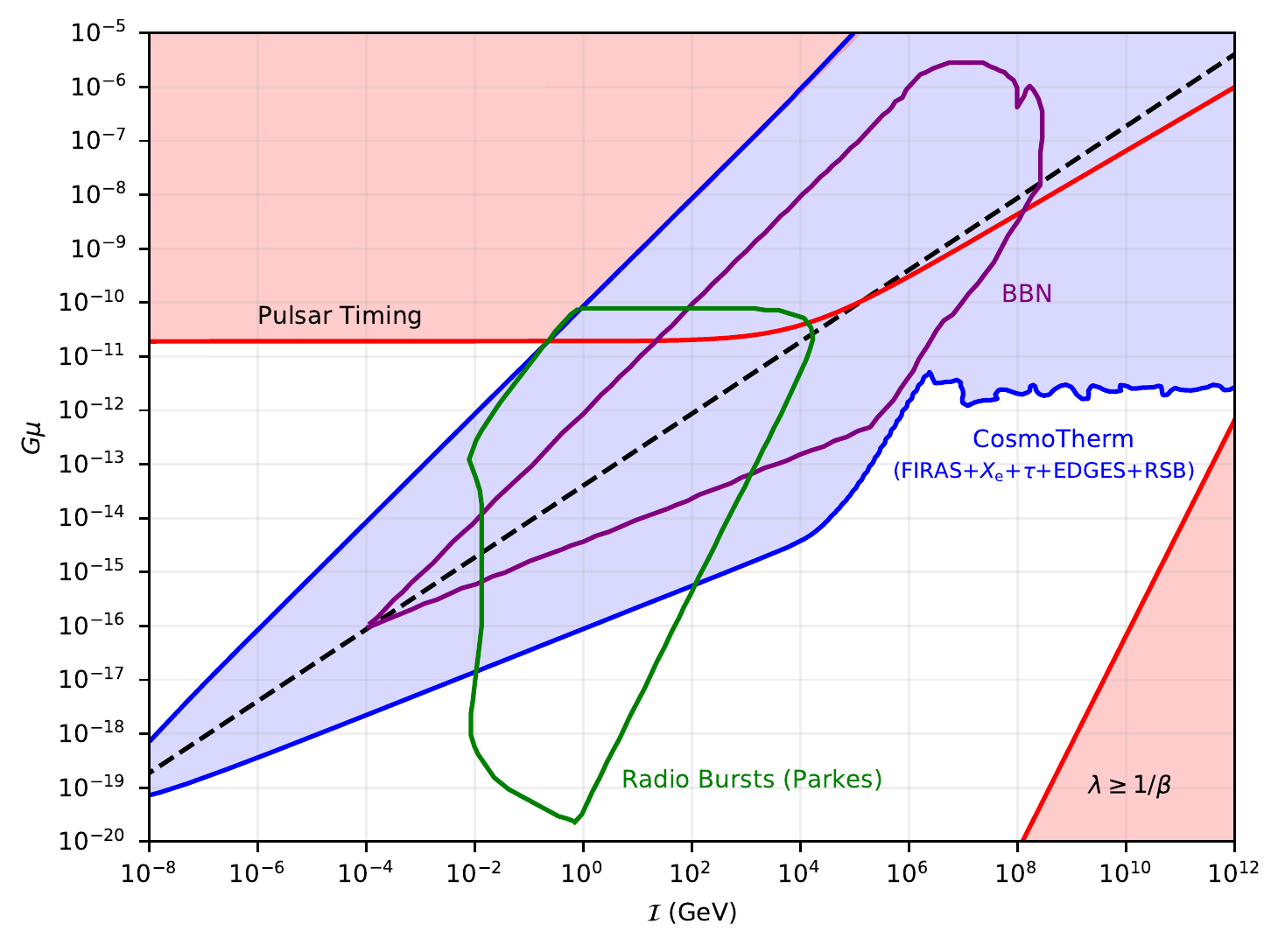}
\caption{A summary plot of the constraints analyzed in this work, as well as limits obtained by \citet{Miyamoto} on pulsar timing array measurements, big-bang nucleosynthesis, and transient radio bursts which we didn't directly consider in our work. The work of \citet{Miyamoto} uses a slightly different string loop model, so their constraints should be viewed as approximate boundaries when compared with ours.}
\label{fig:FinalConstraints}
\end{figure*}
%%%%%%%%%%%%%%%%%%%%%%%%%%%%%%%%%%%%%%%%%%%%%%%%%%

In this work, we have investigated the spectral signatures produced by a network of superconducting strings in place at early times ($z_{\rm form} \gg 10^7$). We have made improvements to the approximate analytic understanding of the spectral distortions produced by this network, and strengthened these results by performing full numerical solutions to the thermalization problem using \texttt{CosmoTherm}. 

Analytically, we have refined the previous estimates on primordial distortion signatures made by \citet{Tashiro} and \citet{Miyamoto} by including a non-trivial contribution to the $\mu$-distortion coming from strong entropy injections. This negative $\mu$ contribution can be seen by comparing Fig.~\ref{fig:drhoConstraint}, which neglects the entropy injection, to Fig.~\ref{fig:drhodNConstraint}. Additionally, in Sect.~\ref{sec:numericalImplementation} we also develop a more sophisticated analytic formalism for describing the instantaneous spectrum of photons produced by such a string network, given in Eq.~\eqref{eq:radSpecMat} and Eq.~\eqref{eq:matSpec}. These can readily be applied to other physical scenarios to produce more robust analytic estimates.

We then developed a simple module in \texttt{CosmoTherm} to handle the injection and processing of the cosmic string source function. The analytic approximations were compared against the full numerical solution in Fig.~\ref{fig:SDconstraints}, where we find the analytics to be a conservative underestimate of the full constraining power of spectral distortions. 
The extra constraining power achieved by \texttt{CosmoTherm} is the result of being able to perform a full spectral analysis of the \COBEF data (and the \PIXIE forecast). Previous estimates relied on constraining the specific distortion parameters, $\mu$ and $y$. With the full spectral data from \texttt{CosmoTherm}, we are able to go beyond this by comparing the string induced spectrum to the residuals of \COBEF, yielding constraints on non-$\mu$ and non-$y$ type distortions. This full shape spectral analysis will generically increase the constraining power of spectral distortions to models of exotic energy/entropy injection, when compared against the simple analytic estimates.  

With the numerical implementation, we also gain access to precise spectral information at virtually any redshift ($z \lesssim 10^7$), which allows us to easily and efficiently derive constraints from other datasets. In the left panel of Fig.~\ref{fig:AllConstraints}, we show $2\sigma$ constraints from \COBEF \citep{Fixsen}, CMB anisotropies \citep{Planck2018params}, the radio synchrotron background \citep{ARCADE2011, DT2018}, and the optical depth to reionization as measured by the \citet{Planck2018params}. Additionally, we utilize the EDGES \citep{Edges2018} datapoint as a strict lower limit on $\delta T_{\rm b}$ at $z \simeq 18$ when combining all of our constraints. This analysis also presents an update and more robust treatment compared to other analytic estimates presented by \citet{Miyamoto}, and \citet{TashiroIonization}.

In \citet{Acharya2023} it was recently discovered that the presence of a sufficiently steep soft photon background (spectral index $\gtrsim 2.5$ at $\nu \lesssim 1$ GHz) can cause significant heating of gas in the late universe, leading to an increase in the spin temperature $T_{\rm s}$. In turn, this dampens the amplitude of the brightness temperature at cosmic dawn ($\delta T_{\rm b}$), relaxing the constraints derived from EDGES. \citet{Brandenberger2019} derived strong limits on the high current region of the string parameter space using an analytic approximation which neglected this new effect as well as overestimated the photon flux due to omission of radiative transfer effects. We find that by including this effect, constraints driven by EDGES disappear.

A summary of our constraints, as well as other datasets that we did not include, can be seen in Fig.~\ref{fig:FinalConstraints}. As with most models of BSM physics, there exists some degree of theoretical uncertainty that can be difficult to quantify. The form of the string loop density distribution is well established from Nambu-Goto simulations for non-superconducting strings \citep{NGsim5}, but to our knowledge, large-scale simulations with superconducting strings have not been performed \citep[though recent progress has been made in ][]{Rybak2023}. Another simplifying assumption is that all loops carry the same, time independent current from their formation until their eventual decay. Current generation and dissipation on these loops will ultimately depend on the local environment in which they propagate, and modelling of this is beyond the scope of our work. We also note that \citet{Miyamoto} include an additional channel for the string decay through plasma dissipation. This is a frictional effect which again depends on the dynamics of the local environment, which we do not model. Based upon the work of these authors, it appears that this effect may become important for very low string tensions ($G\mu \lesssim 10^{-18}$), and so we advise caution in the interpretation of our constraints at that level.

In addition to these BSM related uncertainties, we have treated the injection of energy and photons in an approximate manner (see Sect.~\ref{sec:Xe_treat}). Similarly, significant uncertainties exist in the modeling of reionization and the 21cm signal, although the latter do not drive any constraint here. Finally, our treatment of the CMB anisotropy likelihood (see Sect.~\ref{sec:CMB_Xe_like}) had the goal of quickly exploring the range of models without a significant computational burden. Similarly, marginalization over spectral distortion foregrounds will have to be more carefully considered, in particular when assessing the constraints from future CMB spectrometers. We leave these improvements to future work, anticipating that some of the details may change, while leaving the broad conclusions unaltered.

To conclude, superconducting cosmic strings offer an interesting and well-motivated model that can probe particle physics from the top down. By hunting for their signatures in different cosmological and astrophysical datasets, we learn more about the phase transitions that may (or may not) have taken place in the very early universe. \texttt{CosmoTherm} is a powerful and flexible tool capable of uncovering the many spectral nuances of not just cosmic strings, but virtually any scenario that injects energy or entropy into the background. Here, we have focused on the derivation of constraints for cosmic strings, but we plan to apply this toolbox to a wider array of physics beyond the standard model in the future.

\section*{Acknowledgements}

We would like to acknowledge the initial work of Jiten Dhandha in setting up the 21 cm and reionization modules in {\tt CosmoTherm}. We would also like to thank Richard Battye and Wenzer Qin for helpful comments and discussions on the draft.
This work was supported by the ERC Consolidator Grant {\it CMBSPEC} (No.~725456).
JC was furthermore supported by the Royal Society as a Royal Society University Research Fellow at the University of Manchester, UK (No.~URF/R/191023).
BC would also like to acknowledge support from an NSERC-PDF.

%%%%%%%%%%%%%%%%%%%%%%%%%%%%%%%%%%%%%%%%%%%%%%%%%%
\section*{Data Availability}

The data underlying in this article are available in this article and can further be made available on request.


\bibliographystyle{mnras}
\bibliography{SPCSrefs,noteRefs,Lit,DCBHrefs}

%%%%%%%%%%%%%%%%%%%%%%%%%%%%%%%%%%%%%%%%%%%%%%%%%%

%%%%%%%%%%%%%%%%% APPENDICES %%%%%%%%%%%%%%%%%%%%%

\appendix

\section{Entropy Injection Approximations}
\label{sec:entropyInjApprox}
In order to compute the entropy injection from a network of cosmic string loops, it was necessary for us to introduce the functions $\mathcal{J}_{\rm r}(l,\kappa)$ and $\mathcal{J}_{\rm m}(l_{\rm a}, l_{\rm b}, \kappa)$ in Eqs. (\ref{eq:Jr}) and (\ref{eq:Jm}) respectively. Here, we investigate the asymptotic regimes of these functions.
These are also useful when describing the average emission spectra caused by cosmic strings.

For $\kappa l \ll 1$, $\mathcal{J}_{\rm r}$ reduces to a simpler integral which can be expressed in terms of a hypergeometric function as
%%%%%%%%%%%%%%%%%%%%%%%%%%%%%%%%%%%%%%%%%%%%%%%%%%
\begin{align}
\label{eq:F_int}
\mathcal{F}_{\rm r}(l)&=
\mathcal{J}_{\rm r}(l, 0)=
\int_{0}^{l}  
\frac{{l'}^{1/3}\id l'}{(1+l')^{5/2}}
=\frac{3}{4} l^{4/3}\,_{2}F_1\left(\frac{4}{3},\frac{5}{2},\frac{7}{3},-l\right)
\nonumber\\
&\approx 
\frac{\mathcal{F}_\infty}{1+(l_{\rm c}/l)^{7/6}}
\end{align}
%%%%%%%%%%%%%%%%%%%%%%%%%%%%%%%%%%%%%%%%%%%%%%%%%%
with 
%%%%%%%%%%%%%%%%%%%%%%%%%%%%%%%%%%%%%%%%%%%%%%%%%%
$\mathcal{F}_\infty=\Gamma\left[\frac{7}{6}\right]\Gamma\left[\frac{7}{3}\right]/\sqrt{\pi}
\approx 0.6232$ and $l_{\rm c}=[(7/6)\,\mathcal{F}_\infty]^{-6/7}\approx 1.314$.
%
%%%%%%%%%%%%%%%%%%%%%%%%%%%%%%%%%%%%%%%%%%%%%%%%%%
\begin{figure}
\includegraphics[width=\columnwidth]{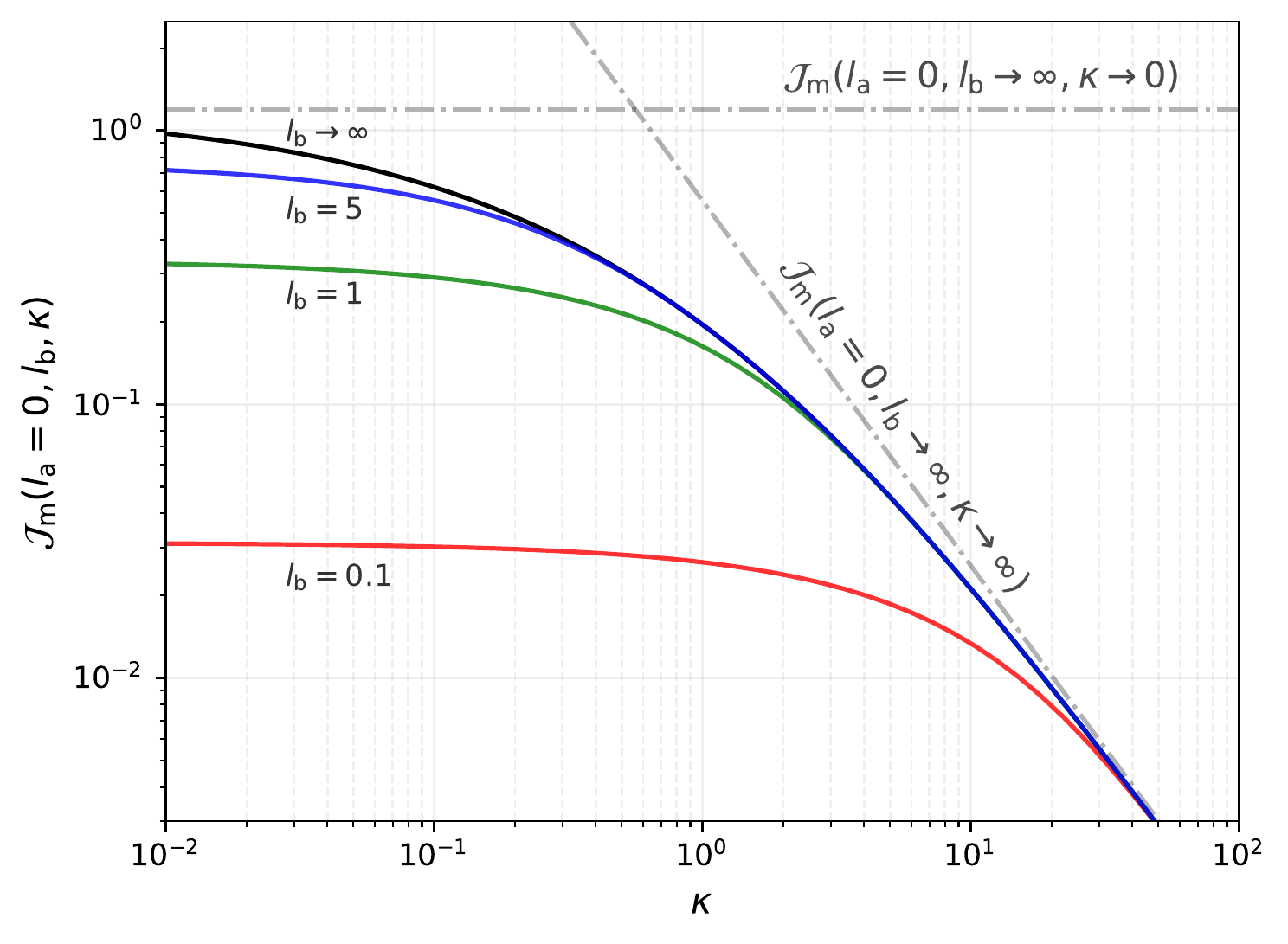}
\caption{The form of $\mathcal{J}_{\rm m}(l_{\rm a},l_{\rm b},\kappa)$ with $l_{\rm a} = 0$ and varying values of $l_{\rm b}$. Setting $l_{\rm a} = 0$ is equivalent to requiring that all radiation loops have fully evaporated. In other words, we are likely deep in matter radiation when this is satisfied. Approximate forms for small and large $\kappa$ can be found in Eq.~\eqref{eq:F_int_m} and Eq.~\eqref{eq:G_int_m}, respectively. For matter loops, $l_{\rm b} = 1/\lambda$, meaning small values of $l_{\rm b}$ may also be interpreted as the large $\lambda$ regime.}
\label{fig:JMplot}
\end{figure}
%%%%%%%%%%%%%%%%%%%%%%%%%%%%%%%%%%%%%%%%%%%%%%%%%%
The approximation was obtained by inspecting the asymptotic form of the solution at large $l$. The approximation does not capture the limit $l\ll 1$ very well, which is $\mathcal{F}_{\rm r}\approx (3/4)\,
l^{4/3}$; however, usually this regime does not contribute much and is only relevant to large $\lambda$. A slightly improved approximation that is valid to better than $1\%$ ($10\%$) precision at $0.1\lesssim l$ ($0.0002\lesssim l$) is
%%%%%%%%%%%%%%%%%%%%%%%%%%%%%%%%%%%%%%%%%%%%%%%%%%
\begin{align}
\label{eq:Fr}
\mathcal{F}_{\rm r}(l)
&\approx 
\frac{\mathcal{F}_\infty}{1+(\alpha_{\rm f}\, l_{\rm c}/l)^{(7/6)\,\beta_{\rm f} +\gamma_{\rm f}\ln(l_{\rm c}/l)}}
\end{align}
%%%%%%%%%%%%%%%%%%%%%%%%%%%%%%%%%%%%%%%%%%%%%%%%%%
with $\alpha_{\rm f}=0.9044$, $\beta_{\rm f}=0.9871$ and $\gamma_{\rm f}=0.0164$, obtained from fitting the function. In numerical applications, we simply pre-tabulate $\mathcal{F}_{\rm r}$ for $10^{-2}\leq l \leq 10^2$ and use fourth order Taylor expansions outside of that domain.

In the opposite limit where $\kappa l \gg 1$, one finds
%%%%%%%%%%%%%%%%%%%%%%%%%%%%%%%%%%%%%%%%%%%%%%%%%%
\begin{align}
\label{eq:Jinf}
\mathcal{G}_{\rm r}(l,\kappa) &= \mathcal{J}_{\rm r}(l,\kappa \rightarrow \infty) 
\approx 
\frac{1}{ \kappa^{1/3}} \int_0^l \frac{{\rm d}l'}{(1+l')^{5/2}} \frac{{\rm e}^{-\kappa l'}}{\Gamma[2/3]} \nonumber \\
&\approx \frac{3}{4\Gamma[2/3]} \frac{1}{\kappa^{4/3}}
\end{align}
%%%%%%%%%%%%%%%%%%%%%%%%%%%%%%%%%%%%%%%%%%%%%%%%%%
For large values of $\kappa$ we find that $\mathcal{G}_r(l,\kappa)$ quickly goes to zero as all of the produced photons are rapidly absorbed by the plasma. Figure~\ref{fig:JRplotCombined} shows the validity of these approximations.

We can do similar approximations for the matter loops. For $\kappa=0$, we have $\mathcal{J}_{\rm m}(l_a, l_b,0)=\mathcal{F}_{\rm m}(l_b)-\mathcal{F}_{\rm m}(l_a)$ with
%%%%%%%%%%%%%%%%%%%%%%%%%%%%%%%%%%%%%%%%%%%%%%%%%%
\begin{align}
\label{eq:F_int_m}
\mathcal{F}_{\rm m}(l)&=
\int_{0}^{l}  
\frac{{l'}^{1/3}\id l'}{(1+l')^{2}}
=\frac{3}{4} l^{4/3}\,_{2}F_1\left(\frac{4}{3},2,\frac{7}{3},-l\right)
\nonumber \\
&=\frac{1}{3}\ln\left(\frac{1+l^{1/3}}{\sqrt{1-l^{1/3}+l^{2/3}}}\right)
-\frac{l^{1/3}}{1+l}
\nonumber \\
&\qquad \qquad 
+\frac{1}{\sqrt{3}}\left[\frac{\pi}{2}+\tan^{-1}\left(\frac{l^{1/3}-2}{\sqrt{3}\,l^{1/3}}\right)\right].
\end{align}
%%%%%%%%%%%%%%%%%%%%%%%%%%%%%%%%%%%%%%%%%%%%%%%%%%
This function will be extremely important for the computation of the emission spectrum in the matter dominated era.

\begin{figure}
\includegraphics[width=\columnwidth]{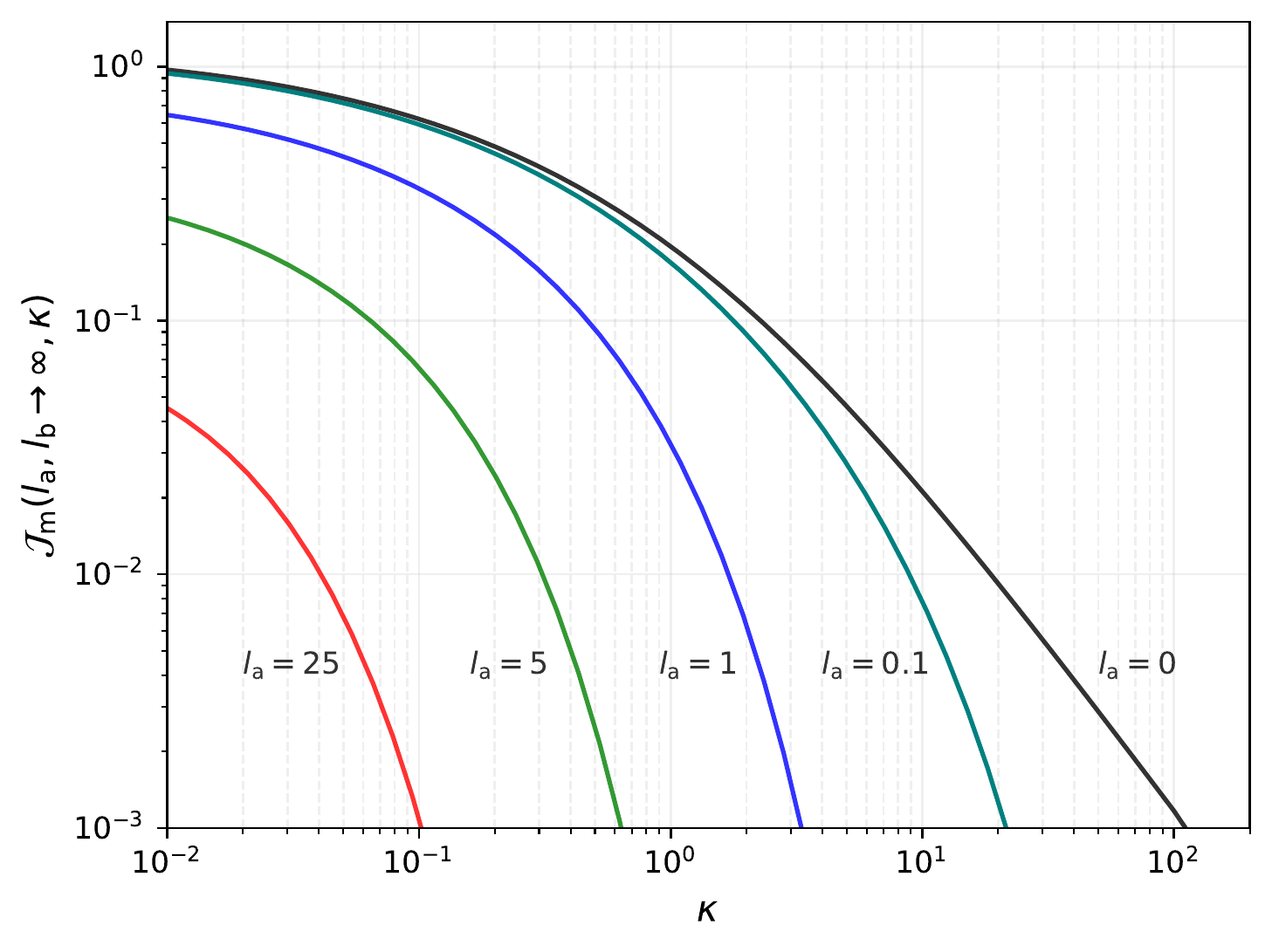}
\caption{The form of $\mathcal{J}_{\rm m}(l_{\rm a}, l_{\rm b},\kappa)$ with $l_{\rm b} \rightarrow \infty $ (equivalently, $\lambda \rightarrow 0$). At $t_{\rm eq}$ we have $l_{\rm a} = l_{\rm b}$, while afterwards $l_{\rm a}$ steadily decreases as the loops shrink in size. At constant $\kappa$, the emission spectrum steadily rises until $t = t_{\rm end}$ when $l_{\rm a} = 0$. In reality, $\kappa$ increases as $l_{\rm a}$ decreases. Figure~\ref{fig:windowFunc} shows the redshift dependence of the window function which acts as a modulation on $\mathcal{J}_{\rm m}(l_{\rm a},l_{\rm b}, \kappa$).}
\label{fig:JMplotLA}
\end{figure}

Finally, for $\kappa \rightarrow \infty$ it is straightforward to show that $\mathcal{J}_{\rm m}(l_{\rm a}, l_{\rm b}, \kappa \rightarrow \infty) \approx \mathcal{G}_{\rm m}(l_{\rm b}, \kappa) - \mathcal{G}_{\rm m}(l_{\rm a},\kappa)$, where
%%%%%%%%%%%%%%%%%%%%%%%%%%%%%%%%%%%%%%%%%%%%%%%%%%
\begin{align}
\label{eq:G_int_m}
\mathcal{G}_{\rm m}(l, \kappa)
&\approx 
\frac{1}{ \kappa^{1/3}} \int_0^l \frac{{\rm d}l'}{(1+l')^{2}} \frac{{\rm e}^{-\kappa l'}}{\Gamma[2/3]} \nonumber \\
&\approx \frac{3}{4\Gamma[2/3]} \frac{1}{\kappa^{4/3}} \left(1 + \mathcal{O}\left(\frac{1}{\kappa}\right) - \frac{{\rm e}^{-l \kappa}}{(1+l)^2} \right)
\end{align}
%%%%%%%%%%%%%%%%%%%%%%%%%%%%%%%%%%%%%%%%%%%%%%%%%%
This allows a final simplification
%%%%%%%%%%%%%%%%%%%%%%%%%%%%%%%%%%%%%%%%%%%%%%%%%%
\begin{align}
\mathcal{J}_{\rm m}(l_{\rm a}, l_{\rm b}, \kappa \rightarrow \infty)
&\approx 
\frac{3}{4\Gamma[2/3]} \frac{1}{\kappa^{4/3}} \left(\frac{{\rm e}^{-l_{\rm a} \kappa}}{(1+l_{\rm a})^2} - \frac{{\rm e}^{-l_{\rm b} \kappa}}{(1+l_{\rm b})^2} \right).
\end{align}
%%%%%%%%%%%%%%%%%%%%%%%%%%%%%%%%%%%%%%%%%%%%%%%%%%
We show these expressions in Fig.~\ref{fig:JMplot}. In Fig.~\ref{fig:JMplotLA} we illustrate the evolution of $\mathcal{J}_{\rm m}(l_{\rm a},l_{\rm b}, \kappa)$ for small values of $\lambda$ soon after $t_{\rm eq}$. In general, $l_{\rm a}$ and $l_{\rm b}$ are related through
%%%%%%%%%%%%%%%%%%%%%%%%%%%%%%%%%%%%%%%%%%%%%%%%%%
\begin{align}
l_{\rm a} = \begin{dcases} \frac{t_{\rm eq}}{t} (l_{\rm b} + 1) - 1 \hspace{20mm} (t_{\rm eq} < t \leq t_{\rm end})\\
0 \hspace{37.5mm} (t_{\rm end} < t),
\end{dcases}
\end{align}
%%%%%%%%%%%%%%%%%%%%%%%%%%%%%%%%%%%%%%%%%%%%%%%%%%
where $l_{\rm b} = 1/\lambda$ at all times $t \geq t_{\rm eq}$. 
%%%%%%%%%%%%%%%%%%%%%%%%%%%%%%%%%%%%%%%%%%%%%%%%%%
%%%%%%%%%%%%%%%%%%%%%%%%%%%%%%%%%%%%%%%%%%%%%%%%%%

\section{Particle Physics to Astrophysics dictionary} 
\label{sec:dictionary}

Astrophysical environments are often the ideal locations to probe particle physics beyond the standard model. To simplify the notational burden, Particle physicists prefer to work in natural units where $\hbar = c = k_{\rm b} = 1$. This allows for a simplification in which all units can be expressed in GeV to some power. To recover real physical units at the end of a computation, one needs only to insert appropriate values of $\hbar$, $c$, and $k_{\rm b}$ until the desired units are acquired. Most of the results we present are expressed in a dimensionless form, simplifying the interpretation. To get a more physical understanding of the enormous impact these strings can have on the astrophysical and cosmological observables, we present a simple conversion to units that may be more familiar to an astrophysicist in Table \ref{tab:table}.

\begin{table} 
	\centering
	\caption{Conversion table for physical parameters of interest in our work. The value of $\mu$ is chosen such that $G\mu = 10^{-11}$. We also define $L_{\rm c} = \Gamma G\mu t_0$, the radius at which a loop existing today would decay within one Hubble time. For the definitions of $L_{\rm c}$ and $P_{\rm g}$, $P_{\gamma}$, we use $G\mu = 10^{-11}$ and $\mathcal{I} = 10^4 \,\, \textrm{GeV}$.}
	\label{tab:table}
	\begin{tabular}{lccr} % four columns, alignment for each
		\hline
		   & Particle Physics & SI & Astrophysics\\
		\hline
		$\mu$ & $1.5 \times 10^{27} \,\, \textrm{GeV}^2$ & $1.4 \times 10^{16} \,\, \textrm{kg/m}$ & $ 208 \,\, M_{\sun}/\textrm{pc}$ \\
		$\mathcal{I}$ & $10^4 \,\, \textrm{GeV}$ & $8.0 \times 10^9 \,\, \textrm{Amps}$ & ---------------- \\
		$P_{\rm g} \simeq P_{\gamma}$  & $3.0 \times 10^{18} \,\, \textrm{GeV}^2$ & $7.3 \times 10^{32}$ Watts & $7.3 \times 10^{39} \,\, \textrm{erg/s}$\\
            $L_{\rm c}$  & $2.4 \times 10^{33} \,\, \textrm{GeV}^{-1}$ & $4.7 \times 10^{17} \,\, \textrm{m}$ & $15.4 \,\, \textrm{pc}$\\
		\hline
	\end{tabular}
\end{table}

%%%%%%%%%%%%%%%%%%%%%%%%%%%%%%%%%%%%%%%%%%%%%%%%%%

% Don't change these lines
\bsp	% typesetting comment
\label{lastpage}
\end{document}